\documentclass[11pt,a4paper]{article}
\def\dlev{0}
\usepackage[scaled]{uarial}

\usepackage[T1]{fontenc}
\usepackage{minibox}
\usepackage{amsmath}
\usepackage{graphicx}
\usepackage{color}
\usepackage{soul}
\usepackage{subfigure}
\usepackage{setspace}
\usepackage[usenames,dvipsnames]{xcolor}
\usepackage{tikz}
\usepackage{setspace}
\usepackage[colorlinks=false,pagebackref=false,urlcolor=Emerald,citecolor=blue,linkcolor=red,hypertexnames=false]{hyperref}
\usepackage{xr}
\usepackage{rotating}
\ifnum\dlev=0
\usepackage[a4paper, includefoot, top=2.5cm, bottom=2.2cm, left=2.8cm, right=1.5cm]{geometry}

\newcommand{\ea}[1]{{}#1}
\newcommand{\ed}[1]{}
\newcommand{\lls}[1]{}
\newcommand{\lle}[1]{}
\fi\ifnum\dlev>0
\usepackage[normalem]{ulem}
\usepackage{soul}
\usetikzlibrary{calc}
\usetikzlibrary{decorations.pathmorphing}
\newcommand{\ea}[1]{\textcolor{red}{#1}}
\newcommand{\ed}[1]{\textcolor{red}{\sout{#1}}}
\newcommand{\lls}[1]{\linelabel{#1-start}}
\newcommand{\lle}[1]{\linelabel{#1-end}}
\fi
\newcommand{\ear}[2]{\lls{#1}\ea{#2}\lle{#1}}

\newcommand{\edge}[1][]{\mathbin{\tikz [baseline=-0.25ex,#1] \draw [-,#1] (0pt,0.4ex) -> (1.1em,0.4ex);}}
\newcommand{\noedge}{\hspace{1.6em}}
\newcommand{\subs}[1]{_{\mathrm{#1}}}
\newcommand{\sups}[1]{^{\mathrm{#1}}}

\newcommand{\eq}[1]{\begin{equation}#1\end{equation}}
\newcommand{\eqn}[1]{\eq{#1\nonumber}}
\newcommand{\eqa}[1]{\begin{eqnarray}#1\end{eqnarray}}
\newcommand{\refeq}[1]{Eq \ref{eq-#1}}

\newcommand{\reftab}[1]{\textbf{Table \ref{tab-#1}}}
\newcommand{\Reftab}[1]{\reftab{#1}}
\newcommand{\refig}[1]{\textbf{Figure \ref{fig-#1}}}
\newcommand{\Refig}[1]{\refig{#1}}
\newcommand{\refssec}[1]{\textbf{Section \ref{ssec-#1}}}
\newcommand{\Refssec}[1]{\refssec{#1}}
\newcommand{\Refsup}[1]{\textbf{Supplementary Material \ref{ssec-#1}}}
\newcommand{\cm}[1]{}
\newcommand*{\vcenteredhbox}[1]{\begingroup
\setbox0=\hbox{#1}\parbox{\wd0}{\box0}\endgroup}

\def\LLR{\mathrm{LLR}}
\def\H{{\mathcal H}}
\def\Hn{\H\subs{null}}
\def\Ha{\H\subs{alt}}
\def\var{\mathrm{Var}}
\def\D{{\mathcal D}}
\def\pkg{Findr}
\def\tnamea{relevance}
\def\tnameb{controlled}
\def\Tnamea{Relevance}
\def\Tnameb{Controlled}

\graphicspath{{figures/}{./}{../figures/}}
\def\titleorig{Efficient and accurate causal inference with hidden confounders from genome-transcriptome variation data}
\def\github{https://github.com/lingfeiwang/findr}
\def\authors{{\large
  Lingfei Wang and Tom Michoel\footnote{Corresponding author. Email: Tom.Michoel@roslin.ed.ac.uk}

  Division of Genetics and Genomics, The Roslin Institute, The University of Edinburgh, Easter Bush, Midlothian EH25 9RG, UK \\
}}

\usepackage{a4wide}
\usepackage[super,numbers,sort&compress]{natbib}
\usepackage{natbib}
\makeatletter

\renewcommand\@biblabel[1]{#1.}

\makeatother

\usepackage{lineno}

\begin{document}

\begin{spacing}{1}\LARGE \textbf{\titleorig}\\[-8mm]\end{spacing}

\authors

\section*{Abstract}

Mapping gene expression as a quantitative trait using whole genome-sequencing and transcriptome analysis allows to discover the functional consequences of genetic variation. We developed a novel method and ultra-fast software Findr for higly accurate causal inference between gene expression traits using cis-regulatory DNA variations as causal anchors, which improves current methods by taking into account hidden confounders and weak regulations. Findr outperformed existing methods on the DREAM5 Systems Genetics challenge and on the prediction of microRNA and transcription factor targets in human lymphoblastoid cells, while being nearly a million times faster. Findr is publicly available at \github.

\section*{Author summary}

Understanding how genetic variation between individuals determines variation in observable traits or disease risk is one of the core aims of genetics. It is known that genetic variation often affects gene regulatory DNA elements and directly causes variation in expression of nearby genes. This effect in turn cascades down to other genes via the complex pathways and gene interaction networks that ultimately govern how cells operate in an ever changing environment. In theory, when genetic variation and gene expression levels are measured simultaneously in a large number of individuals, the causal effects of genes on each other can be inferred using statistical models similar to those used in randomized controlled trials. We developed a novel method and ultra-fast software Findr which, unlike existing methods, takes into account the complex but unknown network context when predicting causality between specific gene pairs. Findr's predictions have a significantly higher overlap with known gene networks compared to existing methods, using both simulated and real data. Findr is also nearly a million times faster, and hence the only software in its class that can handle modern datasets where the expression levels of ten-thousands of genes are simultaneously measured in hundreds to thousands of individuals.

\section{Introduction}
Genetic variation in non-coding genomic regions, including at loci associated with complex traits and diseases identified by genome-wide association studies, predominantly plays a gene-regulato\-ry role\cite{albert2015role}. Whole genome and transcriptome analysis of natural populations has therefore become a common practice to understand how genetic variation leads to variation in phenotypes\cite{civelek2014systems}. The number and size of studies mapping genome and transcriptome variation has surged in recent years due to the advent of high-throughput sequencing technologies, and ever more expansive catalogues of expression-associated DNA variants, termed expression quantitative trait loci (eQTLs), are being mapped in humans, model organisms, crops and other species\cite{albert2015role,Lappalainen:2013,ardlie2015genotype,franzen2016}. Unravelling the causal hierarchies between DNA variants and their associated genes and phenotypes is now the key challenge to enable the discovery of novel molecular mechanisms, disease biomarkers or candidate drug targets from this type of data\cite{schadt2009,talukdar2016}.

It is believed that genetic variation can be used to infer the causal directions of regulation between coexpressed genes, based on the principle that genetic variation causes variation in nearby gene expression and acts as a causal anchor for identifying downstream genes \cite{rockman2008reverse,li2010critical}. Although numerous statistical models have been proposed for causal inference with genotype and gene expression data from matching samples\cite{schadt2005integrative, Chen:2007, aten2008using, Millstein:2009, neto2013modeling, Millstein:2016}, no software implementation in the public domain is efficient enough to handle the volume of contemporary datasets, hindering any attempts to evaluate their performances. Moreover, existing statistical models rely on a conditional independence test which assumes that no hidden confounding factors affect the coexpression of causally related gene pairs. However gene regulatory networks are known to exhibit redundancy\cite{gitter2009backup} and are organized into higher order network motifs\cite{alon2007}, suggesting that confounding of causal relations by known or unknown common upstream regulators is the rule rather than the exception. \ear{3-1-2}{Moreover, it is also known that the conditional independence test is susceptible to variations in relative measurement errors between genes \cite{rockman2008reverse,li2010critical}, an inherent feature of both microarray and RNA-seq based expression data\cite{ritchie2015limma}.}

To investigate and address these issues, we developed \pkg{} (Fast Inference of Networks from Directed Regulations), an ultra-fast software package that incorporates existing and novel statistical causal inference tests. The novel tests were designed to take into account the presence of unknown confounding effects, and were evaluated systematically against multiple existing methods using both simulated and real data.

\section{Results}

\subsection{Findr incorporates existing and novel causal inference tests}

Findr performs six likelihood ratio tests involving pairs of genes (or exons or transcripts) $A$, $B$, and an eQTL $E$ of $A$ (\Reftab{tests}, \Refssec{lrt}). \pkg{} then calculates Bayesian posterior probabilities of the hypothesis of interest being true based on the observed likelihood ratio test statistics (denoted $P_i$, $i=0$ to $5$, $0\leq P_i\leq 1$, \Refssec{bi}). For this purpose, \pkg{} utilizes newly derived analytical formulae for the null distributions of the likelihood ratios of the implemented tests (\Refssec{null}, \Refig{eg4}). This, together with efficient programming, resulted in a dramatic speedup compared to the standard computationally expensive approach of generating random permutations. The six posterior probabilities are then combined into the traditional causal inference test, our new causal inference test, and separately a correlation test that does not incorporate genotype information (\Refssec{testse}). Each of these tests \ear{1-1-1}{verifies whether the data arose from a specific subset of ($E$, $A$, $B$) relations (\Reftab{tests}) among the full hypothesis space of all their possible interactions, and} results in a probability of a causal interaction $A\to B$ being true, which can be used to rank predictions according to significance or to reconstruct directed networks of gene regulations by keeping all interactions exceeding a probability threshold.

\subsection{Traditional causal inference fails in the presence of hidden confounders and weak regulations}
\pkg's computational speed allowed us to systematically evaluate traditional causal inference methods for the first time. We obtained five datasets with 999 samples simulated from synthetic gene regulatory networks of 1,000 genes with known genetic architecture from the DREAM5 Systems Genetics Challenge (\Refssec{datasets}), and subsampled each dataset to observe how performance depends on sample size (\Refssec{eval}). The correlation test ($P_0$) does not incorporate genotype information and was used as a benchmark for performance evaluations in terms of areas under the receiver operating characteristic (AUROC) and precision-recall (AUPR) curves (\Refssec{eval}). The traditional method\cite{Chen:2007} combines the secondary ($P_2$) and independence ($P_3$) tests sequentially (\Reftab{tests}, \Refssec{testse}), and was evaluated by comparing $P_2$ and $P_2P_3$ separately against the correlation test. Both the secondary test alone and the traditional causal inference test combination were found to \emph{underperform} the correlation test (\Refig{dcomb}A,B). Moreover, the inclusion of the conditional independence test \emph{worsened} inference accuracy, more so with increasing sample size (\Refig{dcomb}A,B) and increasing number of regulations per gene (\Refsup{consist}). Similar performance drops were also observed for the Causal Inference Test (CIT) \cite{Millstein:2009,Millstein:2016} software, which also is based on the conditional independence test (\Refig{cit}).

We believe that the failure of traditional causal inference is due to an elevated false negative rate (FNR) coming from two sources. First, the secondary test is less powerful in identifying weak interactions than the correlation test. In a true regulation $E\rightarrow A\rightarrow B$, the secondary linkage ($E\rightarrow B$) is the result of two direct linkages chained together, and is harder to detect than either of them. The secondary test hence picks up fewer true regulations, and consequently has a higher FNR. Second, the conditional independence test is counter-productive in the presence of hidden confounders (i.e. common upstream regulators). \lls{3-3-1}In such cases, even if $E\rightarrow A\rightarrow B$ is genuine,  the conditional independence test will find $E$ and $B$ to be still correlated after conditioning on $A$ \ea{due to a collider effect} (\Refig{indep})\ea{\cite{Cole:2010}}.\lle{3-3-1} Hence the conditional independence test only reports positive on $E\rightarrow A\rightarrow B$ relations without confounder, further raising the FNR. This is supported by the observation of worsening performance with increasing sample size (where confounding effects become more distinguishable) and increasing number of regulations per gene (which leads to more confounding).

\ear{3-1-1}{To further support this claim, we examined the inference precision among the top predictions from the traditional test, separately for gene pairs directly unconfounded or confounded by at least one gene (\Refssec{eval}). Compared to unconfounded gene pairs, confounded ones resulted in significantly more false positives among the top predictions (\refig{dcomb}C). Furthermore, the vast majority of real interactions fell outside the top 1\% of predictions (i.e.\ had small posterior probability)  [92\% (651/706) for confounded and 86\% (609/709) for unconfounded interactions,  \refig{dcomb}C]. Together, these results again showed the failure of the traditional test on confounded interactions and its high false negative rate overall.}

\subsection{\pkg{} accounts for weak secondary linkage, allows for hidden confounders, and outperforms existing methods on simulated data}

\begin{figure}[p]
\begin{center}
\begin{tabular}{p{0em}p{0.43\linewidth}p{0em}p{0.43\linewidth}}
\vspace{0pt}\textbf{{\large A}} &\vspace{0pt}\includegraphics[width=\linewidth]{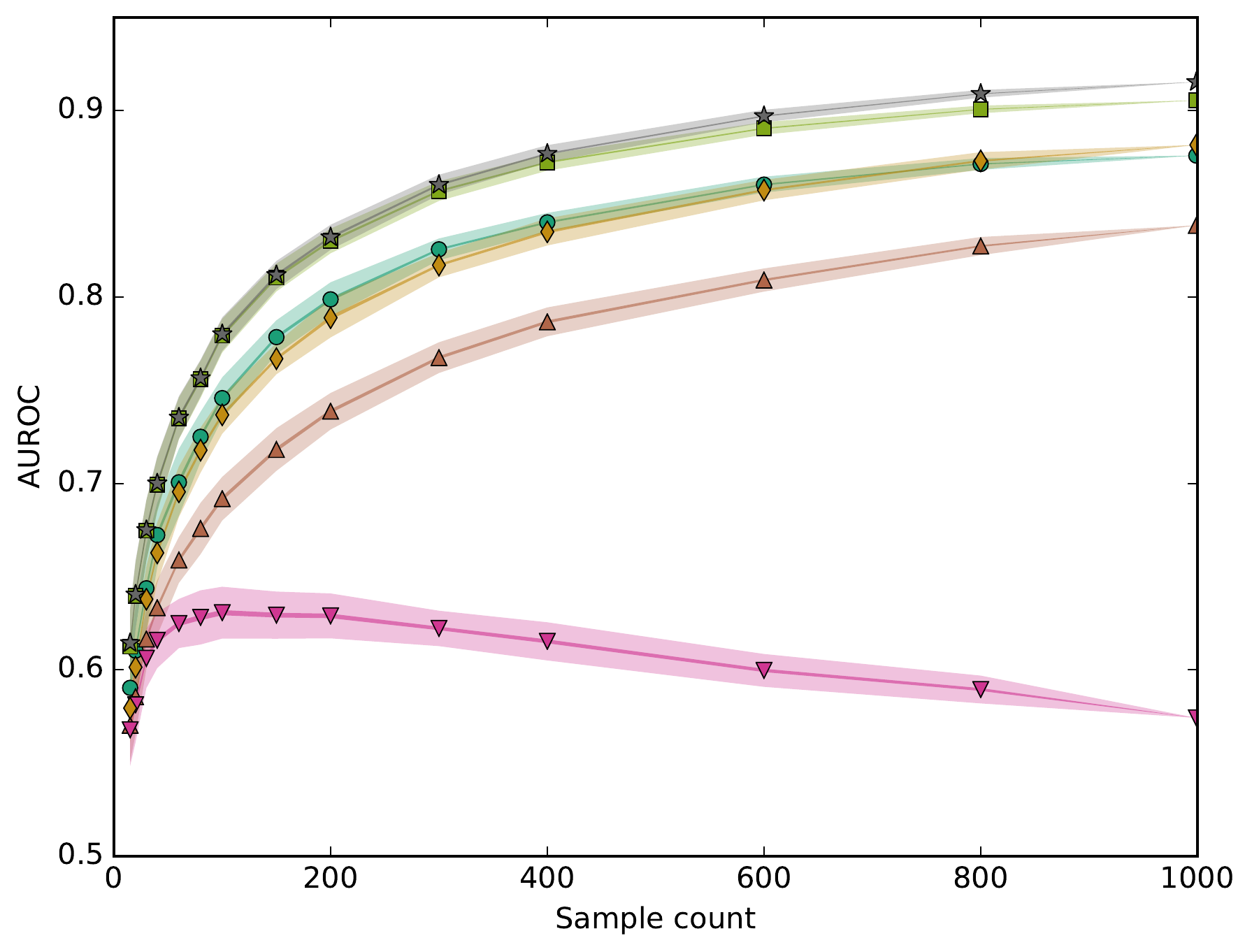}&
\vspace{0pt}\textbf{{\large B}} &\vspace{0pt}\includegraphics[width=\linewidth]{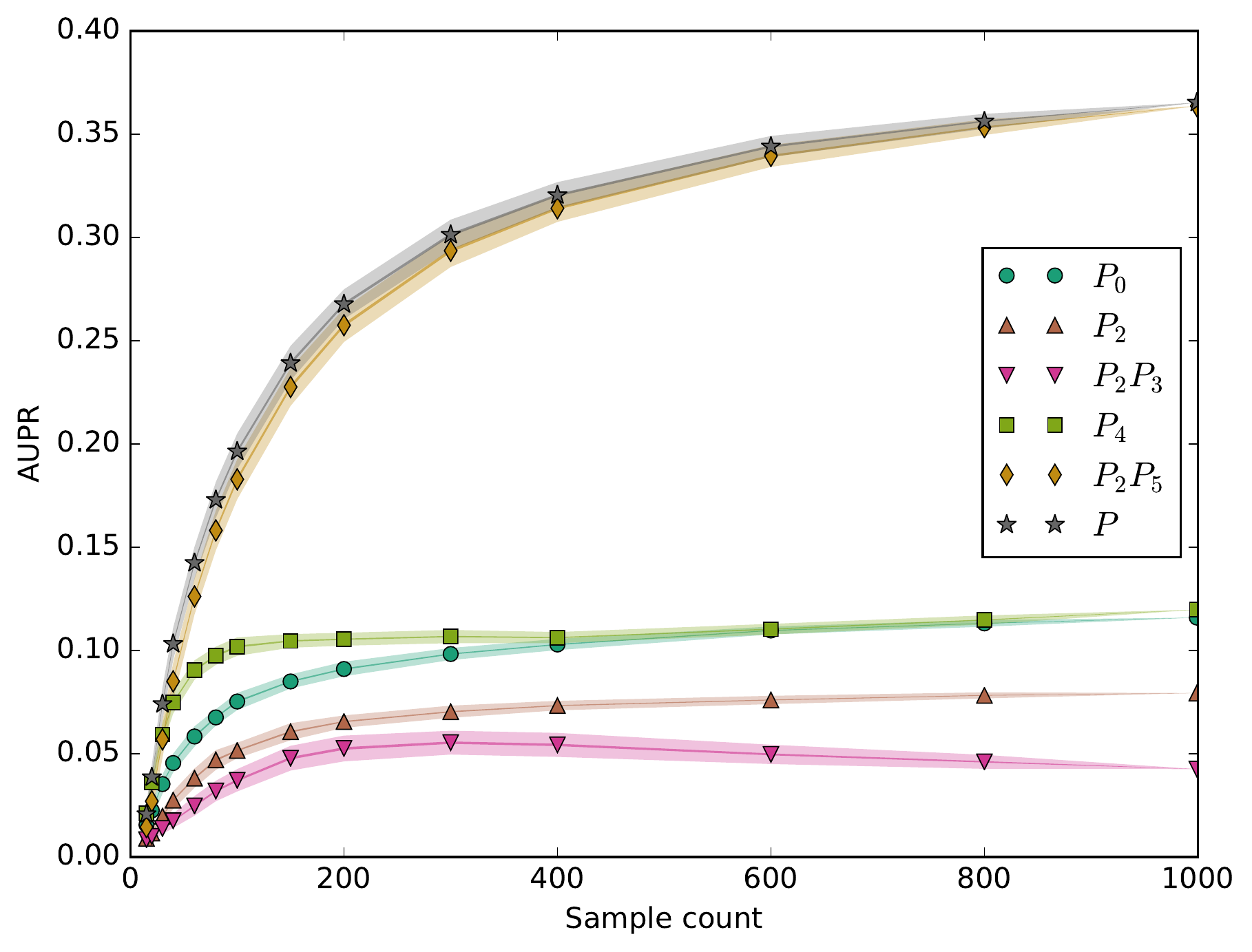}\\
\vspace{0pt}\textbf{{\large C}} &\vspace{0pt}\includegraphics[width=\linewidth]{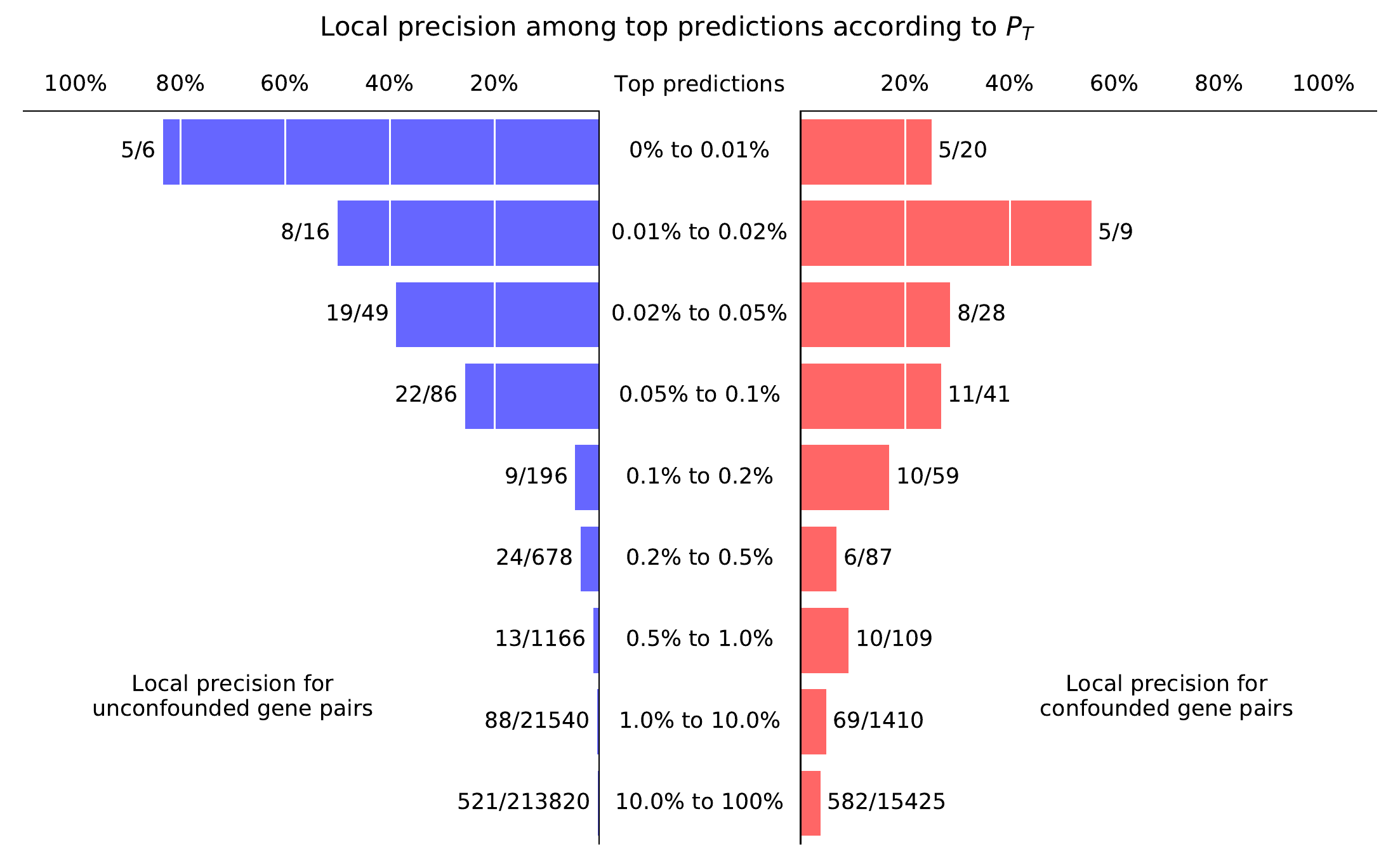}&
\vspace{0pt}\textbf{{\large D}} &\vspace{0pt}\includegraphics[width=\linewidth]{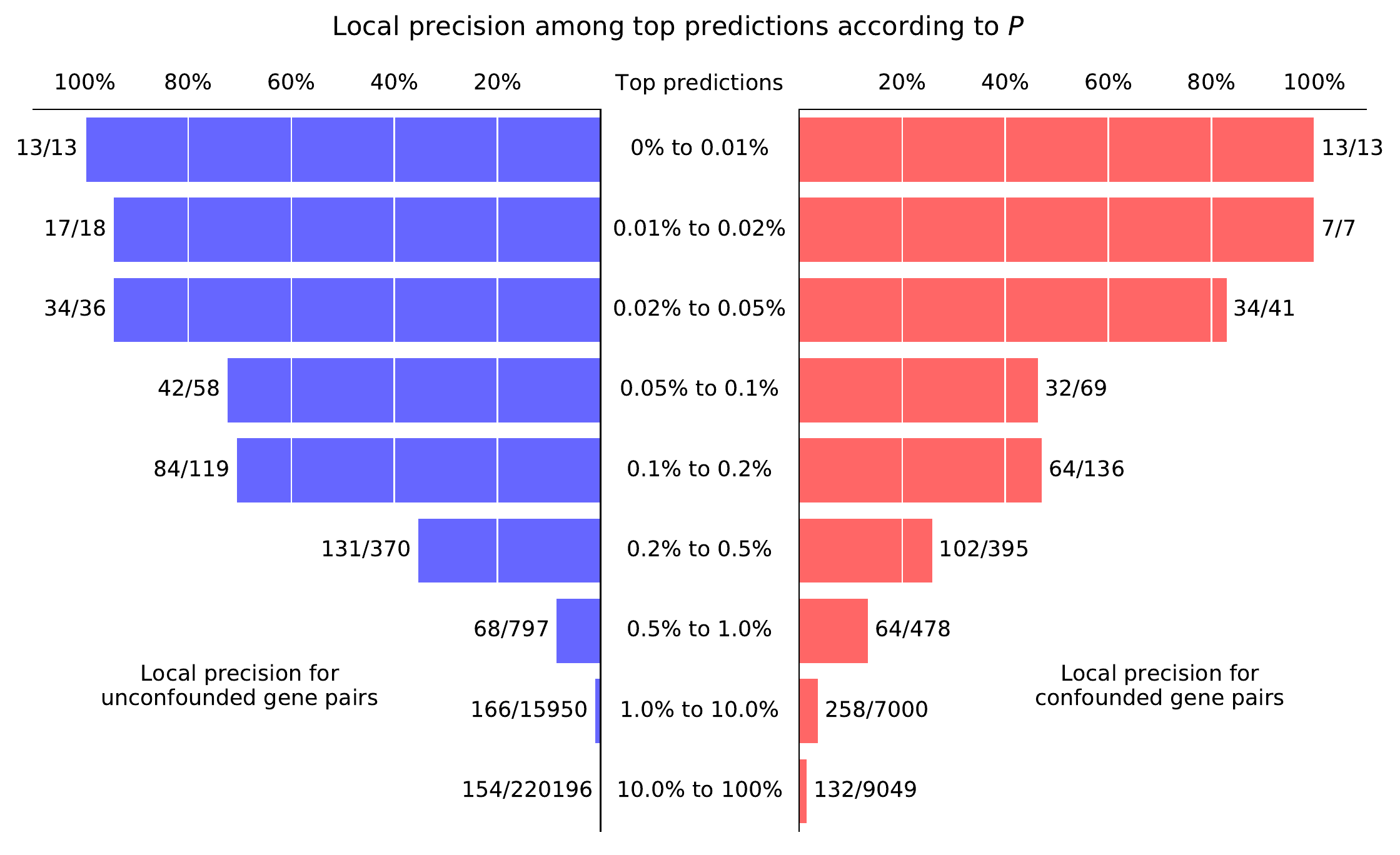}\\
\vspace{0pt}\textbf{{\large E}} &\vspace{0pt}\includegraphics[width=\linewidth]{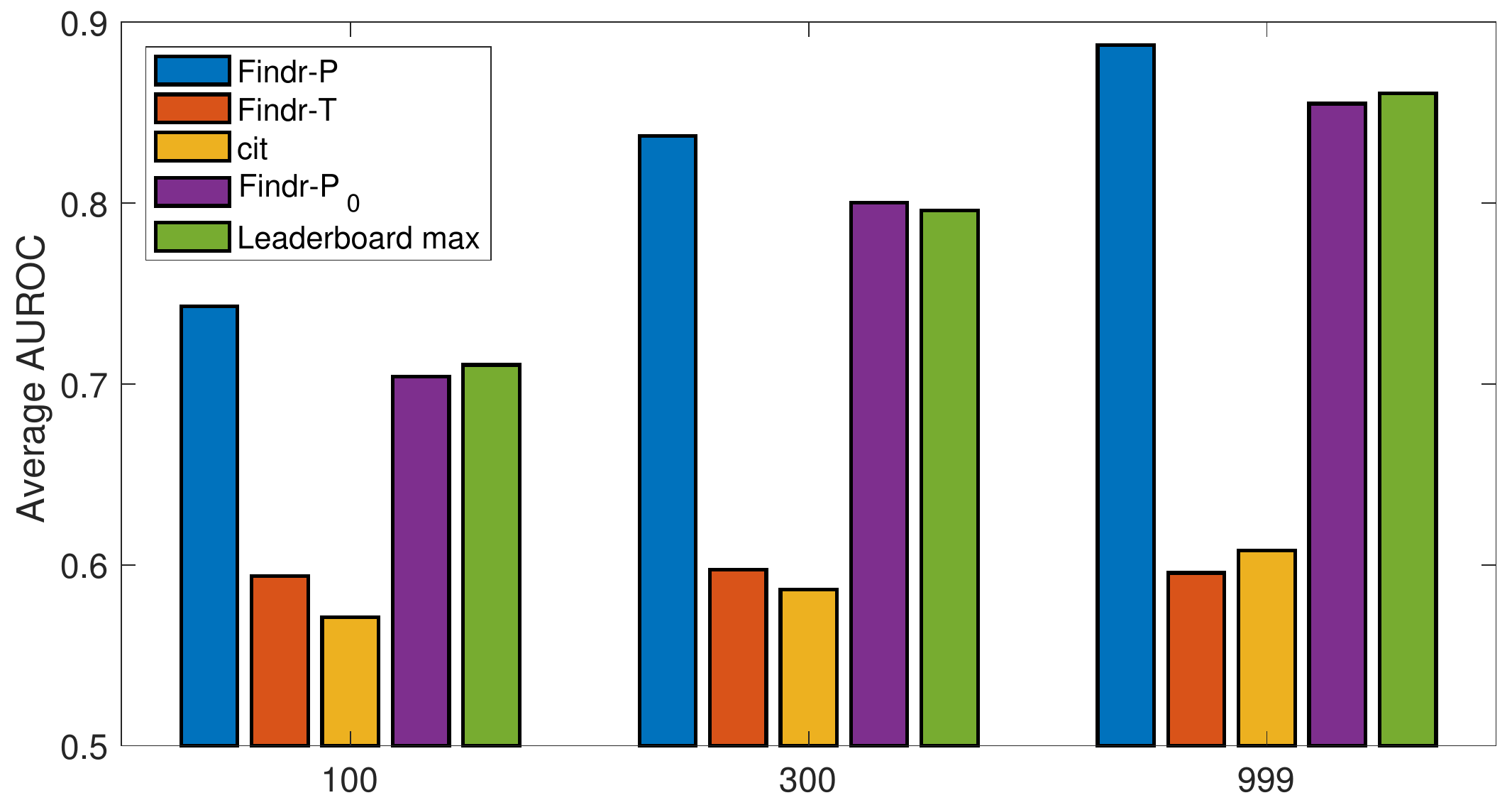}&
\vspace{0pt}\textbf{{\large F}} &\vspace{0pt}\includegraphics[width=\linewidth]{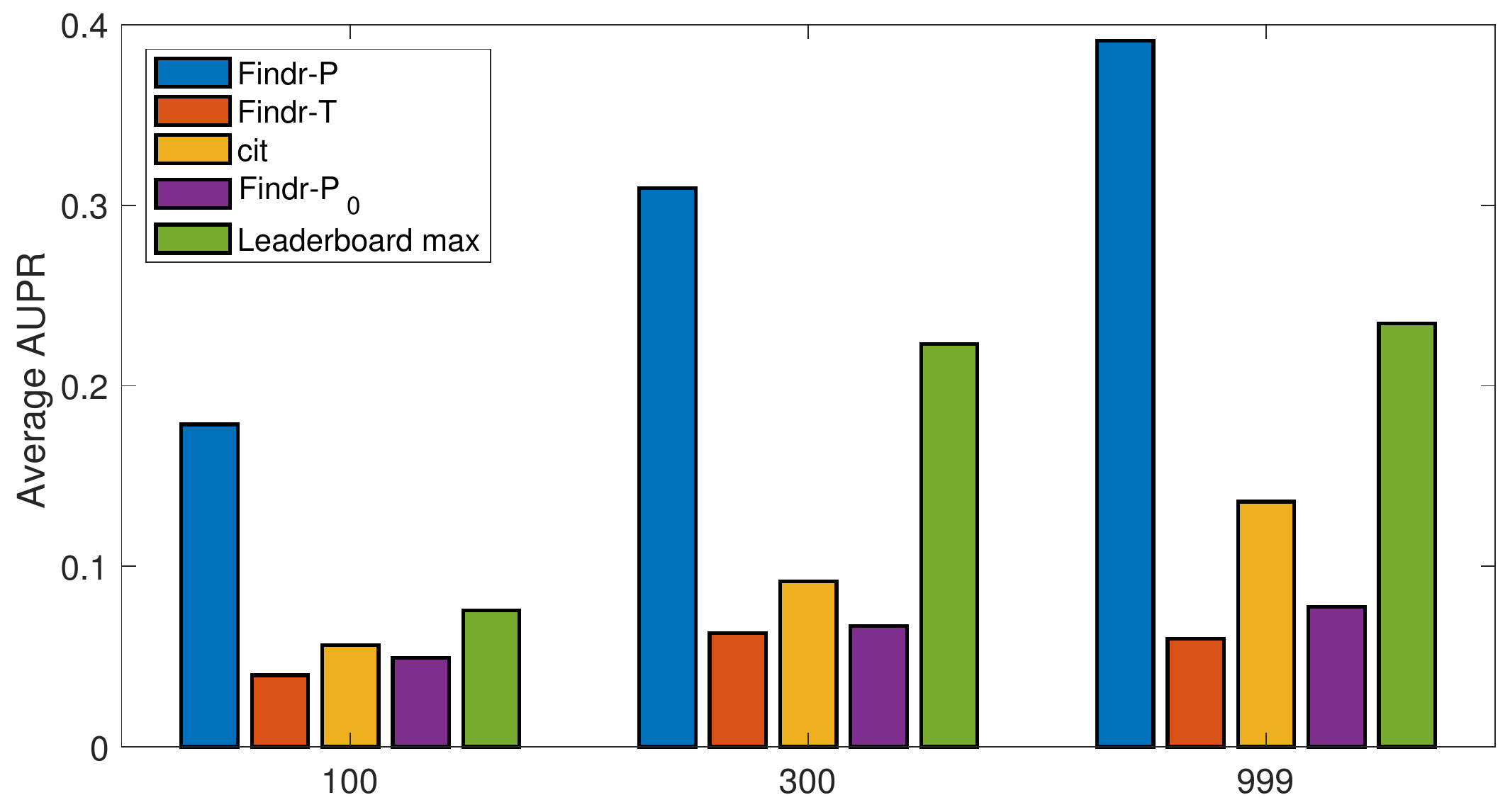}
\end{tabular}
\end{center}
\caption{\textbf{\pkg{} achieves best prediction accuracy on the DREAM5 Systems Genetics Challenge.} (\textbf{A}, \textbf{B}) The mean AUROC (\textbf{A})  and AUPR (\textbf{B}) on subsampled data are shown for traditional ($P_2$, $P_2P_3$) and newly proposed ($P_4$, $P_2P_5$, $P$) causal inference tests against the baseline correlation test ($P_0$). Every marker corresponds to the average AUROC or AUPR at specific sample sizes. Random subsampling at every sample size was performed 100 times. Half widths of the lines and shades are the standard errors and standard deviations respectively. $P_i$ corresponds to test $i$ numbered in \Reftab{tests}; $P$ is the new composite test (\Refssec{testse}). This figure is for dataset 4 of \ea{the} DREAM challenge. For results on other datasets of the same challenge, see \Refig{dcomb-other}. \ea{(\textbf{C}, \textbf{D}) Local precision of top predictions for the traditional (\textbf{C}) and novel (\textbf{D}) tests for dataset 4 of the DREAM challenge. Numbers next to each bar ($x/y$) indicate the number of true regulations ($x$) and the total number of gene pairs ($y$) within the respective range of prediction scores. For results on other datasets, see \refig{cmpbars}. } (\textbf{E}, \textbf{F}) The average AUROC (\textbf{E}) and AUPR (\textbf{F}) over 5 DREAM datasets with respectively 100, 300 and 999 samples are shown for \pkg{}'s new (Findr-$P$), traditional (Findr-$P_T$), and correlation (Findr-$P_0$) tests, for CIT and for the best scores on the DREAM challenge leaderboad. For individual results on all 15 datasets, see \Reftab{dream}. \label{fig-dcomb}}
\end{figure}

To overcome the limitations of traditional causal inference methods, \pkg{} incorporates two additional tests (\Reftab{tests} and \Refssec{lrt}). The \tnamea{} test ($P_4$) verifies that $B$ is not independent from $A$ and $E$ simultaneously and is more sensitive for picking up weak secondary linkages than the secondary linkage test. The \tnameb{} test ($P_5$) ensures that the correlation between $A$ and $B$ cannot be fully explained by $E$, i.e.\ excludes pleiotropy.  The same subsampling analysis revealed that $P_4$ performed best in terms of AUROC, and AUPR with small sample sizes, whilst the combination $P_2P_5$ achieved highest AUPR for larger sample sizes (\Refig{dcomb}A,B). Most importantly, both tests consistently outperformed the correlation test ($P_0$), particularly for AUPR. This demonstrates conclusively in a comparative setting that the inclusion of genotype data indeed can improve regulatory network inference. These observations are consistent across all five DREAM datasets (\Refig{dcomb-other}).

We combined the advantages of $P_4$ and $P_2P_5$ by averaging them in a composite test ($P$) (\Refssec{testse}), which outperformed $P_4$ and $P_2P_5$ at all sample sizes (\Refig{dcomb} and \Refig{dcomb-other}) and hence was appointed as Findr's new test for causal inference. Findr's new test ($P$) \ea{obtained consistently higher levels of local precision (i.e.\ one minus local FDR) on confounded and unconfounded gene pairs compared to \pkg's traditional causal inference test ($P_T$) (\refig{dcomb}C,D), and} outperformed \ea{the}\ed{\pkg's} traditional \ed{causal inference }test ($P_T$), correlation test ($P_0$), CIT, and every participating method of the DREAM5 Systems Genetics Challenge (\Refssec{datasets}) in terms of AUROC and AUPR on all 15 datasets (\Refig{dcomb}E,F, \Reftab{dream}\ea{, \Refig{dreampr}}). 

\ear{3-1-5}{Specifically, \pkg's new test was able to address the inflated FNR of the traditional method due to confounded interactions. It performed almost equally well on confounded and unconfounded gene pairs and, compared to the traditional test, significantly fewer real interactions fell outside the top 1\% of predictions (55\% vs.\ 92\% for confounded and 45\% vs.\ 86\% for unconfounded interactions, \refig{dcomb}D, \refig{cmpbars}).}

\subsection{\ea{The conditional independence test incurs false negatives for unconfounded regulations due to measurement error}\label{ssec-condfn}}

\ea{The traditional causal inference method based on the conditional indepedence test results in false negatives for confounded interactions, whose effect was shown signficant for the simulated DREAM datasets. However, the traditional test surprisingly reported more confounded gene pairs than the new test in its top predictions (albeit with lower precision), and correspondingly fewer unconfounded gene pairs (\Refig{dcomb}C,D, \refig{cmpbars}).}

\ea{We hypothesized that this inconsistency originated from yet another source of false negatives, where measurement error can confuse the conditional independence test. Measurement error in an upstream variable (called $A$ in \Reftab{tests}) does not affect the expression levels of its downstream targets, and hence a more realistic model for gene regulation is $E\rightarrow A\sups{(t)}\rightarrow B$ with $A\sups{(t)}\rightarrow A$, where the measured quantities are $E$, $A$, and $B$, but the true value for $A$, noted $A\sups{(t)}$, remains unknown. When the measurement error (in $A\sups{(t)}\rightarrow A$) is significant, conditioning on $A$ instead of $A\sups{(t)}$ cannot remove all the correlation between $E$ and $B$ and would therefore report false negatives for unconfounded interactions as well. This effect has been previously studied, for example in epidemiology as the ``spurious appearance of odds-ratio heterogeneity''\cite{Greenland:1980}.}

\begin{figure}[p]
\begin{center}
\begin{tabular}{p{0em}p{0.43\linewidth}p{0em}p{0.43\linewidth}}
\vspace{0pt}\textbf{{\large A}} &\vspace{0pt}\includegraphics[width=\linewidth]{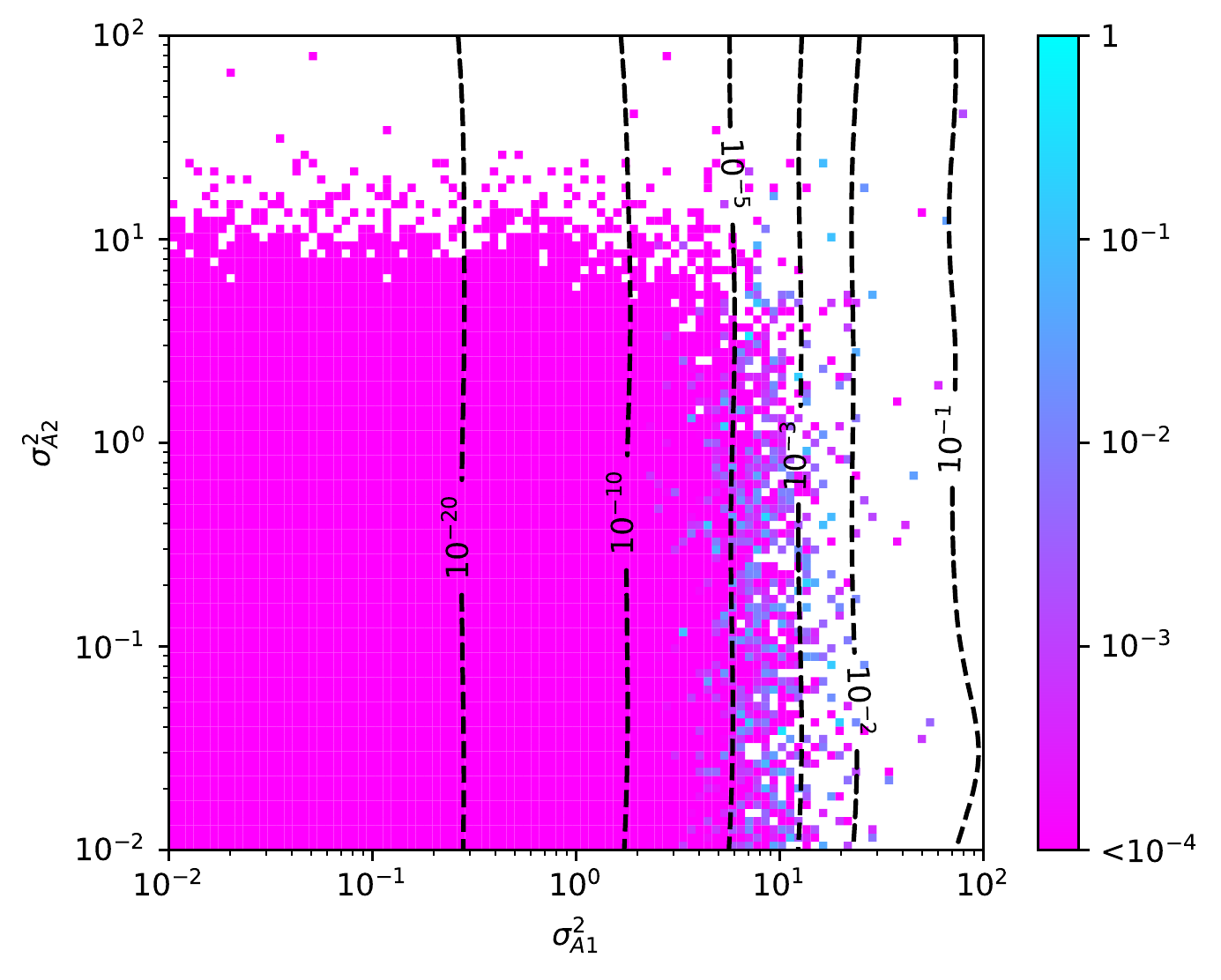}&
\vspace{0pt}\textbf{{\large B}} &\vspace{0pt}\includegraphics[width=\linewidth]{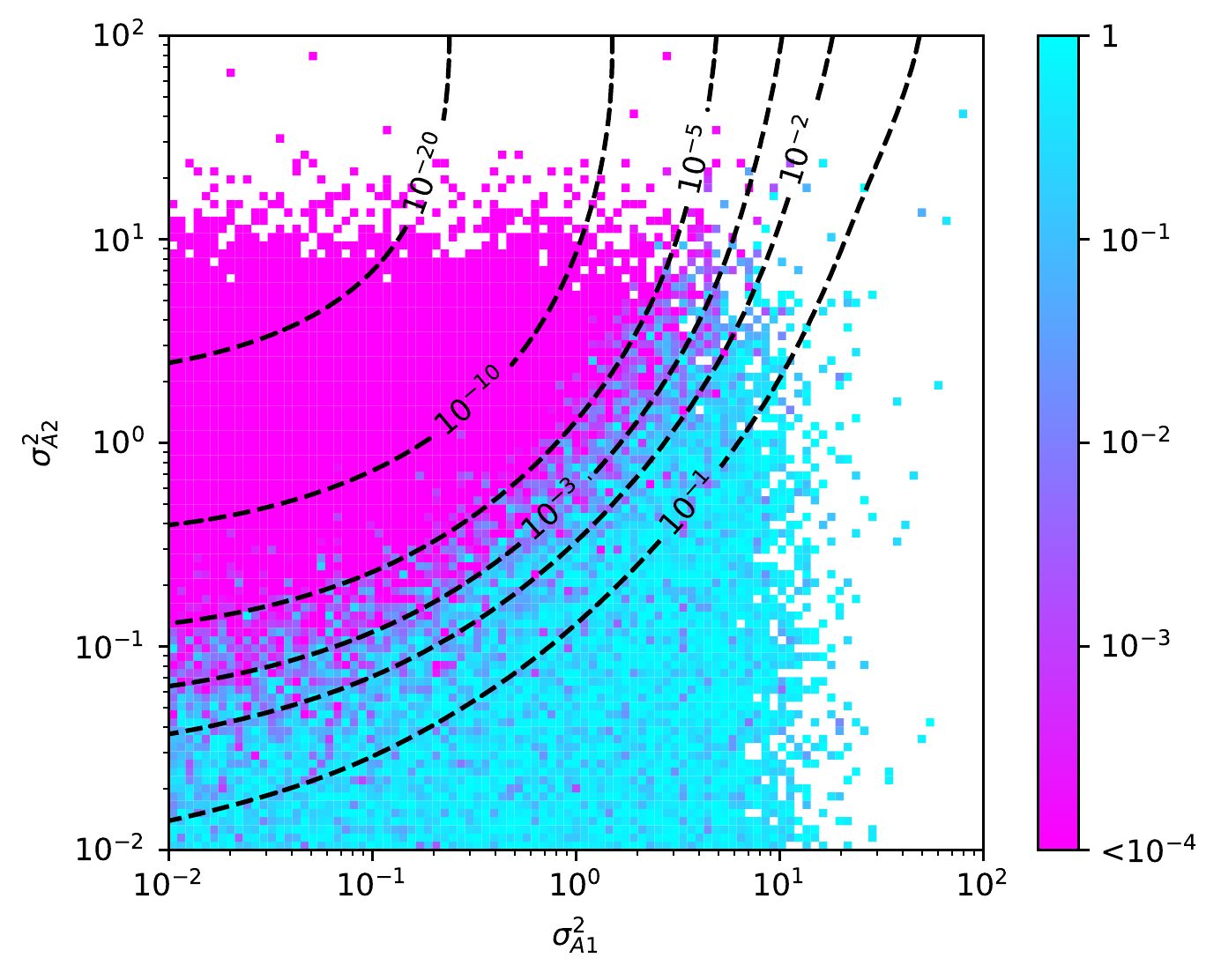}\\
\vspace{0pt}\textbf{{\large C}} &\vspace{0pt}\includegraphics[width=\linewidth]{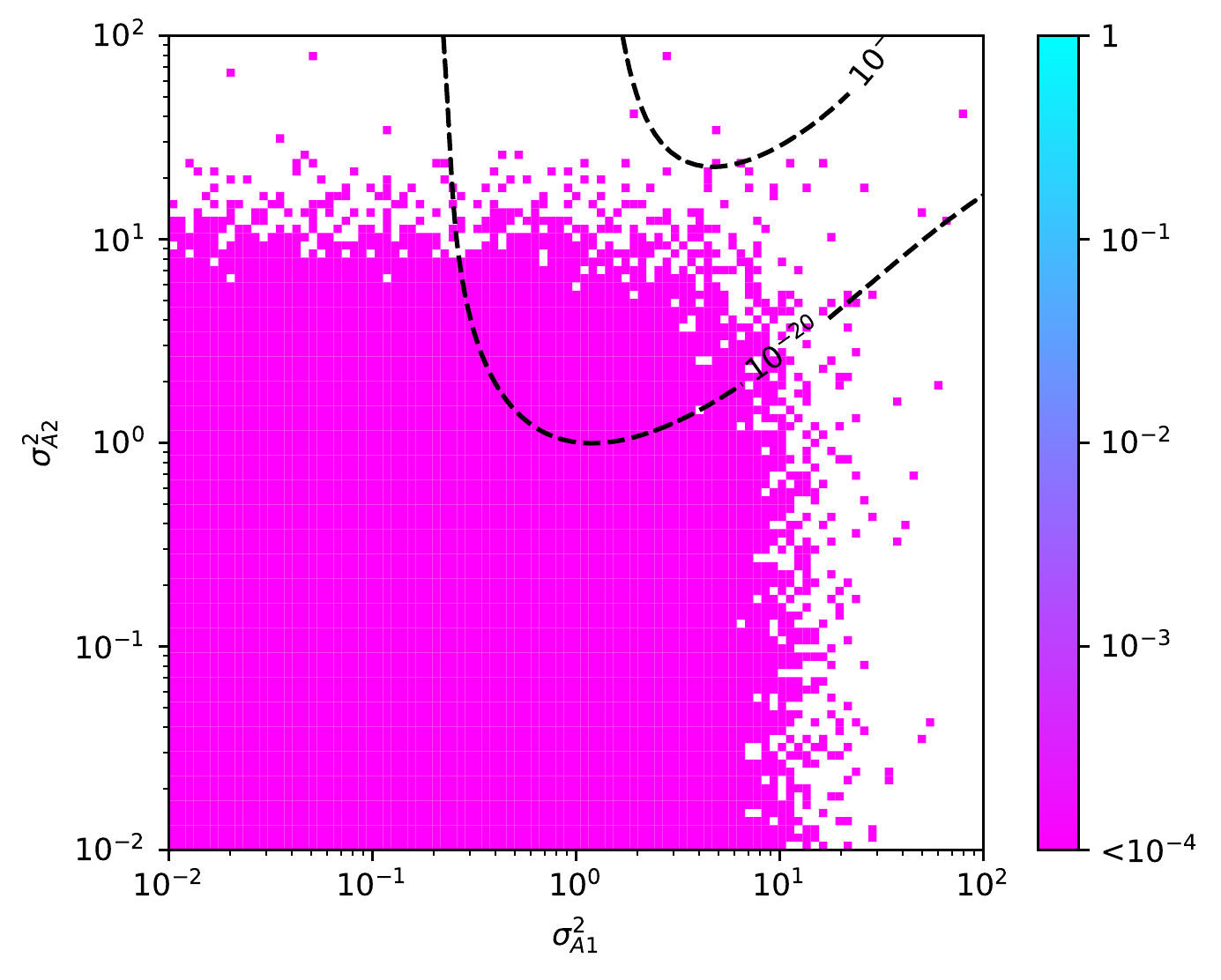}&
\vspace{0pt}\textbf{{\large D}} &\vspace{0pt}\includegraphics[width=\linewidth]{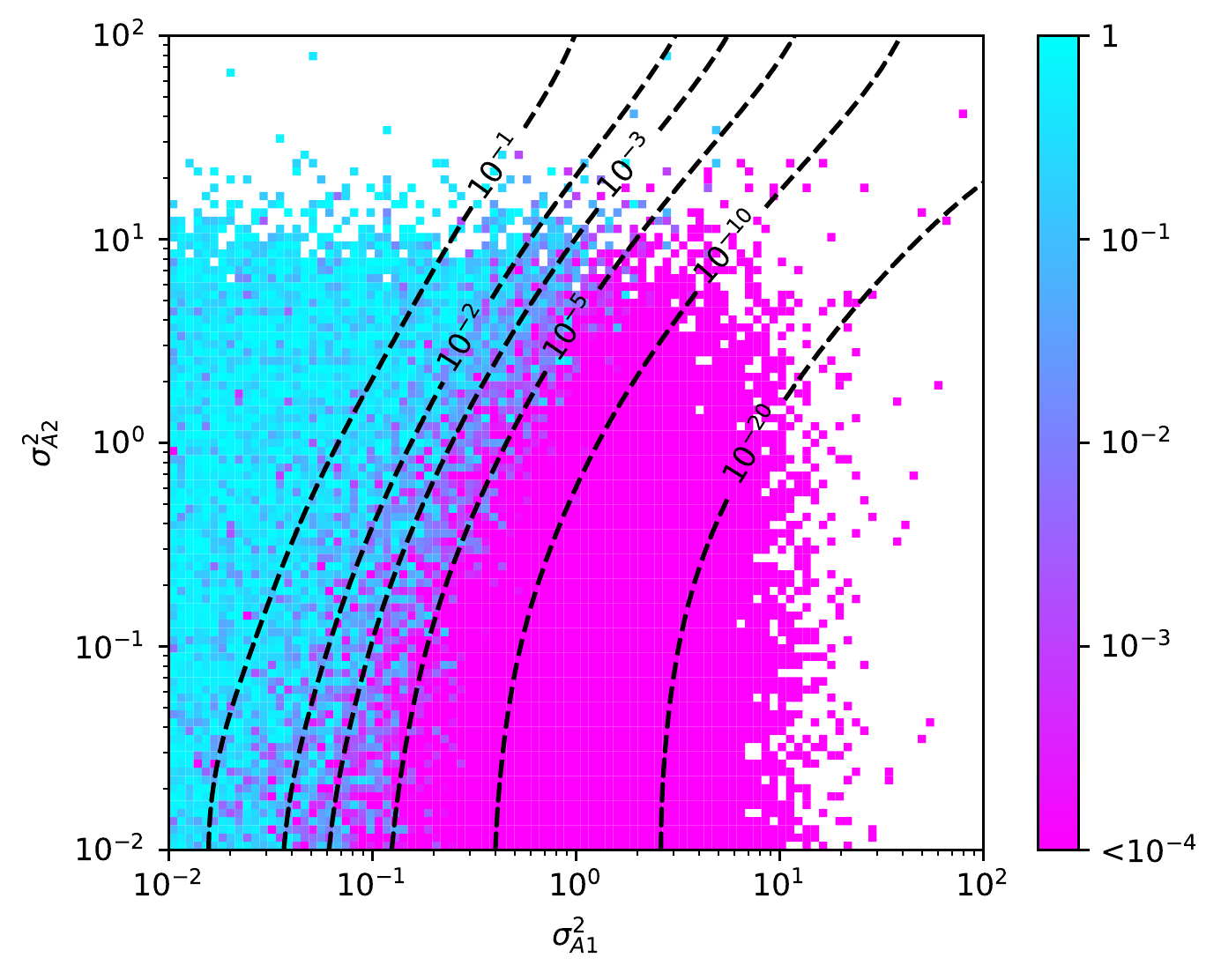}
\end{tabular}
\end{center}
\caption{\ea{\textbf{The conditional independence test yields false negatives for unconfounded regulations in the presence of even minor measurement errors.} Null hypothesis p-values of the secondary linkage (\textbf{A}), conditional independence (\textbf{B}), relevance (\textbf{C}), and controlled (\textbf{D}) tests are shown on simulated data from the ground truth model $E\rightarrow A\sups{(t)}\rightarrow B$ with $A\sups{(t)}\rightarrow A$. $A\sups{(t)}$'s variance coming from $E$ is set to one, $\sigma_{A1}^2$ is $A\sups{(t)}$'s variance from other sources and $\sigma_{A2}^2$ is the variance due to measurement noise. A total of 100 values from $10^{-2}$ to $10^2$ were taken for $\sigma_{A1}^2$ and $\sigma_{A2}^2$ to form the $100\times100$ tiles. Tiles that did not produce a significant eQTL relation $E\rightarrow A$ with p-value $\le10^{-6}$ were discarded. Contour lines are for the log-average of smoothened tile values. Note that for the conditional independence test (\textbf{B}), the true model corresponds to the null hypothesis, i.e.\ small (purple) p-values correspond to \emph{false negatives}, whereas for the other tests the true model corresponds to the alternative hypothesis, i.e.\ small (purple) p-values correspond to \emph{true positives}  (cf.\ \reftab{tests}). For details of the simulation and results from other parameter settings, see \refssec{sim} and \refig{sims} respectively.}\label{fig-sim}}
\end{figure}

\ea{We verified our hypothesis with a simple simulation (\Refssec{sim}). In a typical scenario with 300 samples from a monoallelic species, minor allele frequency 0.1, and a third of the total variance of $B$ coming from $A\sups{(t)}$, the conditional independence test reported false negatives (likeilihood ratio p-value $\ll1$, i.e.\ rejecting the null hypothesis of conditional indepencence, cf.\ \Reftab{tests}) as long as measurement error contributed more than half of $A$'s total unexplained variance (\refig{sim}B). False negatives occurred at even weaker measurement errors, when the sample sizes were larger or when stronger $A\rightarrow B$ regulations were assumed (\refig{sims}).}

\ea{This observation goes beyond the well-known problems that arise from a large measurement error in all variables, which acts like a hidden confounder\cite{li2010critical}, or from a much larger measurement error in $A$ than $B$, which can result in $B$ becoming a better measurement of $A\sups{(t)}$ than $A$ itself \cite{rockman2008reverse}. In this simulation, the false negatives persisted even if $E\rightarrow A$ was observationally much stronger than $E\rightarrow B$, such as when $A$'s measurement error was only $10\%$ ($\sigma_{A1}^2=0.1$) compared to up to $67\%$ for $B$ (\refig{sim}B). This suggested a unique and mostly neglected source of false negatives that would not affect other tests. Indeed, the secondary, relevance, and controlled tests were much less sensitive to measurement errors (\refig{sim}A,C,D).}

\subsection{\pkg{} outperforms traditional causal inference and machine learning methods on microRNA target prediction}

\begin{figure}[p]
  \centering
  \begin{minipage}[t]{.7\linewidth}
    \textbf{A}\\
    \includegraphics[width=\linewidth]{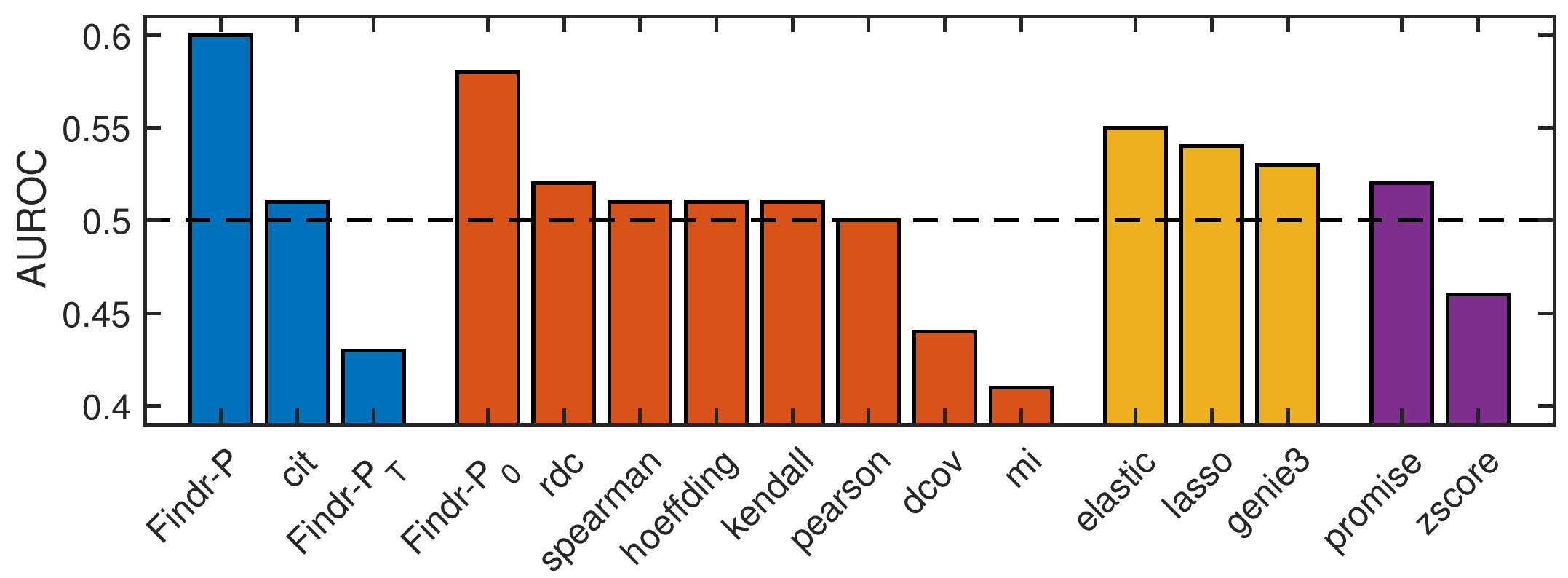}
     \textbf{B}\\
     \includegraphics[width=\linewidth]{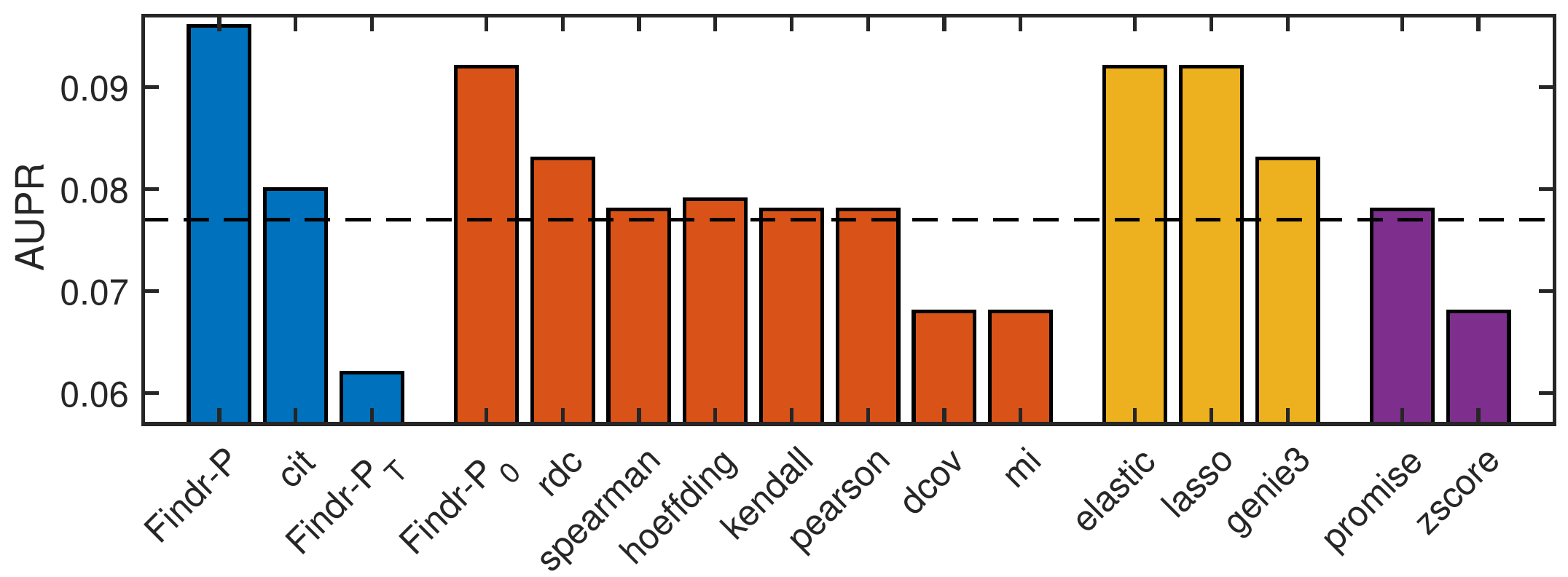}
     \textbf{C}\\
     \includegraphics[width=\linewidth]{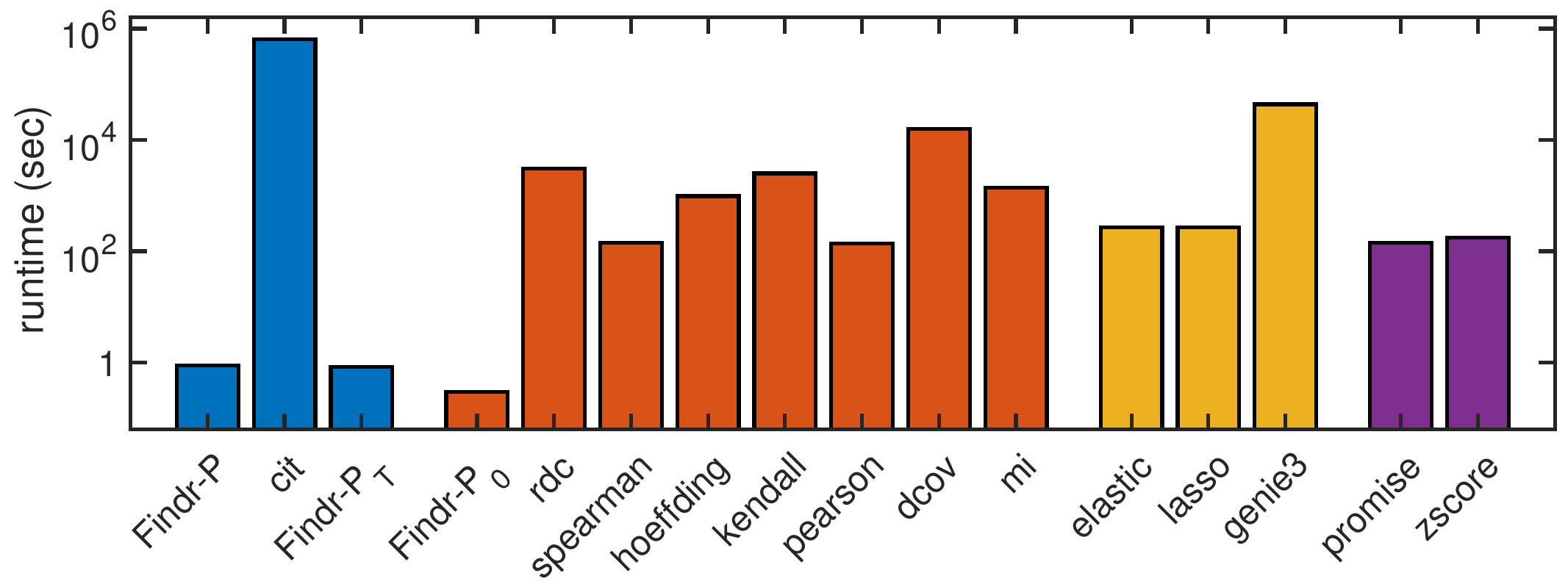}
  \end{minipage}\hfill
  \caption{\textbf{\pkg{} achieves highest accuracy and speed on the prediction of miRNA target genes from the Geuvadis data.} Shown are the AUROC (\textbf{A}), AUPR (\textbf{B}) and runtime (\textbf{C}) for 16 miRNA target prediction methods. Methods are colored by type: blue, genotype-assisted causal inference methods; red, pairwise correlation methods; yellow, multivariate regression methods; purple, other methods. Dashed lines are the AUROC and AUPR from random predictions. For method details, see \Refsup{vag-sup}.}
  \label{fig-mirna}
\end{figure}

In order to evaluate \pkg{} on a real dataset, we performed causal inference on miRNA and mRNA sequencing data in lymphoblastoid cell lines from 360 European individuals in the Geuvadis study \cite{Lappalainen:2013} (\Refssec{datasets}). We first tested 55 miRNAs with reported significant cis-eQTLs against 23,722 genes. Since miRNA target predictions from sequence complimentarity alone result in high numbers of false positives, prediction methods based on correlating miRNA and gene expression profiles are of great interest \cite{Huang:2007}. Although miRNA target prediction using causal inference from genotype and gene expression data has been considered\cite{Su:2011}, it remains unknown whether the inclusion of genotype data improves existing expression-based methods.  To compare \pkg{} against the state-of-the-art for expression-based miRNA target prediction, we used miRLAB, an integrated database of experimentally confirmed human miRNA target genes with a uniform interface to predict targets using twelve methods, including linear and non-linear, pairwise correlation and multivariate regression methods \cite{Le:2015}.  We were able to infer miRNA targets with 11/12 miRLAB methods, and also applied the GENIE3 random forest regression method \cite{Huynh-Thu:2010}, CIT, and the three tests in \pkg: the new ($P$) and traditional ($P_T$) causal inference tests and the correlation test ($P_0$) (\Refsup{vag-sup}). \pkg's new test achieved highest AUROC and AUPR among the 16 methods attempted\lls{1-4-1}\lls{2-m4-1}\ea{. In particular, \pkg's new test significantly outperformed the traditional test and CIT, the two other genotype-assisted methods}, while also being over 500,000 times faster \ea{than CIT}  (\Refig{mirna}, \Reftab{Geuvadis}, \Refig{Geuvadis}). \lls{2-2-1}\ea{\pkg's correlation test outperformed all other methods not using genotype information, including correlation, regression, and random forest methods, and was 500 to 100,000 times faster (\Refig{mirna}, \Reftab{Geuvadis}, \Refig{Geuvadis}). This further illustrates the power of the Bayesian gene-specific background estimation method implemented in all \pkg's tests (\refssec{bi}).}\lle{1-4-1}\lle{2-2-1}\lle{2-m4-1}

\subsection{\pkg{} predicts transcription factor targets with more accurate FDR estimates}

We considered 3,172 genes with significant cis-eQTLs in the Geuvadis data \cite{Lappalainen:2013} (\Refssec{datasets}) and inferred regulatory interactions to the 23,722 target genes using \pkg's traditional ($P_T$), new ($P$) and correlation ($P_0$) tests, and CIT. Groundtruths of experimentally confirmed causal gene interactions in human, and mammalian systems more generally, are of limited availability and mainly concern transcription or transcription-associated DNA-binding factors (TFs). Here we focused on a set of 25 TFs in the set of eQTL-genes for which either differential expression data following siRNA silencing (6 TFs) or TF-binding data inferred from ChIP-sequencing and/or DNase footprinting (20 TFs) in a lymphoblastoid cell line (GM12878) was available \cite{cusanovich2014functional} (\Refssec{datasets}). AUPRs and AUROCs did not exhibit substantial differences, other than modest improvement over random predictions (\Refig{bar}). To test for enrichment of true positives among the top-ranked predictions, which would be missed by global evaluation measures such as AUPR or AUROC, we took advantage of the fact that \pkg's probabilities are empirical local precision \ed{(i.e.\ one minus local FDR)} estimates for each test (\Refssec{bi}), and assessed how estimated local precisions of new, traditional, and correlation tests reflected the actual precision. Findr's new test correctly reflected the precision values at various threshold levels, and was able to identify true regulations at high precision control levels (\Refig{LCL}). However, the traditional test significantly underestimated precision due to its elevated FNR. This lead to a lack of predictions at high precision thresholds but enrichment of true regulations at low thresholds, essentially nullifying the statistical meaning of its output probability $P_T$. On the other hand, the correlation test significantly overestimated precisions because it is unable to distinguish causal, reversed causal or confounded interactions, which raises its FDR. The same results were observed when alternative groundtruth ChIP-sequencing networks were considered (\Refig{bar}, \Refig{encode}).

\section{Discussion}
We developed a highly efficient, scalable software package \pkg{} (Fast Inference of Networks from Directed Regulations) implementing novel and existing causal inference tests. Application of Findr on real and simulated genome and transcriptome variation data showed that our novel tests, which account for weak secondary linkage and hidden confounders \ea{at the potential cost of an increased number of false positives}, resulted in a significantly improved performance to predict known gene regulatory interactions \ea{compared to existing methods, particularly traditional methods based on conditional independence tests, which had highly elevated false negative rates}.

\ea{\lls{1-3-1}Causal inference using eQTLs as causal anchors relies on crucial assumptions which have been discussed in-depth elsewhere \cite{li2010critical,rockman2008reverse}. Firstly, it is assumed that genetic variation is always causal for variation in gene expression, or quantitative traits more generally, and is independent of any observed or hidden confounding factors. Although this assumption is valid for randomly sampled individuals, caution is required when this is not the case (e.g.\ case-control studies). Secondly, measurement error is assumed to be independent and comparable across variables. Correlated measurement error acts like a confounding variable, whereas a much larger measurement error in the source variable $A$ than the target variable $B$ may lead to an inversion of the inferred causal direction. \lls{3-1-3} The conditional independence test in particular relies on the unrealistic assumptions that hidden confounders and measurement errors are absent, the violation of which incurs false negatives and a failure to correctly predict causal relations, as shown throughout this paper.\lle{1-3-1}\lle{3-1-3}}

\lls{1-3-2}\ed{Among the six likelihood ratio tests performed in \pkg{}, given the eQTL variable, only the conditional independence test is asymmetric between the two genes considered (\Reftab{tests}). The}\ea{Although the} new\ea{ly proposed} test avoids the elevated FNR from the conditional independence test, \ea{it is not without its own limitations. Unlike the conditional independence test, the relevance and controlled tests (\Reftab{tests}) are symmetric between the two genes considered.} \ea{Therefore the} \ed{and therefore its} direction of causality \ea{in the new test} arises predominantly from using a different eQTL when testing the reverse interaction, potentially leading to a higher FDR as a minor trade-off.\lle{1-3-2} \ea{About 10\% of } \ed{Coexpressed genes typically do not lie close to each other on the genome and hence their} cis-regulatory eQTLs \ed{will indeed be different and not linked.} \ea{are linked (as \textit{cis}-eQTLs) to the expression of more than one gene\cite{tong2017shared}.}  In the\ea{se} \ea{cases, it appears} \ed{rare instances where coexpressed genes do share cis-eQTLs, it is more likely} that the shared \ea{\textit{cis}-}eQTL regulates the genes independently\ea{\cite{tong2017shared}}, which \ed{is assessed} \ea{in \pkg\ is accounted for }by the `controlled' test (\Reftab{tests}). When causality between genes and phenotypes or among phenotypes is tested, sharing or linkage of (e)QTLs can be more \ed{frequent} \ea{common}. Resolving causality in these cases may require the use of \pkg{}'s conservative, traditional causal inference test, in conjunction with the new test.

In this paper we have addressed the challenge of pairwise causal inference, but to reconstruct the actual pathways and networks that affect a phenotypic trait, two important limitations have to be considered. First, linear pathways propagate causality, and may thus appear as densely connected sets of triangles in pairwise causal networks. Secondly, most genes are regulated by multiple upstream factors, and hence some true edges may only have a small posterior probability unless they are considered in an appropriate multivariate context. The most straightforward way to address these issues would be to model the real directed interaction network as a Bayesian network with sparsity constraints. A major advantage of Findr is that it outputs probability values which can be directly incorporated as prior edge probabilities in existing network inference softwares.

In conclusion, \pkg{} is a highly efficient and accurate open source software tool for causal inference from large-scale genome-transcriptome variation data.  Its nonparametric nature ensures robust performances across datasets without parameter tuning, with easily interpretable output in the form of accurate precision and FDR estimates. \pkg{} is able to predict causal interactions in the context of complex regulatory networks where unknown upstream regulators confound traditional conditional independence tests, and more generically in any context with discrete or continuous causal anchors.

\section{Methods}

\subsection{Datasets\label{ssec-datasets}}
We used the following datasets/databases for evaluating causal inference methods:
\begin{enumerate}
\item Simulated genotype and transcriptome data of synthetic gene regulatory networks from the DREAM5 Systems Genetics challenge A (DREAM for short) \cite{DREAM:5}, generated by the SysGenSIM software \cite{Pinna:2011}. DREAM provides 15 sub-datasets, obtained by simulating 100, 300, and 999 samples of 5 different networks each, containing 1000 genes in every sub-dataset but more regulations for sub-datasets with higher numbering. In every sub-dataset, each gene has exactly one matching genotype variable.  $25\%$ of the genotype variables are cis-expression Quantitative Trait Loci (eQTL), defined in DREAM as: their variation changes the expression level of the corresponding gene directly. The other $75\%$ are trans-eQTLs, defined as: their variation affects the expression levels of only the \emph{downstream targets} of the corresponding gene, but not the gene itself. Because the identities of cis-eQTLs are unknown, we calculated the P-values of genotype-gene expression associations with kruX \cite{Qi:2014}, and kept all genes with a P-value less than $1/750$ to filter out genes without cis-eQTL. For the subsampling analysis (detailed in \refssec{eval}), we restricted the evaluation to the prediction of target genes from these cis-genes only, in line with the assumption that Findr and other causal inference methods require as input a list of genes whose expression is significantly associated with at least one cis-eQTL. For the full comparison of Findr to the DREAM leaderboard results, we predicted target genes for all genes, regardless of whether they had a cis-eQTL.

\item Genotype and transcriptome sequencing data on 465 human lymphoblastoid cell line samples from the Geuvadis project \cite{Lappalainen:2013} consisting of the following data products:
\begin{itemize}
\item Genotype data \ea{(ArrayExpress accession E-GEUV-1)}\cite{GEUV:1}.
\item Gene quantification data for 23722 genes from nonredundant unique samples and after quality control and normalization \ea{(ArrayExpress accession E-GEUV-1)}\cite{GEUV:2}.
\item Quantification data of miRNA, with the same standard as gene quantification data \ea{(ArrayExpress accession E-GEUV-2)}\cite{GEUV:3}.
\item Best eQTLs of mRNAs and miRNAs \ea{(ArrayExpress accessions E-GEUV-1 and E-GEUV-2)}\cite{GEUV:4,GEUV:5}.
\end{itemize}

We restricted our analysis to 360 European samples which are shared by gene and miRNA quantifications. Excluding invalid eQTLs from the Geuvadis analysis, such as single-valued genotypes, 55 miRNA-eQTL pairs and 3172 gene-eQTL pairs were retained.

\item For validation of predicted miRNA-gene interactions, we extracted  the ``strong'' ground-truth table from miRLAB \cite{Le:2015,miRLAB:1}, which contains experimentally confirmed miRNA-gene regulations from the following databases: TarBase \cite{Vergoulis:2012}, miRecords \cite{Xiao:2009}, miRWalk \cite{Dweep:2011}, and miRTarBase \cite{Hsu:2014}. The intersection of the Geuvadis and ground-truth table contains 20 miRNAs and 1054 genes with 1217 confirmed regulations, which are considered for prediction validation. Interactions that are present in the ground-truth table are regarded as true while others as false.

\item For verification of predicted gene-gene interactions, we obtained differential expression data following siRNA silencing of 59 transcription-associated factors (TFs) and DNA-binding data of 201 TFs for 8872 genes in a reference lymphoblastoid cell line (GM12878) from \cite{cusanovich2014functional}. Six siRNA-targeted TFs, 20 DNA-binding TFs, and 6,790 target genes without missing differential expression data intersected with the set of 3172 eQTL-genes and 23722 target genes in Geuvadis and were considered for validation. We reproduced the pipeline of \cite{cusanovich2014functional} with the criteria for true targets as having a False Discovery Rate (FDR) $<0.05$ from R package \emph{qvalue} for differential expression in siRNA silencing, or having at least 2 TF-binding peaks within 10kb of their transcription start site. We also obtained the filtered proximal TF-target network from \cite{gerstein2012architecture}, which had 14 TFs and 7,000 target genes in common with the Geuvadis data.
\end{enumerate}

\subsection{General inference algorithm\label{ssec-algorithm}}
Consider a set of observations sampled from a mixture distribution of a null and an alternative hypothesis. For instance in gene regulation, every observation can correspond to expression levels of a pair of genes wich are sampled from a bivariate normal distribution with zero (null hypothesis) or non-zero (alternative hypothesis) correlation coefficient. In \pkg{}, we predict the probability that any sample follows the alternative hypothesis with the following algorithm (based on and modified from \cite{Chen:2007}):
\begin{enumerate}
\item \lls{3-8-1}For robustness against outliers, we convert every continuous variable into standard normally distributed $N(0,1)$ values using a rank-based inverse normal transformation across all samples. We name this step as \emph{supernormalization}.\lle{3-8-1}
\item We propose a null and an alternative hypothesis for every likelihood ratio test (LRT) of interest where, by definition, the null hypothesis space is a subset of the alternative hypothesis. Model parameters are replaced with their maximum likelihood estimators (MLEs) to obtain the log likelihood  ratio (LLR) between the  alternative and null hypotheses (\refssec{lrt}).
\item We derive the analytical expression for the probablity density function (PDF) of the LLR when samples follow the null hypothesis (\refssec{null}).
\item We convert LLRs into posterior probabilities of the hypothesis of interest with the empirical estimation of local FDR\cm{. This is achieved by first estimating the proportion of null hypothesis and then performing Bayesian inference} (\refssec{bi}).
\end{enumerate}
Implementational details can be found in \pkg's source code.

\subsection{Likelihood ratio tests\label{ssec-lrt}}

Consider correlated genes $A$, $B$, and a third variable $E$ upstream of $A$ and $B$, such as a significant eQTL of $A$. The eQTLs can be obtained either \textit{de novo} using eQTL identification tools such as matrix-eQTL \cite{Shabalin:2012} or kruX \cite{Qi:2014}, or from published analyses. Throughout this article, we assume that $E$ is a significant eQTL of $A$, whereas extension to other data types is straightforward. We use $A_i$ and $B_i$ for the expression levels of gene $A$ and $B$ respectively, which are assumed to have gone through the supernormalization in \refssec{algorithm}, and optionally the genotypes of the best eQTL of $A$ as $E_i$, where $i=1,\dots,n$ across samples. Genotypes are assumed to have a total of $n_a$ alleles, so \ea{$E_i\in\{0,\dots,n_a\}$}\ed{$E_i=0,\dots,n_a$}. We define the null and alternative hypotheses for a total of six tests, as shown in \reftab{tests}. LLRs of every test are calculated separately as follows:
\begin{enumerate}\setcounter{enumi}{-1}
\item\textbf{Correlation test:} Define the null hypothesis as $A$ and $B$ are independent, and the alternative hypothesis as they are correlated:
\eq{\Hn\sups{(0)}=A\noedge B,\hspace{4em}\Ha\sups{(0)}=A\edge B.}
The superscript $(0)$ is the numbering of the test. Both hypotheses are modeled with gene expression levels following bivariate normal distributions, as
\eqn{\left(\begin{array}{c}A_i\\B_i\end{array}\right)\sim N\left(\left(\begin{array}{c}0\\0\end{array}\right),\left(\begin{array}{cc}\sigma_{A0}^2&\rho\,\sigma_{A0}\sigma_{B0}\\\rho\,\sigma_{A0}\sigma_{B0}&\sigma_{B0}^2\end{array}\right)\right),}
for $i=1,\dots,n$. The null hypothesis corresponds to $\rho=0$.

Maximum likelihood estimators (MLE) for the model parameters $\rho$, $\sigma_{A0}$, and $\sigma_{B0}$ are
\eq{\hat\rho=\frac{1}{n}\sum_{i=1}^nA_iB_i,\hspace{2em}\hat\sigma_{A0}=\hat\sigma_{B0}=1,\label{eq-cor-rho}}
and the LLR is simply
\eq{\LLR\sups{(0)}=-\frac{n}{2}\ln(1-\hat\rho^2).\label{eq-cor-llr}}
In the absence of genotype information, we use nonzero correlation between $A$ and $B$ as the indicator for $A\rightarrow B$ regulation, giving the posterior probability
\eqn{P(A\edge B)=P(\Ha\sups{(0)}\mid\LLR\sups{(0)}).}
false negative
\item\textbf{Primary (linkage) test:} Verify that $E$ regulates $A$ from $\Ha\sups{(1)}\equiv E\rightarrow A$ and $\Hn\sups{(1)}\equiv E\noedge A$. For $\Ha\sups{(1)}$, we model $E\rightarrow A$ as $A$ follows a normal distribution whose mean is determined by $E$ categorically, i.e.
\eq{A_i\mid E_i\sim N(\mu_{E_i},\sigma_A^2).\label{eq-ba-hm-d1}}
From the total likelihood $p(A\mid E)=\prod_{i=1}^np(A_i\mid E_i)$, we find MLEs for model parameters $\mu_j,j=0,1,\dots,n_a$, and $\sigma_A$, as
\eqn{\hat\mu_j=\frac{1}{n_j}\sum_{i=1}^nA_i\delta_{E_ij},\hspace{2em}\hat\sigma_A^2=1-\sum_{j=0}^{n_a}\frac{n_j}{n}\hat\mu_j^2,}
where $n_j$ is the sample count by genotype category,
\eqn{n_j\equiv\sum_{i=1}^n\delta_{E_ij}.}
The Kronecker delta function is defined as $\delta_{xy}=1$ for $x=y$, and 0 otherwise. When summing over all genotype values ($j=0,\dots,n_a$), we only pick those that exist ($n_j>0$) throughout this article.  Since the null hypothesis is simply that $A_i$ is sampled from a genotype-independent normal distribution, with MLEs of mean zero and standard deviation one due to the supernormalization (\refssec{algorithm}), the LLR for test 1 becomes
\eq{\LLR\sups{(1)}=-\frac{n}{2}\ln\hat\sigma_A^2.\label{eq-ba-hm-llr1}}
By favoring a large $\LLR\sups{(1)}$, we select $\Ha\sups{(1)}$ and verify that $E$ regulates $A$, with
\eqn{P(E\rightarrow A)=P(\Ha\sups{(1)}\mid\LLR\sups{(1)}).}

\item\textbf{Secondary (linkage) test:} The secondary test is identical with the primary test, except it verifies that $E$ regulates $B$. Hence repeat the primary test on $E$ and $B$ and obtain the MLEs:
\eqn{\hat\nu_j=\frac{1}{n_j}\sum_{i=1}^nB_i\delta_{E_ij},\hspace{2em}\hat\sigma_B^2=1-\sum_{j=0}^{n_a}\frac{n_j}{n}\hat\nu_j^2,}
and the LLR as
\eqn{\LLR\sups{(2)}=-\frac{n}{2}\ln\hat\sigma_B^2.}
$\Ha\sups{(2)}$ is chosen to verify that $E$ regulates $B$.

\item\textbf{(Conditional) independence test:} Verify that $E$ and $B$ are independent when conditioning on $A$. This can be achieved by comparing $\Ha\sups{(3)}\equiv B\leftarrow E\rightarrow A\wedge(A\mathrm{\ correlates\ with\ }B)$ against $\Hn\sups{(3)}\equiv E\rightarrow A\rightarrow B$. LLRs close to zero then prefer $\Hn\sups{(3)}$, and ensure that $E$ regulates $B$ only through $A$:
\eqn{P(E\perp B\mid A)=P(\Hn\sups{(3)}\mid\LLR\sups{(3)}).}
For $\Ha\sups{(3)}$, the bivariate normal distribution dependent on $E$ can be represented as
\eqn{\left.\left(\begin{array}{c}A_i\\B_i\end{array}\right)\right|\ E_i\sim N\left(\biggl(\begin{array}{c}\mu_{E_i}\\\nu_{E_i}\end{array}\biggr),\biggl(\begin{array}{cc}\sigma_A^2&\rho\sigma_A\sigma_B\\\rho\sigma_A\sigma_B&\sigma_B^2\end{array}\biggr)\right).}
For $\Hn\sups{(3)}$, the distributions follow \refeq{ba-hm-d1}, as well as
\eqn{B_i\mid A_i\sim N(\rho A_i,\sigma_B^2).}
Substituting parameters $\mu_j,\nu_j,\sigma_A,\sigma_B,\rho$ of $\Ha\sups{(3)}$ and $\mu_j,\rho,\sigma_A,\sigma_B$ of $\Hn\sups{(3)}$ with their MLEs, we obtain the LLR:
\eqa{\LLR\sups{(3)}&=&-\frac{n}{2}\ln\left(\hat\sigma_A^2\hat\sigma_B^2-(\hat\rho+\sigma_{AB}-1)^2\right)\nonumber\\
&&+\frac{n}{2}\ln\hat\sigma_A^2+\frac{n}{2}\ln(1-\hat\rho^2),\label{eq-ba-hm-llr3}}
where
\eqn{\sigma_{AB}\equiv1-\sum_{j=0}^{n_a}\frac{n_j}{n}\hat\mu_j\hat\nu_j,}
and $\hat\rho$ is defined in \refeq{cor-rho}.

\item \textbf{\Tnamea{} test}: Since the indirect regulation $E\rightarrow B$ tends to be weaker than any of its direct regulation components ($E\rightarrow A$ or $A\rightarrow B$), we propose to test $E\rightarrow A\rightarrow B$ with indirect regulation $E\rightarrow B$ as well as the direct regulation $A\rightarrow B$ for stronger distinguishing power on weak regulations. We define $\Ha\sups{(4)}\equiv E\rightarrow A\wedge E\rightarrow B\leftarrow A$ and $\Hn\sups{(4)}\equiv E\rightarrow A\noedge B$. This simply verifies that $B$ is not independent from both $A$ and $E$ simultaneously. In the alternative hypothesis, $B$ is regulated by $E$ and $A$, which is modeled as a normal distribution whose mean is additively determined by $E$ categorically and $A$ linearly, i.e.
\eqn{B_i\mid E_i,A_i\sim N(\nu_{E_i}+\rho A_i,\sigma_B^2).}

We can hence solve its LLR as
\eqn{\LLR\sups{(4)}=-\frac{n}{2}\ln\left(\hat\sigma_A^2\hat\sigma_B^2-(\hat\rho+\sigma_{AB}-1)^2\right)+\frac{n}{2}\ln\hat\sigma_A^2.}

\item \textbf{\Tnameb{} test}: Based on the positives of the secondary test, we can further distinguish the alternative hypothesis $\Ha\sups{(5)}\equiv B\leftarrow E\rightarrow A\wedge A\rightarrow B$ from the null $\Hn\sups{(5)}\equiv B\leftarrow E\rightarrow A$ to verify that $E$ does not regulate $A$ and $B$ independently. Its LLR can be solved as
\eqn{\LLR\sups{(5)}=-\frac{n}{2}\ln\left(\hat\sigma_A^2\hat\sigma_B^2-(\hat\rho+\sigma_{AB}-1)^2\right)+\frac{n}{2}\ln\hat\sigma_A^2\hat\sigma_B^2.}
\end{enumerate}

\subsection{Null distributions for the log-likelihood ratios\label{ssec-null}}
The null distribution of LLR, $p(\LLR\mid\Hn)$, may be obtained either by simulation or analytically. Simulation, such as random permutations from real data or the generation of random data from statistics of real data, can deal with a much broader range of scenarios in which analytical expressions are unattainable. However, the drawbacks are obvious: simulation can take hundreds of times longer than analytical methods to reach a satisfiable precision. Here we obtained analytical expressions of $p(\LLR\mid\Hn)$ for all the tests introduced above.

\begin{enumerate}\setcounter{enumi}{-1}
\item\textbf{Correlation test}: $\Hn\sups{(0)}=A\noedge B$ indicates no correlation between $A$ and $B$. Therefore, we can start from 
\eq{\tilde B_i\sim\mathrm{i.i.d\ }N(0,1).\label{eq-ncor-dist0}}
In order to simulate the supernormalization step, we normalize $\tilde B_i$ into $B_i$ with zero mean and unit variance as:
\eq{B_i\equiv\frac{\tilde B_i-\bar{\tilde B}_i}{\sigma_{\tilde B}},\hspace{3em}\bar{\tilde B}\equiv\frac{1}{n}\sum_{i=1}^n\tilde B_i,\hspace{3em}\sigma_{\tilde B}^2\equiv\frac{1}{n}\sum_{i=1}^n\left(\tilde B_i-\bar{\tilde B}\right)^2.\label{eq-null-norm}}

Transform the random variables $\{\tilde B_i\}$ by defining
\eqa{X_1&\equiv&\frac{1}{\sqrt n}\sum_{i=1}^nA_i\tilde B_i,\\
X_2&\equiv&\frac{1}{\sqrt n}\sum_{i=1}^n\tilde B_i,\\
X_3&\equiv&\left(\sum_{i=1}^n\tilde B_i^2\right)-X_1^2-X_2^2.}
Since $\tilde B_i\sim\mathrm{i.i.d\ }N(0,1)$ (according to \refeq{ncor-dist0}), we can easily verify that $X_1,X_2,X_3$ are independent, and
\eq{X_1\sim N(0,1),\hspace{3em}X_2\sim N(0,1),\hspace{3em}X_3\sim\chi^2(n-2).}
Expressing \refeq{cor-llr} in terms of $X_1,X_2,X_3$ gives
\eq{\LLR\sups{(0)}=-\frac{n}{2}\ln(1-Y),}
in which
\eq{Y\equiv\frac{X_1^2}{X_1^2+X_3}\sim\mathrm{Beta}\Bigl(\frac12,\frac{n-2}2\Bigr)}
follows the Beta distribution.

\lls{3-7-1}We define distribution $\D(k_1,k_2)$ as \ea{the distribution of a random variable $Z=-\frac12\ln(1-Y)$} for $Y\sim\mathrm{Beta}(k_1/2,k_2/2)$, i.e.
\eqn{\ea{Z=}-\frac{1}{2}\ln(1-Y)\sim\D(k_1,k_2).}
The probability density function (PDF) for $\ea{Z}\sim\D(k_1,k_2)$ can be derived as: for $z>0$,\lle{3-7-1}
\eq{p(z\mid k_1,k_2)=\frac{2}{B(k_1/2,k_2/2)}\left(1-e^{-2z}\right)^{(k_1/2-1)}e^{-k_2z},}
and for $z\le0$, $p(z\mid k_1,k_2)=0$. Here $B(a,b)$ is the Beta function. Therefore the null distribution for the correlation test is simply
\eq{\LLR\sups{(0)}/n\sim\D(1,n-2).}

\item\textbf{Primary test}: $\Hn\sups{(1)}=E\noedge A$ indicates no regulation from $E$ to $A$. Therefore, similarly with the correlation test, we start from $\tilde A_i\sim\mathrm{i.i.d\ }N(0,1)$ and normalize them to $A_i$ with zero mean and unit variance.

The expression of $\LLR\sups{(1)}$ then becomes:
\eqn{\LLR\sups{(1)}=-\frac{n}{2}\ln\left(1-\sum_{j=0}^{n_a}\frac{n_j}{n}\frac{\left(\hat{\tilde\mu}_j-\bar{\tilde A}\right)^2}{\sigma_{\tilde A}^2}\right),}
where
\eqn{\hat{\tilde \mu}_j\equiv\frac{1}{n_j}\sum_{i=1}^n\tilde A_i\delta_{E_ij}.}

For now, assume all possible genotypes are present, i.e.\ $n_j>0$ for $j=0,\dots,n_a$. Transform $\{\tilde A_i\}$ by defining
\eqa{X_j&\equiv&\sqrt{n_j}\,\hat{\tilde\mu}_j,\hspace{9em}\mathrm{for\ }j=0,\dots,n_a,\nonumber\\
X_{n_a+1}&\equiv&\Biggl(\sum_{i=1}^n\tilde A_i^2\Biggr)-\Biggl(\sum_{j=0}^{n_a}X_j^2\Biggr).}
Then we can similarly verify that $\{X_i\}$ are pairwise independent, and
\eqa{X_i&\sim& N(0,1),\ \mathrm{for\ }i=0,\dots,n_a,\nonumber\\
X_{n_a+1}&\sim&\chi^2(n-n_a-1).}

Again transform $\{X_i\}$ by defining independent random variables
\eqa{Y_1&\equiv&\sum_{j=0}^{n_a}\sqrt\frac{n_j}{n}\,X_j\sim N(0,1),\nonumber\\
Y_2&\equiv&\left(\sum_{j=0}^{n_a}X_j^2\right)-Y_1^2\sim\chi^2(n_a),\nonumber\\
Y_3&\equiv&X_{n_a+1}\sim\chi^2(n-n_a-1).\nonumber}
Some calculation would reveal
\eqn{\LLR\sups{(1)}=-\frac{n}{2}\ln\left(1-\frac{Y_2}{Y_2+Y_3}\right),}
i.e.
\eqn{\LLR\sups{(1)}/n\sim\D(n_a,n-n_a-1).}

To account for genotypes that do not show up in the samples, define $n_v\equiv\sum_{j\in\{j\mid n_j>0\}}1$ as the number of different genotype values across all samples. Then
\eq{\LLR\sups{(1)}/n\sim\D(n_v-1,n-n_v).\label{eq-ba-null-llr2}}

\item\textbf{Secondary test}: Since the null hypotheses and LLRs of primary and secondary tests are identical, $\LLR\sups{(2)}$ follows the same null distribution as \refeq{ba-null-llr2}.

\item\textbf{Independence test}: The independence test verifies if $E$ and $B$ are uncorrelated when conditioning on $A$, with $\Hn\sups{(3)}=E\rightarrow A\rightarrow B$. For this purpose, we keep $E$ and $A$ intact while randomizing $\tilde B_i$ according to $B$'s correlation with $A$:
\eqn{\tilde B_i\equiv\hat\rho A_i+\sqrt{1-\hat\rho^2}\,X_i,\hspace{2em}X_i\sim\mathrm{i.i.d\ }N(0,1).}
Then $\tilde B_i$ is normalized to $B_i$ according to \refeq{null-norm}. The null distribution of $\LLR\sups{(3)}$ can be obtained with similar but more complex computations from \refeq{ba-hm-llr3}, as
\eq{\LLR\sups{(3)}/n\sim\D(n_v-1,n-n_v-1).}

\item\textbf{\Tnamea{} test}: The null distribution of $\LLR\sups{(4)}$ can be obtained similarly by randomizing $B_i$ according to \refeq{ncor-dist0} and \refeq{null-norm}, as
\eqn{\LLR\sups{(4)}/n\sim\D(n_v,n-n_v-1).\label{eq-null4}}

\item\textbf{\Tnameb{} test}: To compute the null distribution for the \tnameb{} test, we start from
\eq{\tilde B_i=\hat\nu_{E_i}+\hat\sigma_B X_i,\hspace{2em}X_i\sim N(0,1),}
and then normalize $\tilde B_i$ into $B_i$ according to \refeq{null-norm}. Some calculation reveals the null distribution as
\eqn{\LLR\sups{(5)}/n\sim\D(1,n-n_v-1).}
\end{enumerate}

We verified our analytical method of deriving null distributions by comparing the analytical null distribution v.s.\ null distribution from permutation for the \tnamea{} test in \refssec{match}.

\subsection{Bayesian inference of posterior probabilities\label{ssec-bi}}
After obtaining the PDFs for the LLRs from real data and the null hypotheses, we can convert LLR values into posterior probabilities $P(\Ha\mid \LLR)$. We use a similar technique as in \cite{Chen:2007}, which itself was based on a more general framework to estimate local FDRs in genome-wide studies \cite{Storey:2003}. This framework assumes that the real distribution of a certain test statistic forms a mixture distribution of null and alternative hypotheses. After estimating the null distribution, either analytically or by simulation, it can be compared against the real distribution to determine the proportion of null hypotheses, and consequently the posterior probability that the alternative hypothesis is true at any value of the statistic.

To be precise, consider an arbitrary likelihood ratio test. The fundamental assumption is that in the limit $\LLR\rightarrow0^+$, all test cases come from the null hypothesis ($\Hn$), whilst as $\LLR$ increases, the proportion of alternative hypotheses ($\Ha$) also grows. 
The mixture distribution of real LLR values is assumed to have a PDF as
\eqn{p(\LLR)=P(\Hn)p(\LLR\mid\Hn)+P(\Ha)p(\LLR\mid\Ha).}
The priors $P(\Hn)$ and $P(\Ha)$ sum to unity and correspond to the proportions of null and alternative hypotheses in the mixture distribution. For any test $i=0,\dots,5$, Bayes' theorem then yields its posterior probability as
\eq{P(\Ha^{(i)}\mid\LLR^{(i)})=\frac{p(\LLR^{(i)}\mid\Ha^{(i)})}{p(\LLR^{(i)})}P(\Ha^{(i)}).}
Based on this, we can define the posterior probabilities of the selected hypotheses according to \reftab{tests}, i.e.\ the alternative for tests $0, 1, 2, 4, 5$ and the null for test 3 as
\eq{P_i\equiv \left\{\begin{array}{l@{\hspace{5em}}l}
P(\Ha^{(i)}\mid\LLR^{(i)}),&i=0,1,2,4,5,\vspace{1em}\\
P(\Hn^{(i)}\mid\LLR^{(i)}),&i=3.
\end{array}\right.\label{eq-ma-ba-select}}
After obtaining the LLR distribution of the null hypothesis [$p(\LLR\mid\Hn)$], we can determine its proportion [$P(\Hn)$] by aligning $p(\LLR\mid\Hn)$ with the real distribution $p(\LLR)$ at the $\LLR\rightarrow0^+$ side. This provides all the prerequisites to perform Bayesian inference and obtain any $P_i$ from \refeq{ma-ba-select}. 

In practice, PDFs are approximated with histograms. This requires proper choices of histogram bin widths, $P(\Hn)$, and techniques to ensure the conversion from LLR to posterior probability is monotonically increasing and smooth. Implementational details can be found in \pkg{} package and in \refssec{implem}. Distributions can be estimated either separately for every $(E,A)$ pair or by pooling across all $(E,A)$ pairs. In practice, we test on the order of $10^3$ to $10^4$ candidate targets (``$B$'') for every $(E,A)$ such that a separate conversion of LLR values to posterior probabilities is both feasible and recommended, as it accounts for different roles of every gene, especially hub genes, through different rates of alternative hypotheses.

Lastly, in a typical application of \pkg{}, inputs of $(E,A)$ pairs will have been pre-determined as the set of significant eQTL-gene pairs from a genome-wide eQTL associaton analysis. In such cases, we may naturally assume $P_1=1$ for all considered pairs, and skip the primary test.

\subsection{Tests to evaluate\label{ssec-testse}}
Based on the six tests in \refssec{lrt}, we use the following tests and test combinations for the inference of genetic regulations, and evalute them in the results.
\begin{itemize}
\item The correlation test is introduced as a benchmark, against which we can compare other methods involving genotype information. Pairwise correlation is a simple measure for the probability of two genes being functionally related either through direct or indirect regulation, or through coregulation by a third factor. Bayesian inference additionally considers different gene roles. Its predicted posterior probability for regulation is $P_0$.

\item The traditional causal inference test, as explained in \cite{Chen:2007}, suggested that the regulatory relation $E\rightarrow A\rightarrow B$ can be confirmed with the combination of three separate tests: $E$ regulates $A$, $E$ regulates $B$, and $E$ only regulates $B$ through $A$ (i.e.\ $E$ and $B$ become independent when conditioning on $A$). They correspond to the primary, secondary, and independence tests respectively. The regulatory relation $E\rightarrow A\rightarrow B$ is regarded positive only when all three tests return positive. \ear{1-1-2}{The three tests filter the initial hypothesis space of all possible relations between $E$, $A$, and $B$, sequentially to $E\rightarrow A$ (primary test), $E\rightarrow A\wedge E\rightarrow B$ (secondary test), and $E\rightarrow A\rightarrow B\wedge($no confounder for $A$ and $B)$ (conditional independence test). The resulting test is stronger than $E\rightarrow A\rightarrow B$ by disallowing confounders for $A$ and $B$.} So its probability can be broken down as
\eq{P_T\equiv P_1P_2P_3.\label{eq-ba-stat-probtot}}

Trigger \cite{Chen:2007d} is an R package implementation of the method. However, since Trigger integrates eQTL discovery with causal inference, it is not practical for use on modern datasets. For this reason, we reimplemented this method in \pkg, and evaluated it with $P_2$ and $P_2P_3$ separately, in order to assess the individual effects of secondary and independence tests. As discussed above, we expect a set of significant eQTLs and their associated genes as input, and therefore $P_1=1$ is assured and not calculated in this paper or the package \pkg. Note that $P_T$ is the estimated local precision, i.e.\ the probability that tests 2 and 3 are both true. Correspondinly, its local FDR (the probability that one of them is false) is $1-P_T$.

\item  The novel test, aimed specifically at addressing the failures of the traditional causal inference test, combines the tests differently:
\eq{P\equiv\frac{1}{2}(P_2P_5+P_4).\label{eq-ba-p}}
\ea{\lls{1-1-3}Specifically, the first term in \refeq{ba-p} accounts for hidden confounders. The controlled test replaces the conditional independence test and constrains the hypothesis space more weakly, only requiring the correlation between $A$ and $B$ is not entirely due to pleiotropy. Therefore, $P_2P_5$ (with $P_1=1$) verifies the hypothesis that $B\leftarrow E\rightarrow A\wedge(A\not\perp B\mid E)$, a superset of $E\rightarrow A\rightarrow B$.}

\ea{On the other hand, the relevance test in the second term of \refeq{ba-p} addresses weak interactions that are undetectable by the secondary test from existing data ($P_2$ close to 0). This term still grants higher-than-null significance to weak interactions, and verifies that $E\rightarrow A\wedge(E\rightarrow B\vee A\edge B)$, also a superset of $E\rightarrow A\rightarrow B$. In the extreme undetectable limit where $P_2=0$ but $P_4\ne0$, the novel test \refeq{ba-p} automatically reduces to $P=\frac12P_4$, which assumes equal probability of either direction and assigns half of the relevance test probability to $A\rightarrow B$.}

\ea{The composite design of the novel test aims not to miss any genuine regulation whilst distinguishing the full spectrum of possible interactions. \lle{1-1-3}}When the signal level is too weak for tests 2 and 5, we expect $P_4$ to still provide distinguishing power better than random predictions. When the interaction is strong, $P_2P_5$ is then able to pick up true targets regardless of the existence of hidden confounders.
\end{itemize}

\subsection{Evaluation methods\label{ssec-eval}}
\begin{itemize}
\item \textbf{Evaluation metrics:}

\lls{3-9-1}Given the predicted posterior probabilities for every pair $(A,B)$ from any test, or more generically a score from any inference method, we evaluated the predictions against \ea{the direct regulations in the} ground-truth tables \ea{(\refssec{datasets})} with the metrics of Receiver Operating Characteristic (ROC) and Precision-Recall (PR) curves, as well as the Areas Under the ROC (AUROC) and Precision-Recall (AUPR) curves\ea{\cite{stolovitzky2009lessons}}. In particular, AUPR is calculated with the Davis-Goadrich nonlinear interpolation \cite{Davis:2006} with R package \emph{PRROC}.\lle{3-9-1}

\item \textbf{Subsampling:}

In order to assess the effect of sample size on the performances of inference methods, we performed subsampling evaluations. This is made practically possible by the DREAM datasets which contain 999 samples with sufficient variance, as well as the computational efficiency from \pkg{} which makes subsampling computationally feasible. With a given dataset and ground-truth table, the total number of samples $n$, and the number of samples of our actual interest $N<n$, we performed subsampling by repeating following steps $k$ times:
\begin{enumerate}
\item Randomly select $N$ samples out of the total $n$ samples without replacement\label{en-sub-l1}.
\item Infer regulations only based on the selected samples.
\item Compute and record the evaluation metrics of interest (e.g.\ AUROC and AUPR) with the inference results and ground-truths\label{en-sub-l2}.
\end{enumerate}
Evaluation metrics are recorded in every loop, and their means, standard deviations, and standard errors over the $k$ runs, are calculated. The mean indicates how the inference method performs on the metric in average, while the standard deviation reflects how every individual subsampling deviates from the average performance.

\item \lls{3-1-4}\ea{\textbf{Local precision of top predictions separately for confounded and unconfounded gene pairs:}}

\ea{In order to demonstrate the inferential precision among top predictions for any inference test (here the traditional and novel tests separately), we first ranked all (ordered) gene pairs $(A,B)$ according to the inferred significance for $A\rightarrow B$. All gene pairs were split into groups according to their relative significance ranking (9 groups in \refig{dcomb}C,D, as top $0\%$ to $0.01\%$, $0.01\%$ to $0.02\%$, etc). Each group was divided into two subgroups, based on whether each gene pair shared at least one direct upstream regulator gene (confounded) or not (unconfounded), according to the gold standard. Within each subgroup, the local precision was computed as the number of true directed regulations divided by the total number of gene pairs in the subgroup.}\lle{3-1-4}

\end{itemize}

\subsection{\ea{Simulation studies on causal models with measurement error}\label{ssec-sim}}

\ea{We investigated how each statistical test tolerates measurement errors with simulations in a controlled setting. We modelled the causal relation $A\rightarrow B$ in a realistic setup as $E\rightarrow A\sups{(t)}\rightarrow B$ with $A\sups{(t)}\rightarrow A$. $E$ remains as the accurately measured genotype values as the eQTL for the primary target gene $A$. $A\sups{(t)}$ is the true expression level of gene $A$, which is not observable. $A$ is the measured expression level for gene $A$, containing measurement errors. $B$ is the measured expression level for gene $B$.}

\ea{For simplicity, we only considered monoallelic species. Therefore the genotype $E$ in each sample followed the Bernoulli distribution, parameterized by the predetermined minor allele frequency. Each regulatory relation (of $E\rightarrow A\sups{(t)}$, $A\sups{(t)}\rightarrow A$, and $A\sups{(t)}\rightarrow B$) correponded to a normal distribution whose mean was linearly dependent on the regulator variable. In particular, for sample $i$:
\eqa{A\sups{(t)}_i&\sim&N(\widetilde{E_i},\sigma_{A1}^2),\\
A_i&\sim&N(A\sups{(t)}_i,\sigma_{A2}^2),\\
B_i&\sim&N(\widetilde{A\sups{(t)}_i},\sigma_B^2),}
in which $\sigma_{A1}$, $\sigma_{A2}$, and $\sigma_B$ are parameters of the model. Note that $\sigma_B^2$ is $B$'s variance from all unknown sources, including expression level variations and measurement errors. The tilde normalizes the variable into zero mean and unit variance, as:
\eq{\widetilde{X_i}\equiv\frac{X_i-\bar X}{\sqrt{\var(X)}},}
where $\bar X$ and $\var(X)$ are the mean and variance of $X\equiv\{X_i\}$ respectively.}

\ea{Given the five parameters of the model (the number of samples, the minor allele frequency, $\sigma_{A1}$, $\sigma_{A2}$, and $\sigma_B$), we could simulate the observed data for $E$, $A$, and $B$, which were then fed into \pkg{} for tests 2-5 and their p-values of the respective null hypotheses. Supernormalization step was replaced with normalization which merely shifted and scaled variables into zero mean and unit variance.}

\ea{We then chose different configurations on the number of samples, the minor allele frequency, and $\sigma_B$. For each configuration, we varied $\sigma_{A1}$ and $\sigma_{A2}$ in a wide range to obtain a 2-dimensional heatmap plot for the p-value of each test, thereby exploring how each test was affected by measurement errors of different strengths. Only tiles with a significant $E\rightarrow A$ eQTL relation were retained. The same initial random seed was employed for different configurations to allow for replicability.}

\section*{Acknowledgements}

This work was supported by the BBSRC (grant numbers BB/J004235/1 and BB/M020053/1). 

\bibliographystyle{vancouver}
\bibliography{bib,bibsysbiol}

\begin{table}[p]
\center
\caption{Six likelihood ratio tests are performed to test the regulation $A\rightarrow B$, numbered, named, and defined as shown. $E$ is the best eQTL of $A$. Arrows in a hypothesis indicate directed regulatory relations. Genes $A$ and $B$ each follow a normal distribution, whose mean depends additively on its regulator(s), as determined in the corresponding hypothesis. The dependency is categorical on discrete regulators (genotypes) and linear on continuous regulators (gene expression levels). The undirected line represents a multi-variate normal distribution between the relevant variables. In order to identify $A\rightarrow B$ regulation, we select either the null or the alternative hypothesis depending on the test, as shown.\label{tab-tests}}\vspace{1em}
\begin{tabular}{c@{\hspace{1em}}c|c@{\hspace{2em}}c@{\hspace{2em}}c}
Test ID&Test name&\minibox[c]{Null\\(hypothesis)}&\minibox[c]{Alternative\\(hypothesis)}&\minibox[c]{Selected\\hypothesis}\\
\hline
0&Correlation&
\vcenteredhbox{\includegraphics[width=0.1\textwidth]{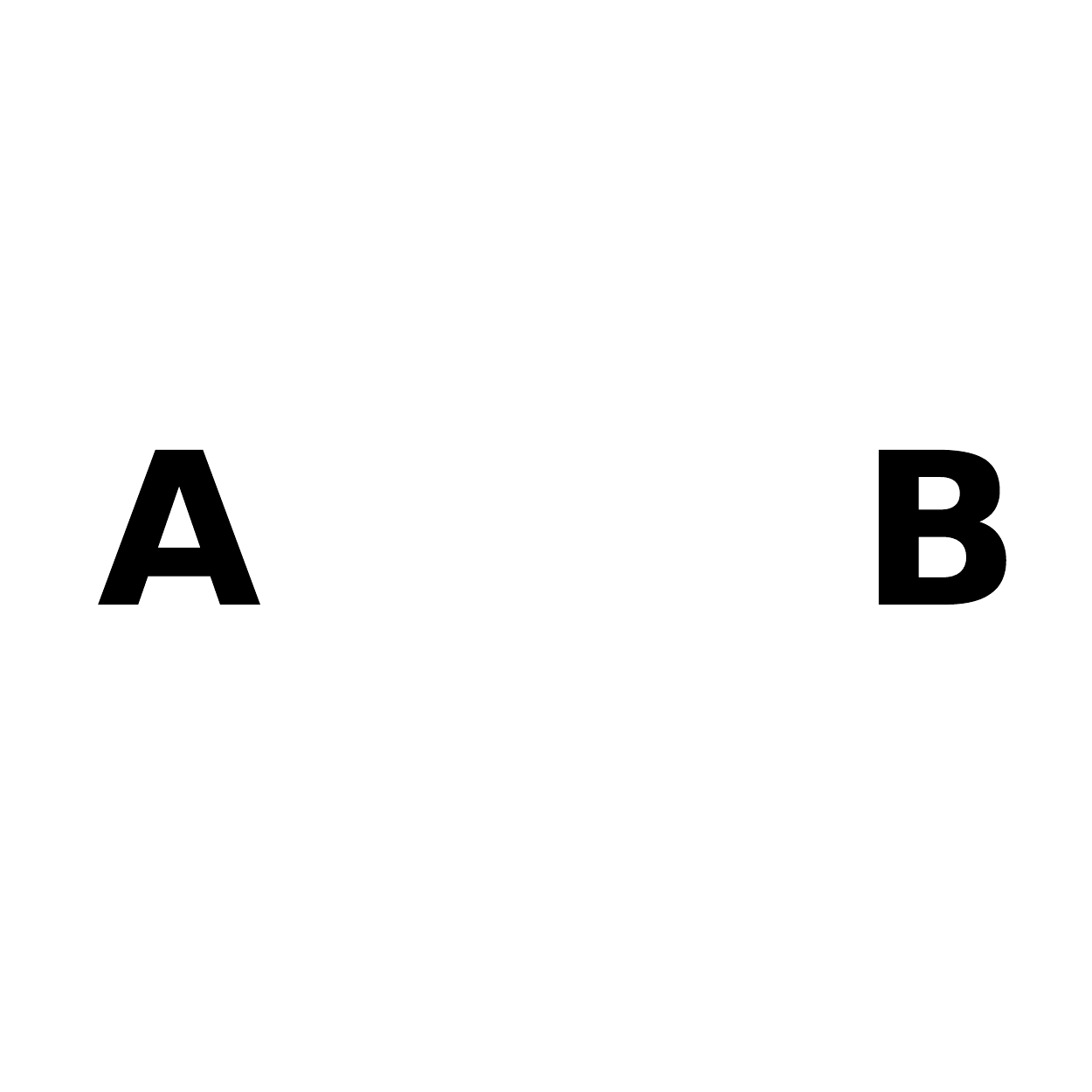}}&
\vcenteredhbox{\includegraphics[width=0.1\textwidth]{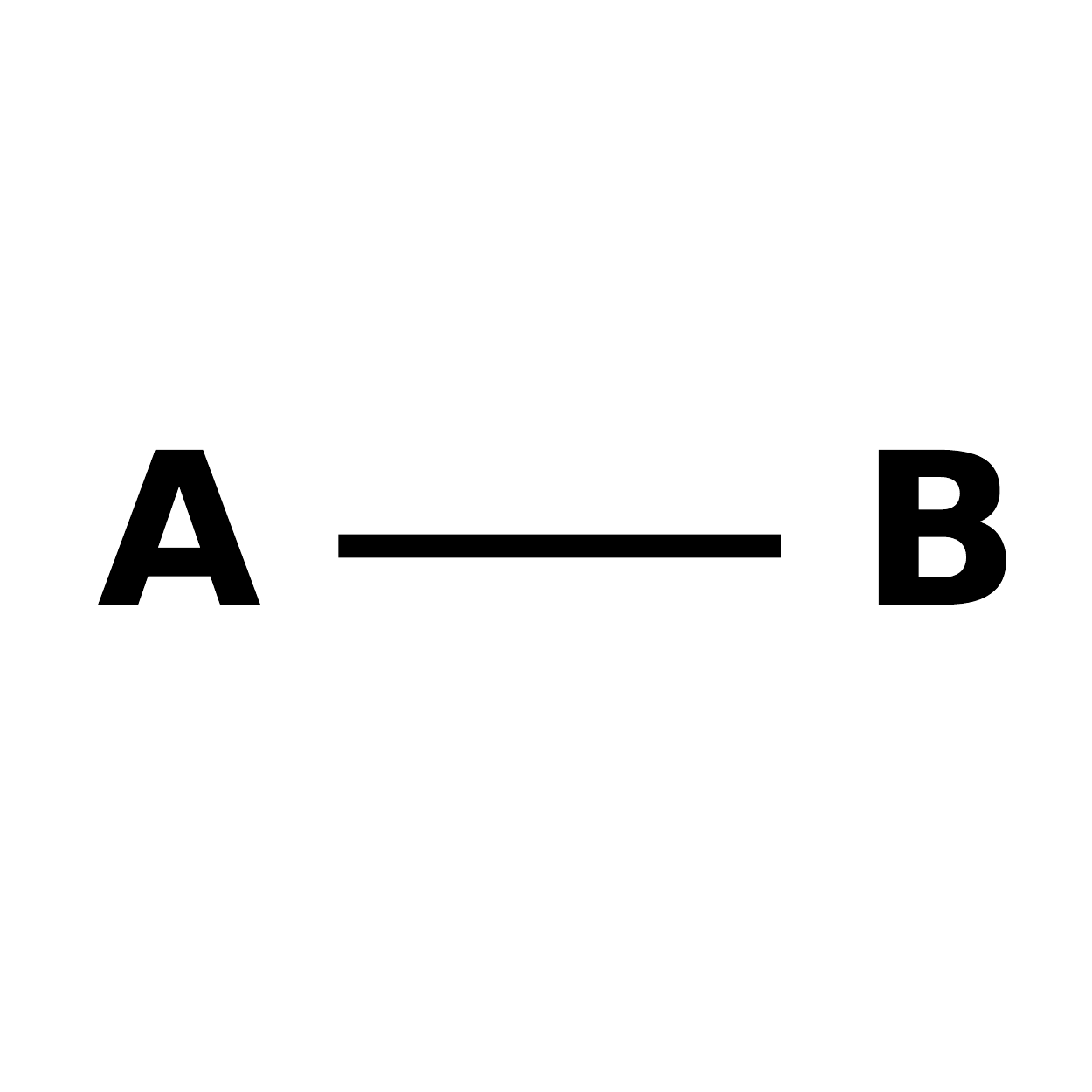}}&Alternative\\
1&\minibox[c]{Primary\\(Linkage)}&
\vcenteredhbox{\includegraphics[width=0.1\textwidth]{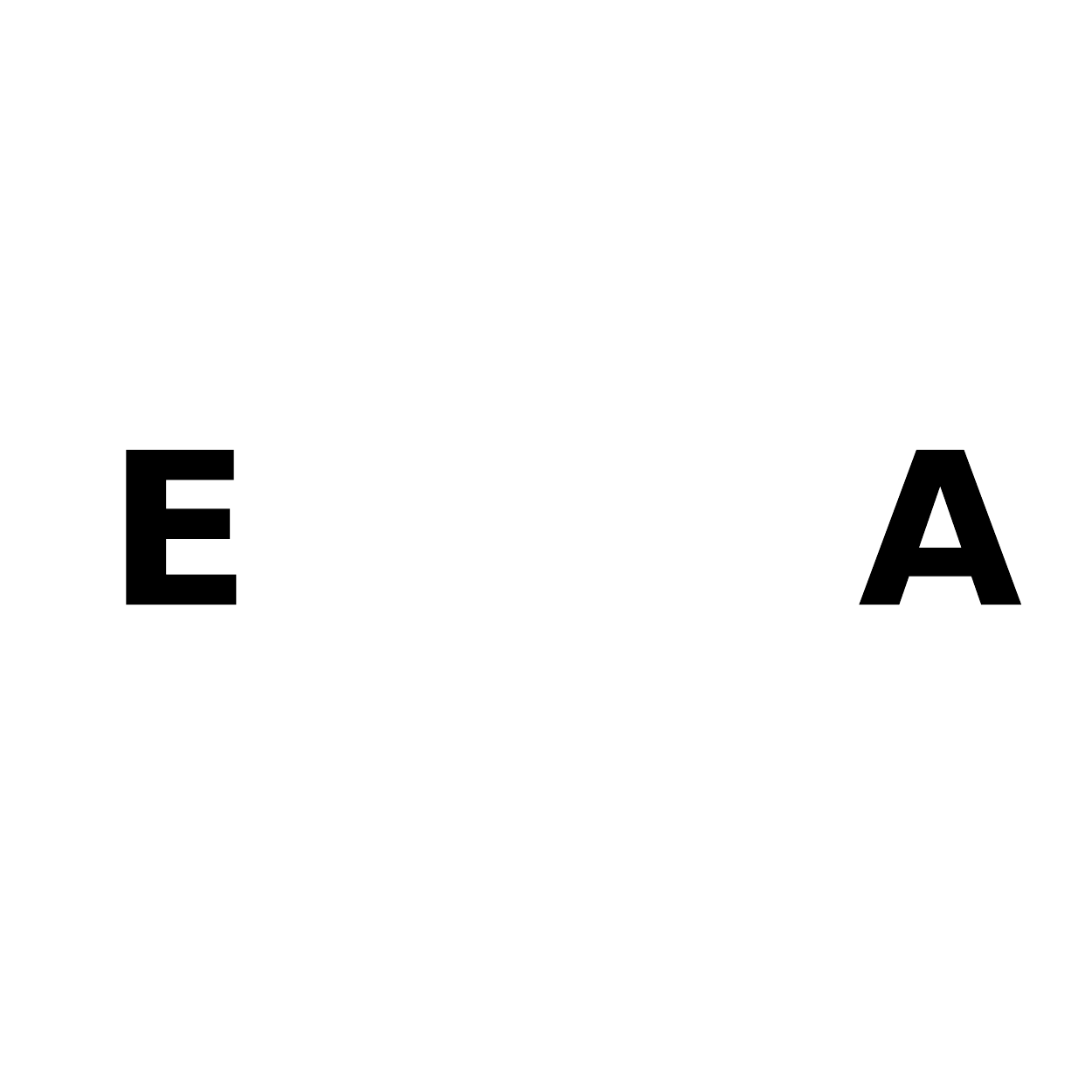}}&
\vcenteredhbox{\includegraphics[width=0.1\textwidth]{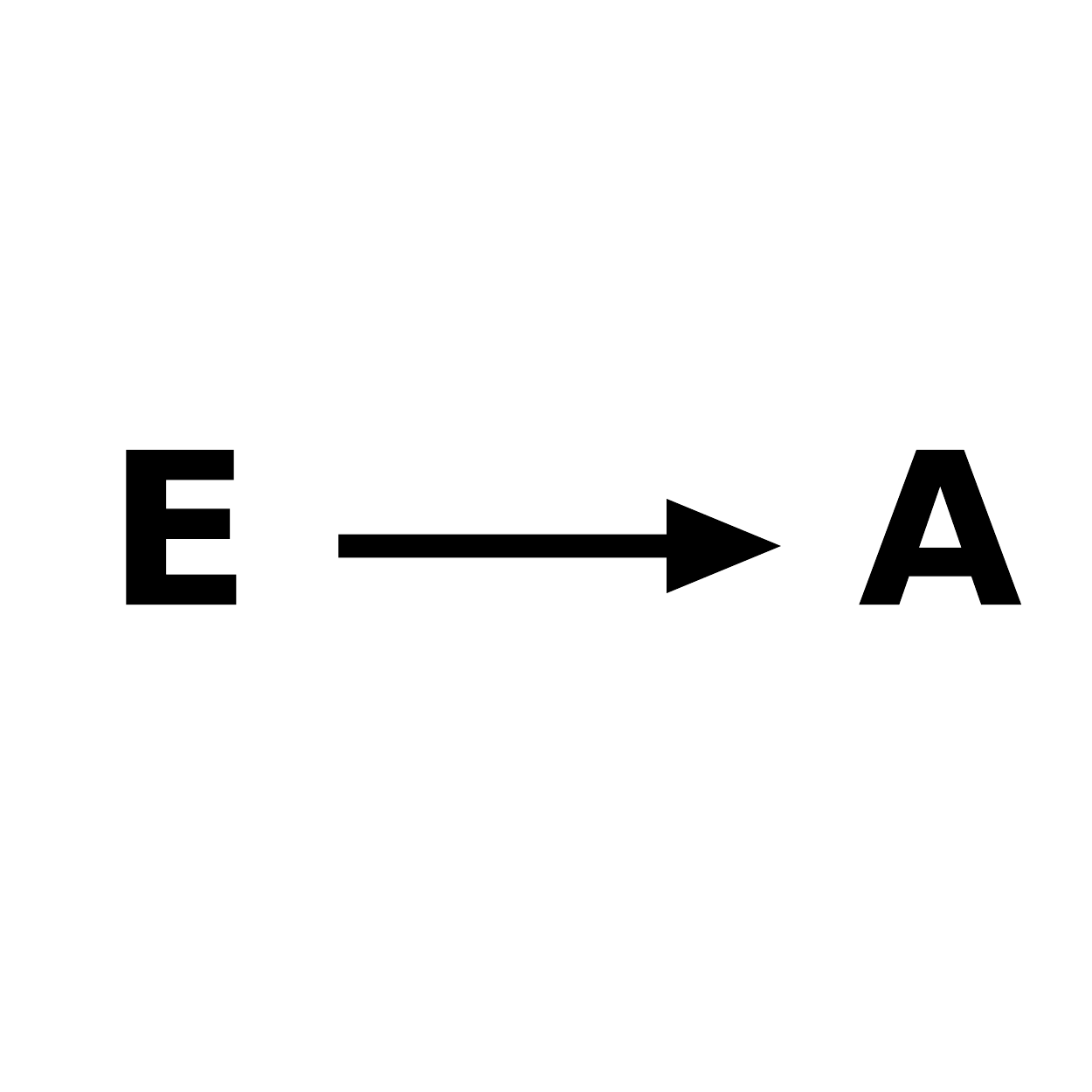}}&Alternative\\
2&\minibox[c]{Secondary\\(Linkage)}&
\vcenteredhbox{\includegraphics[width=0.1\textwidth]{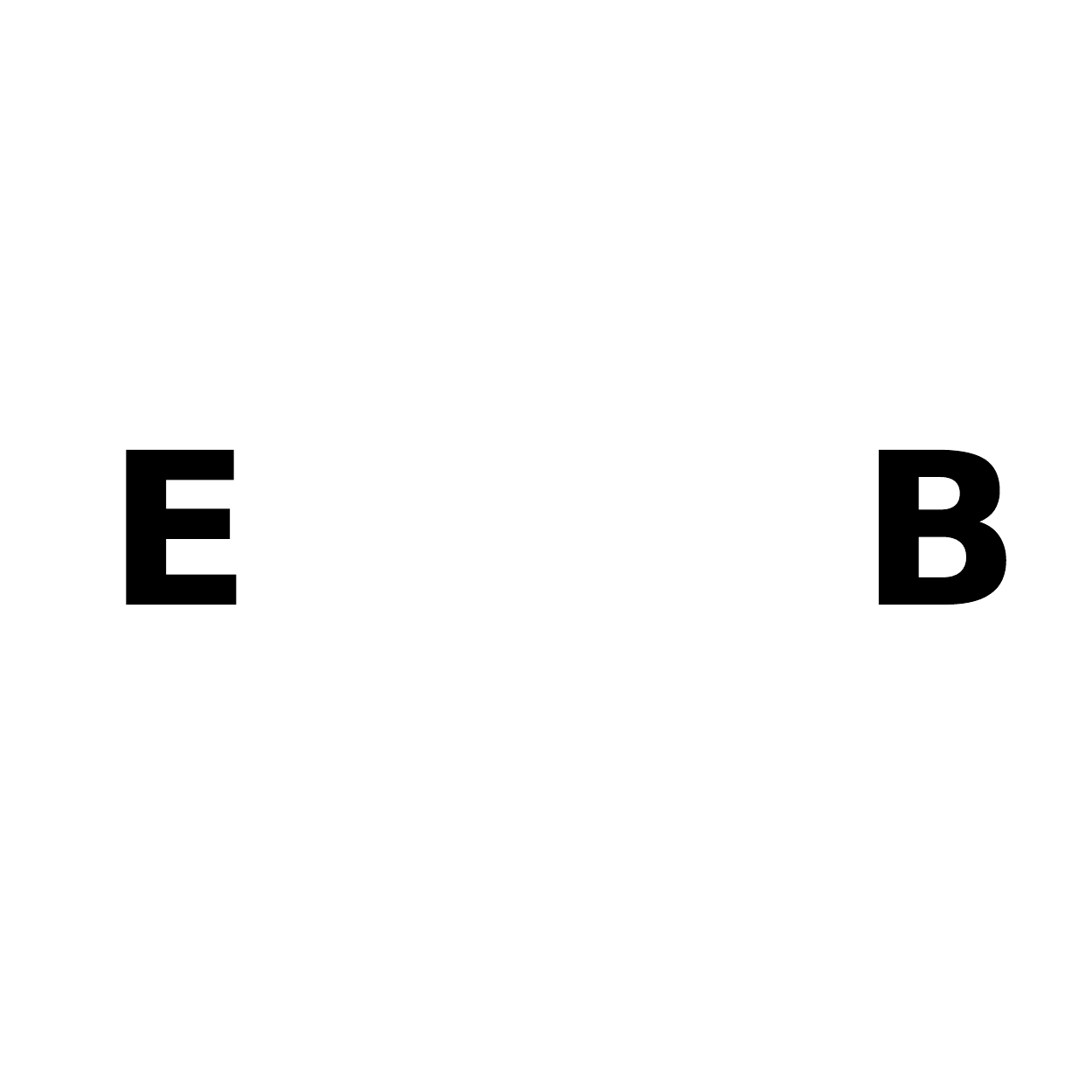}}&
\vcenteredhbox{\includegraphics[width=0.1\textwidth]{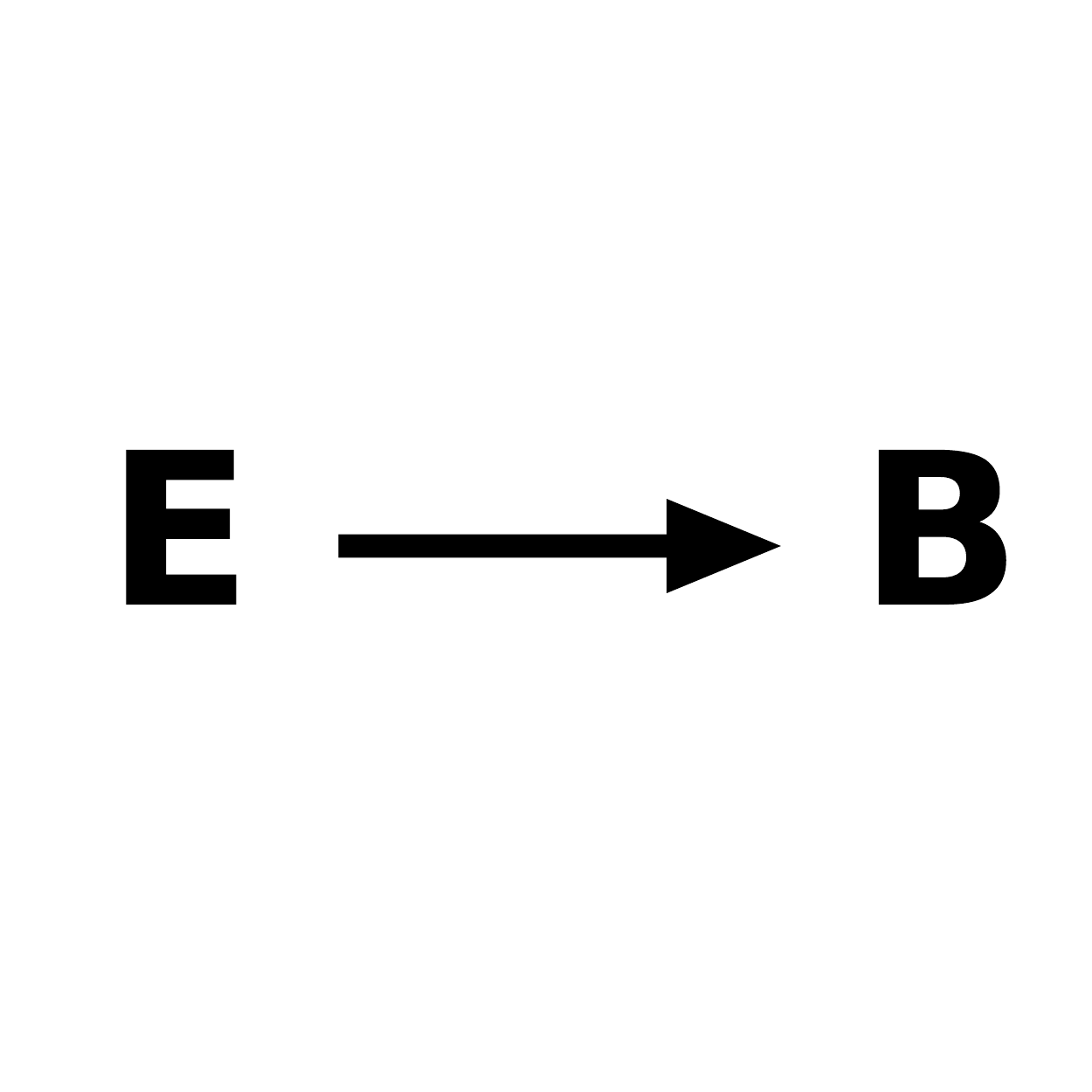}}&Alternative\\
3&\minibox[c]{(Conditional)\\Independence}&
\vcenteredhbox{\includegraphics[width=0.1\textwidth]{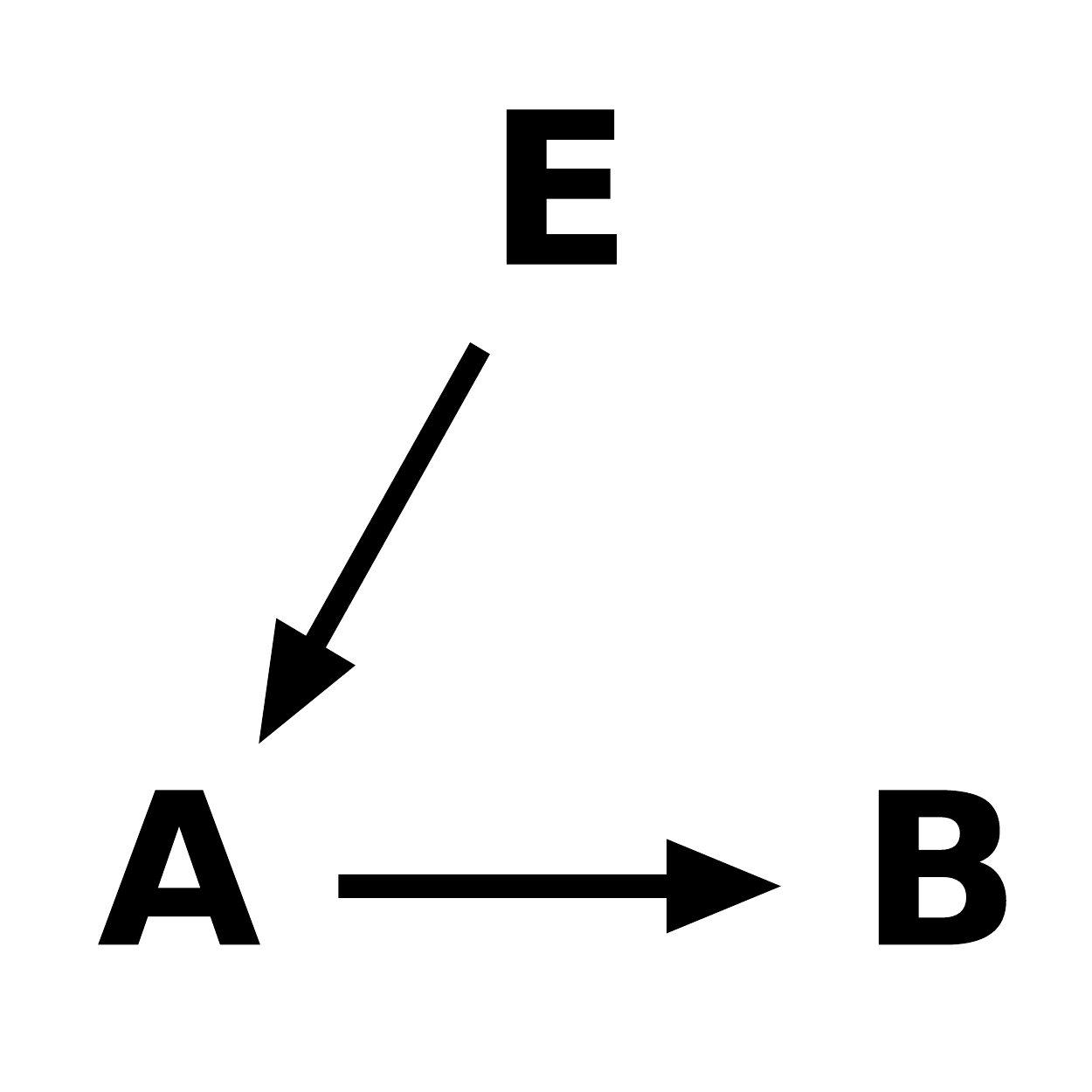}}&
\vcenteredhbox{\includegraphics[width=0.1\textwidth]{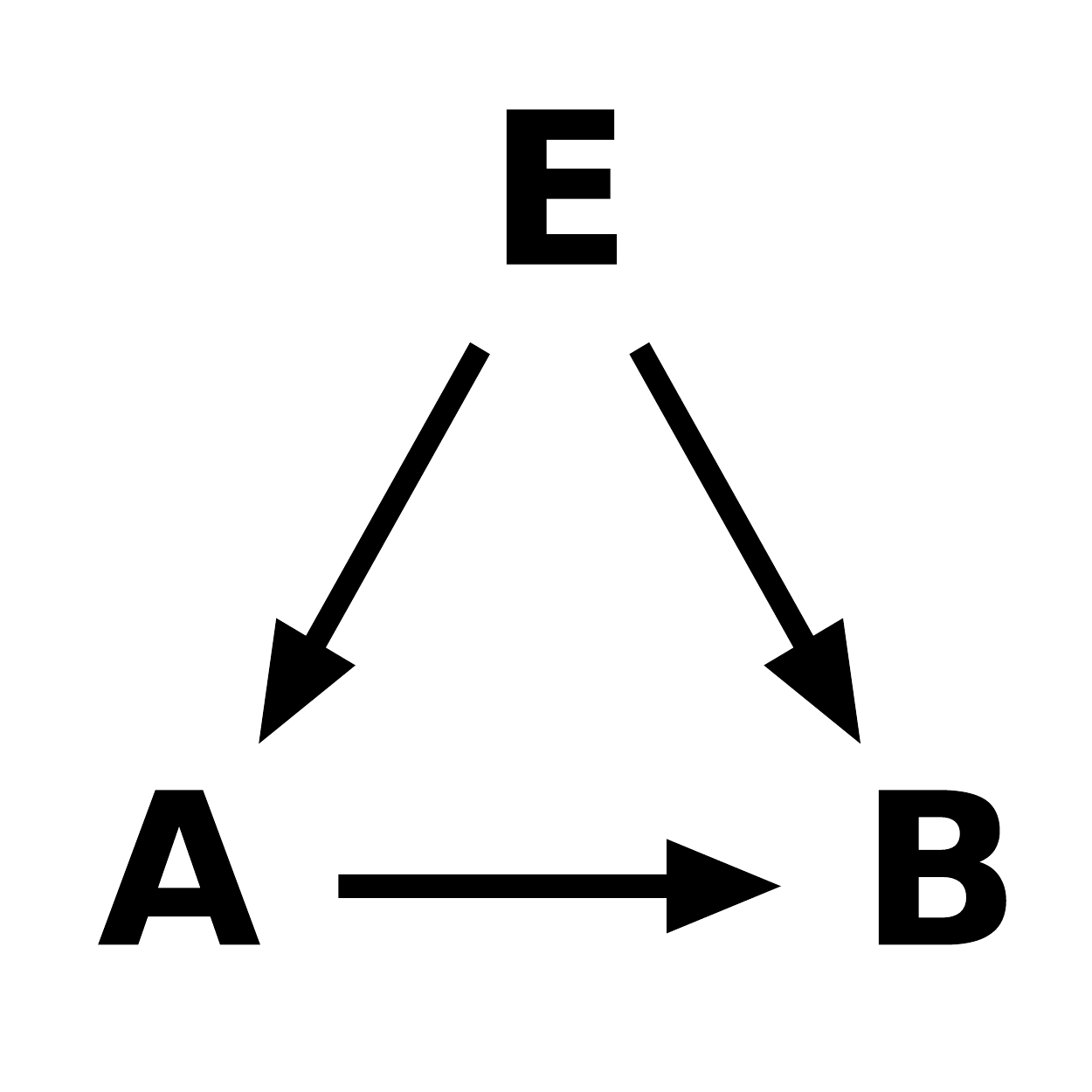}}&Null\\
4&\Tnamea&
\vcenteredhbox{\includegraphics[width=0.1\textwidth]{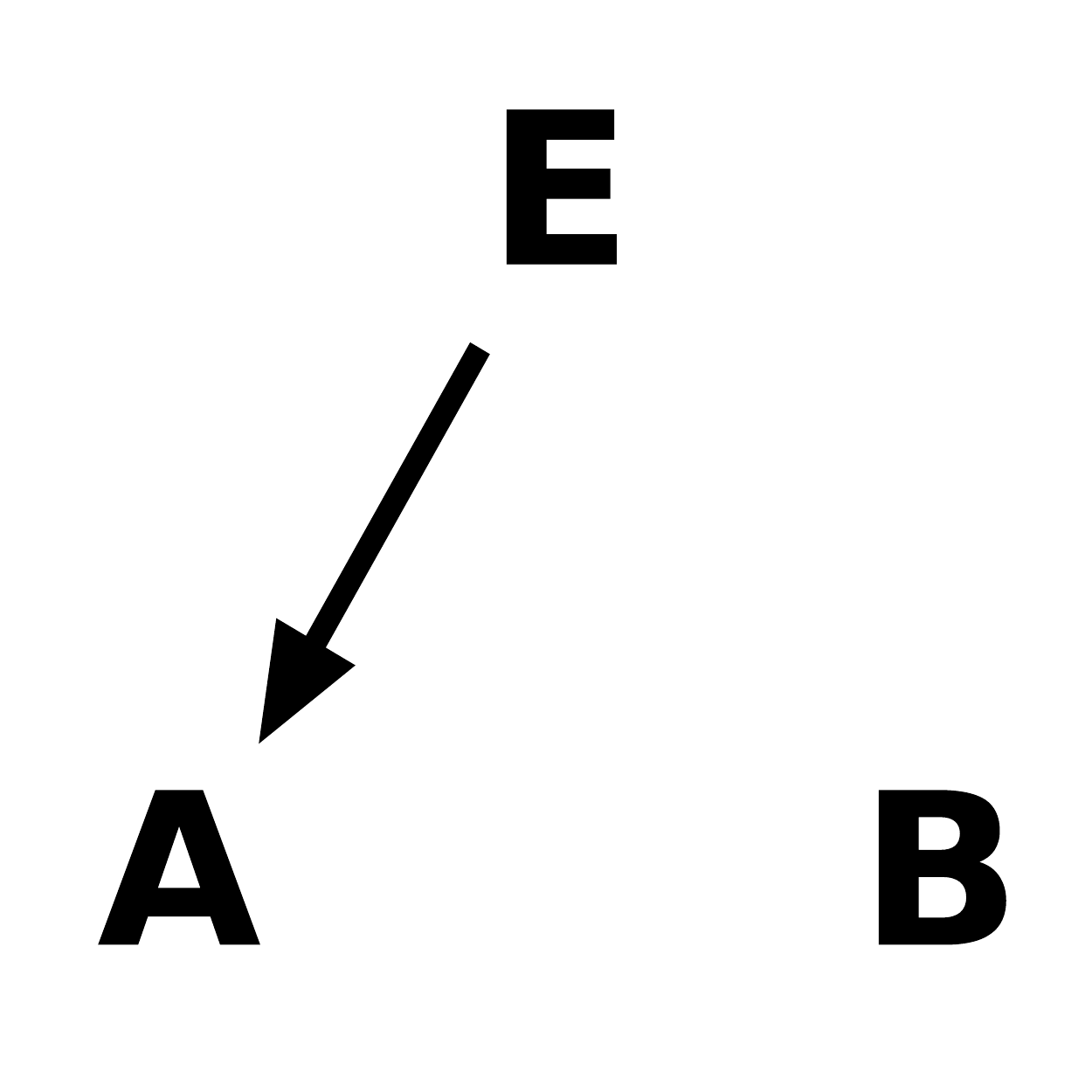}}&
\vcenteredhbox{\includegraphics[width=0.1\textwidth]{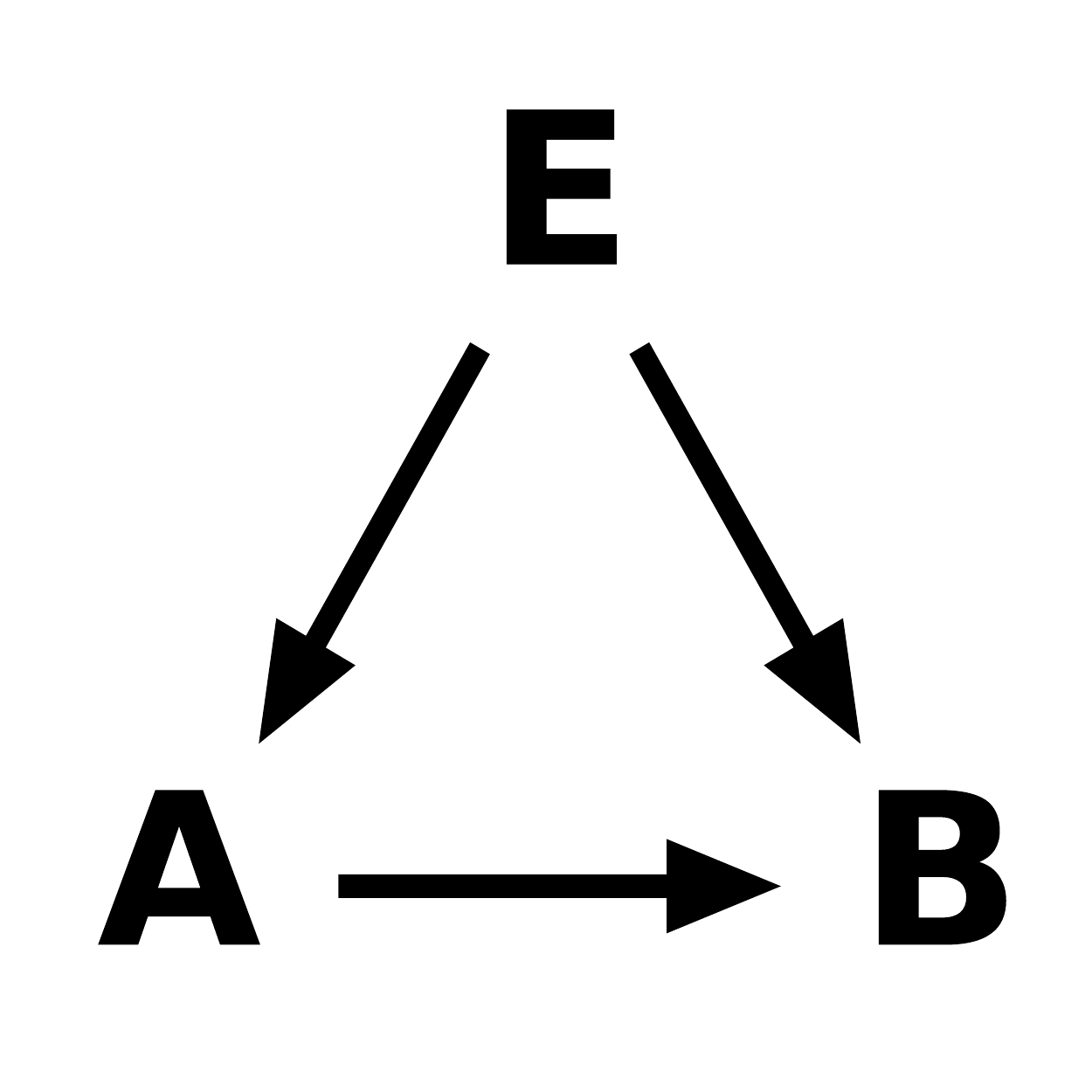}}&Alternative\\
5&\Tnameb&
\vcenteredhbox{\includegraphics[width=0.1\textwidth]{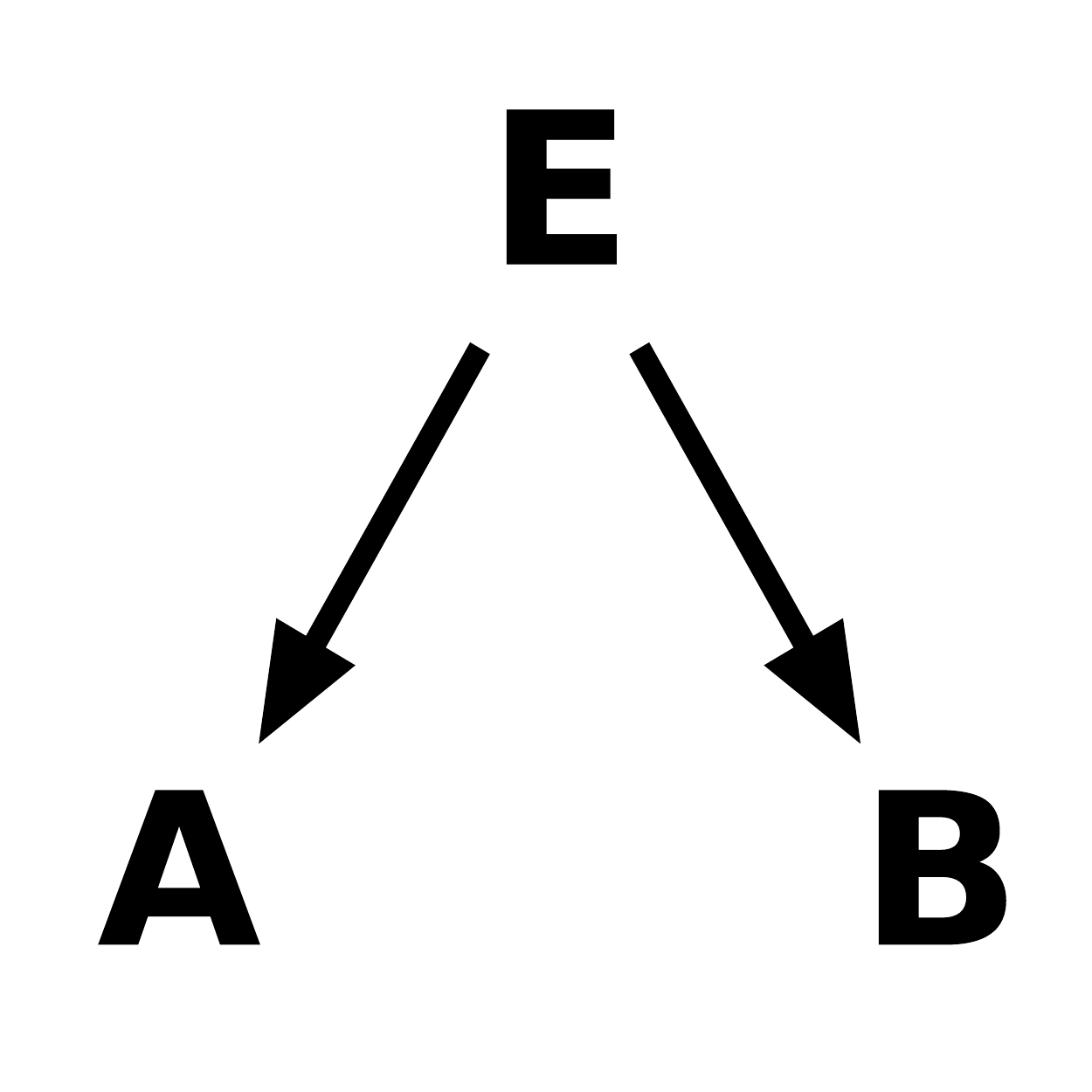}}&
\vcenteredhbox{\includegraphics[width=0.1\textwidth]{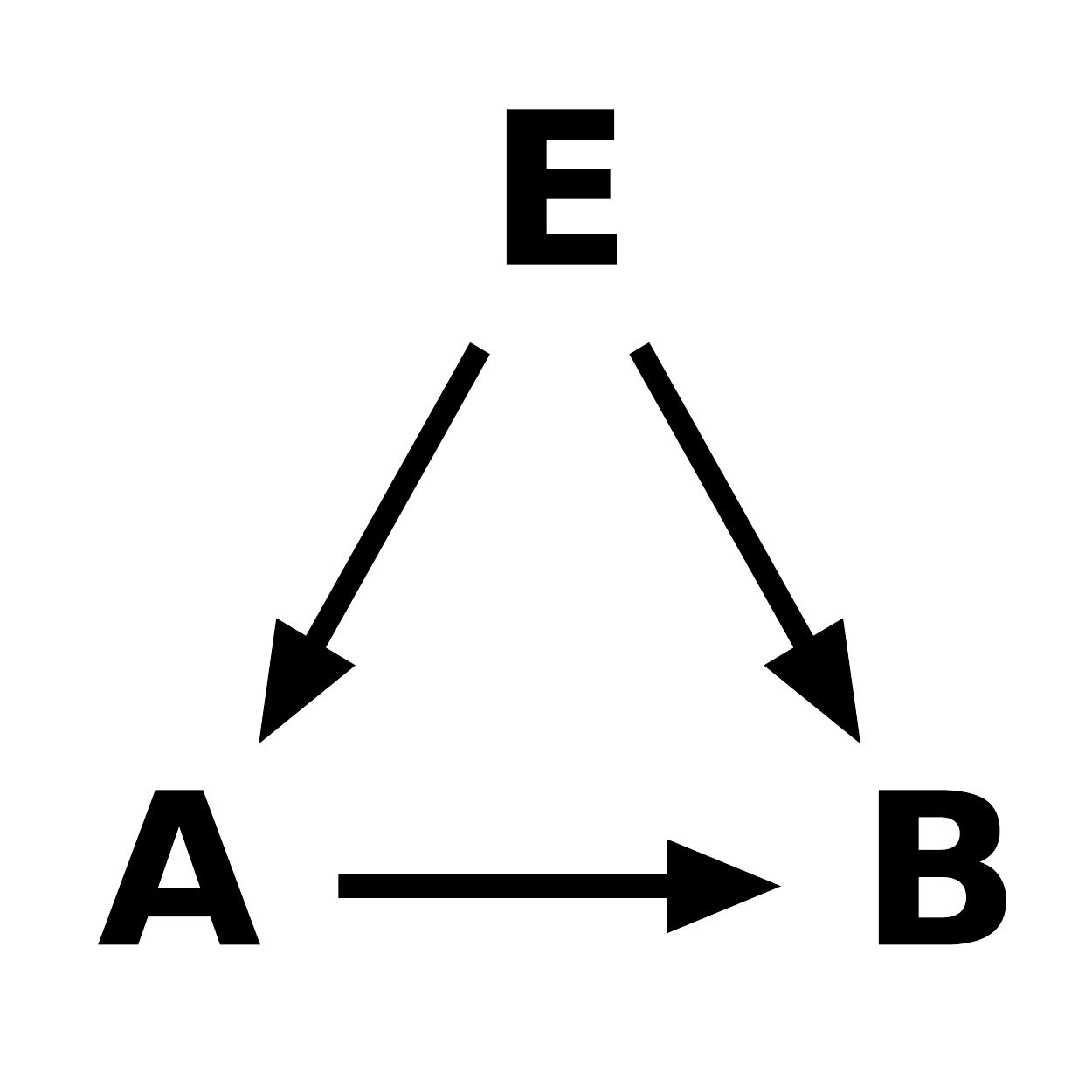}}&Alternative
\end{tabular}
\end{table}

\begin{figure}[p]
  \centering
  \begin{minipage}[t]{.7\linewidth}
     \textbf{A}\\
     \includegraphics[width=\linewidth]{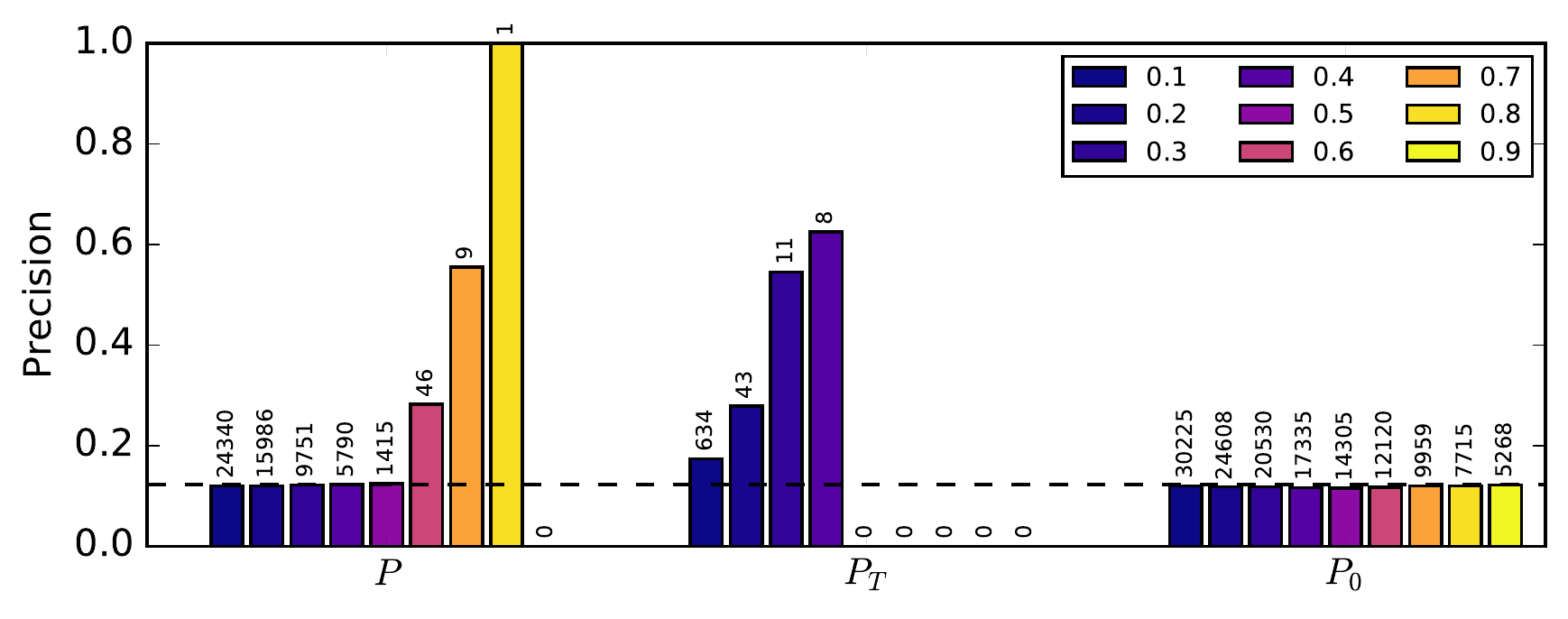}
     \textbf{B}\\
     \includegraphics[width=\linewidth]{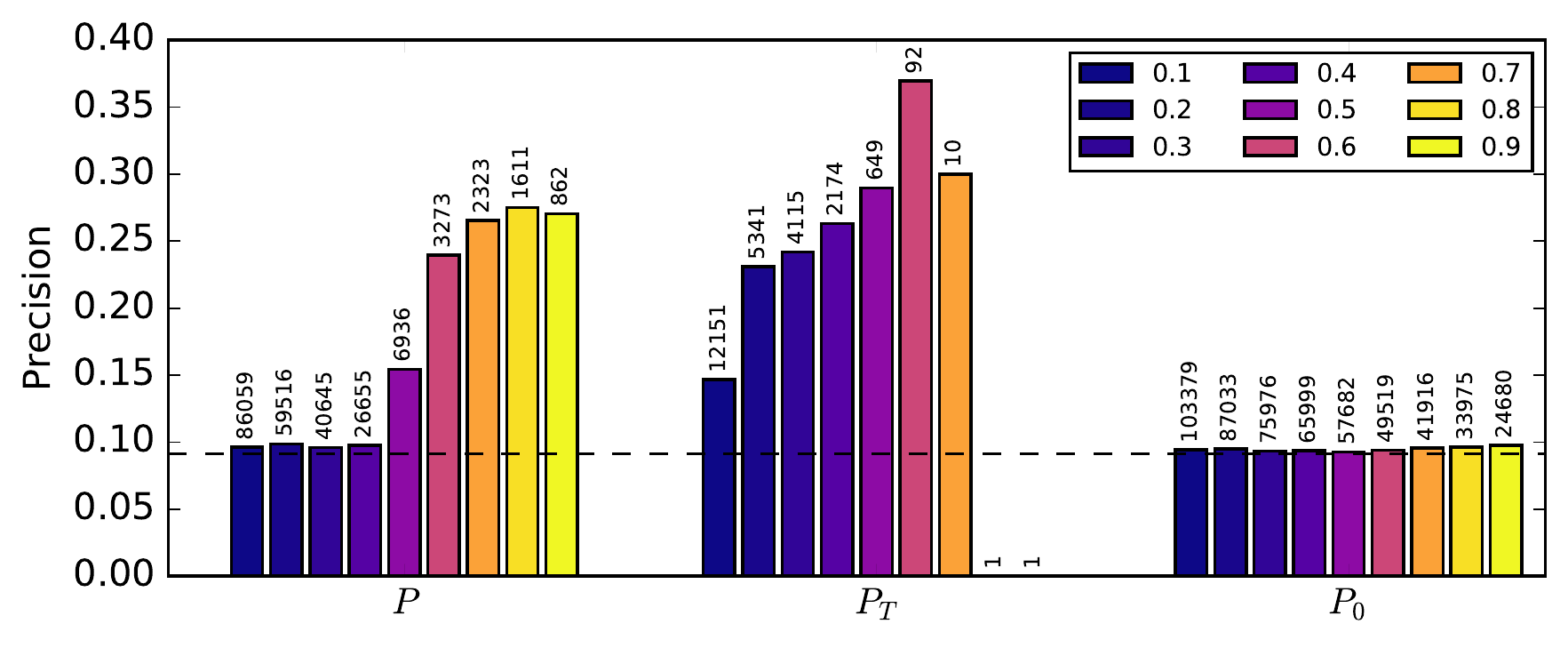}
  \end{minipage}\hfill
  \caption{\textbf{\pkg{} predicts TF targets with more accurate FDR estimates from the Geuvadis data.} The precision (i.e.\ 1-FDR) of TF target predictions is shown at probability cutoffs 0.1 to 0.9 (blue to yellow) with respect to known functional targets from siRNA silencing of 6 TFs (\textbf{A}) and known TF-binding targets of 20 TFs (\textbf{B}). The number above each bar indicates the number of predictions at the corresponding threshold. Dashed lines are precisions from random predictions.}
  \label{fig-LCL}
\end{figure}

\newpage
\appendix
\renewcommand\thefigure{S\arabic{figure}}
\renewcommand\thetable{S\arabic{table}}
\renewcommand\theequation{S\arabic{equation}}
\renewcommand\thesection{S\arabic{section}}
\renewcommand\thesubsection{S\arabic{section}.\arabic{subsection}}


\begin{spacing}{1}\LARGE \textbf{Supplementary Material for ``\titleorig''}\\[-8mm]\end{spacing}

\authors
\setcounter{section}{0}
\setcounter{figure}{0}
\setcounter{table}{0}
\section{Methods}
\subsection{Practical details for Bayesian inference\label{ssec-implem}}
In practice, real PDFs are approximated with histograms. This requires a proper choice of histogram bin widths and counts. We use $\lfloor n_p^\frac{1}{2.5}\rfloor$ bins, capped at 100, where $n_p$ is the total number of points used for generating the histogram. The exponent is chosen for simultaneous precision improvements from higher bin resolution and weaker fluctuation within every bin. The bin widths are also chosen as a smooth transition from uniform sample count for every bin on the $0^+$ side, to uniform bin width on the positive side. (See source code for detail.)

The bin values from real data were then postprocessed to remove empty bins that are between nonempty bins, by filling them with bin values on the positive side. We then aligned the analytical null histogram by intersecting it below the postprocessed real histogram at nonzero bin values. This obtained the ratio of null hypothesis in the mixture distribution. Bayes' theorem then gave raw posterior probability $P(\Ha\mid\LLR)$ at every bin center.

To enforce monotonicity of posterior distribution, we calculated the two following functions. First, starting from raw posterior probability $P(\Ha\mid\LLR)$ at every bin center, set every the posterior at every bin to be no smaller than every value on its the negative side. Second, also starting from raw posterior distribution values $P(\Ha\mid\LLR)$ at every bin center, set every the posterior at every bin to be no larger than every value on its the positive side. A mean is then taken between the two functions to ensure monotonicity whilst minimizing systematic bias.

We then smoothened the monotonic posterior distribution, by convolving its bin differences against a predefined normal filter, after which cumulative sum was calculated to recover the posterior distribution. A normalization was then performed to maintain the span between the previous minimum and maximum. The major purpose of smoothening is to remove duplicate values, especially those introduced during monotonicity enforcement, rather than to obtain a visually smooth function. After smoothening, we performed linear interpolation to obtain the individual post-processed posterior probabilities for each LLR.

More details can be found in the source code.

\section{Results}
\subsection{Iterative conditioning conflicts with local FDR-based probability estimation}
In \cite{Chen:2007}, the authors suggested that the probability of each test should be conditioned on the survival of all preceding tests, i.e.\ that the null distribution of each test should be estimated only on the $(A,B)$ pairs that survive all preceding tests, although this is not implemented in Trigger package.

As an example, we applied the test combination $P_2P_3$ on Geuvadis dataset. After choosing an appropriate threshold probability for secondary test, as suggested in \cite{Chen:2007}, we filtered only the gene pairs positive for secondary test and calculated their real and null LLR distributions of the independence test. Random permutations were applied for null distribution, with high-speed sampling from Metropolis-Hastings algorithm, whose sampling rate is exponentially increasing below secondary test's positivity threshold, and uniformly 1 above that. To balance between efficient sampling and the prevention of being trapped in local maxima, a proper exponential factor of sampling rate ($\approx 1-n_v/n$) can be obtained from the null LLR distribution in \refeq{ba-null-llr2}.

The calculation revealed that the null and real distributions form different shapes at the $\LLR\sups{(3)}\rightarrow0^+$ side, which contradicted with the fundamental assumption of the local FDR-based probability estimation method (\refssec{bi}). On the other hand, the histograms aligned flawlessly without conditioning. We conclude that although appropriate conditioning may enhance statistical power, we are still yet to find a self-consistent approach. Since unconditioned tests have been shown self-consistent and reliable, we do not apply test conditioning in \pkg.

\subsection{Analytical null distribution matches random permutations\label{ssec-match}}
\begin{figure}[h!]
\center
\includegraphics[width=\textwidth]{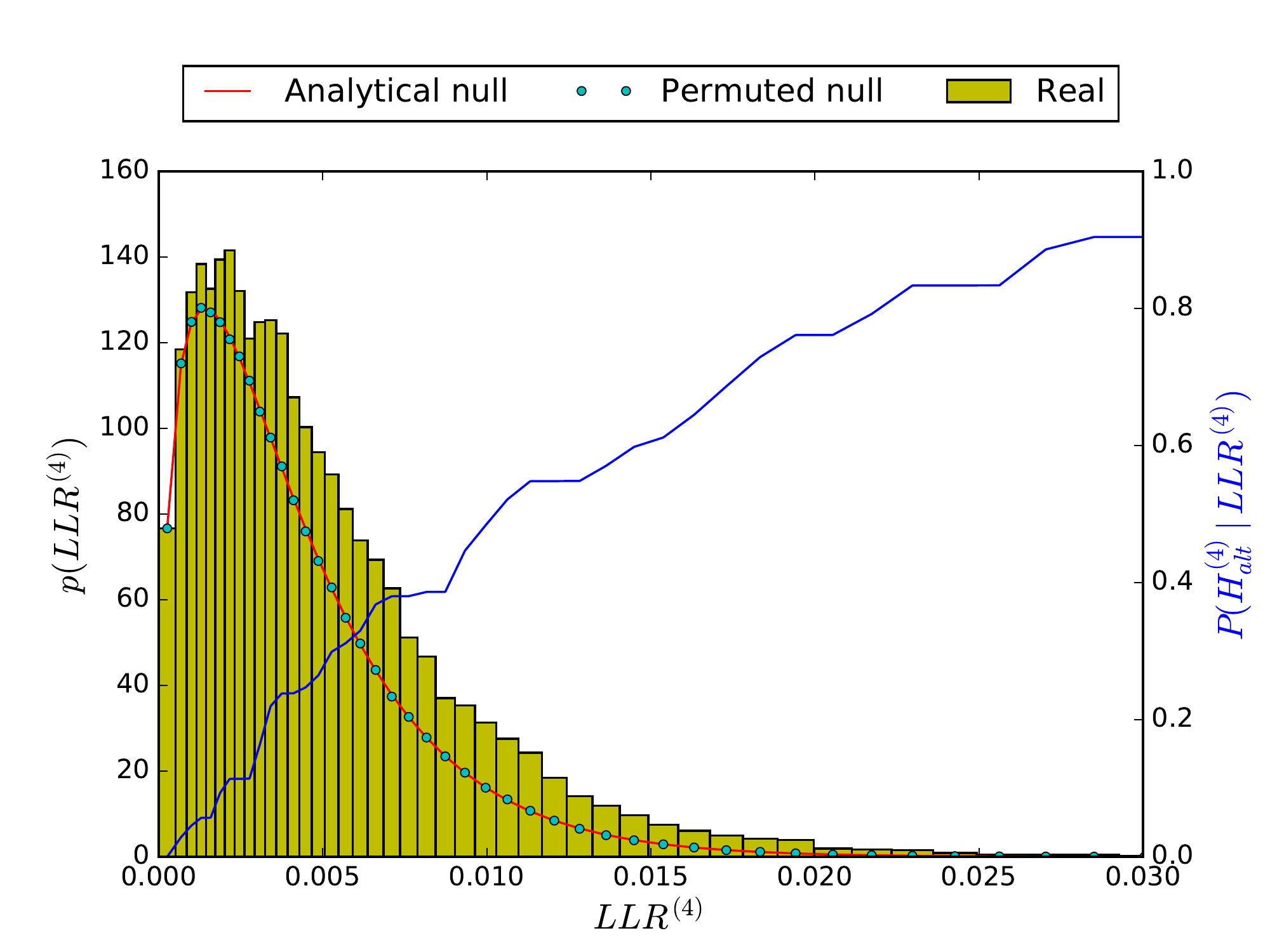}
\caption{LLR distributions of the \tnamea{} test for hsa-miR-200b-3p on 23722 potential targets of Geuvadis dataset. Real, analytical null, and permuted null distributions are demonstrated in the figure, together with the curve of inferred posterior probability of alternative hypothesis. Permutations were randomly conducted on all potential target genes for 100 times. The alignment between analytical and permuted null distributions and the consistent incremental trend of posterior probability  verify our method in deriving analytical null distributions.\label{fig-eg4}}
\end{figure}

An important feature of \pkg{} is the novel derivation of analytical expressions for the distributions of likelihood ratio test statistics under various null distributions (\refssec{null}). We compared the analytical null distributions to empirical null distributions obtained from random simulations. Simulated data were obtained either by permuting sample labels of the independent gene (tests 0,1,2,4), or by simulating expression levels of the gene whilst taking into account existing correlations (tests 3,5). Sufficient simulated data were then fed into the original algorithm to obtain $p(\LLR\mid\Hn)$. As demonstrated with an example in \refig{eg4}, our analytical derivation was confirmed indistinguishable with simulated distribution for miRNA hsa-miR-200b-3p's targets.

More importantly, the analytical result holds for any sample size and does not assume infinite sample sizes ($n\rightarrow\infty$). Indeed, in this asymptotic limit, the LLRs of null distributions reduce to $\chi^2$ distributions, in agreement with Wilks's theorem \cite{Wasserman:2013}. For example, it is easy to confirm: $\lim_{n\rightarrow\infty}2\,\LLR\sups{(1)}\sim\chi^2(n_v-1)$. However, approximating LLR distributions with $\chi^2$ leads to over-estimation of the null PDF at $\LLR\rightarrow0^+$ and under-estimation at $\LLR\rightarrow\infty$. The tilted $p(\LLR|\Hn)$ would then cause systematic over-estimation of $P(\Ha|\LLR)$ for all pairs. For the Geuvadis dataset with 360 samples and $n_v=3$, an over-estimation of $\sim1\%$ can be observed at $\LLR\rightarrow0^+$. This counts an extra $\sim1\%$ of all pairs as the alternative hypothesis, which can be of the same order as the actual percentage of true alternative hypotheses (typically at most a few percent).

\subsection{Subsampling and leaderboard performances of existing and new causal inference methods on DREAM datasets\label{ssec-consist}}

\begin{figure}[h!]
\center
\subfigure{\includegraphics[width=0.37\textwidth]{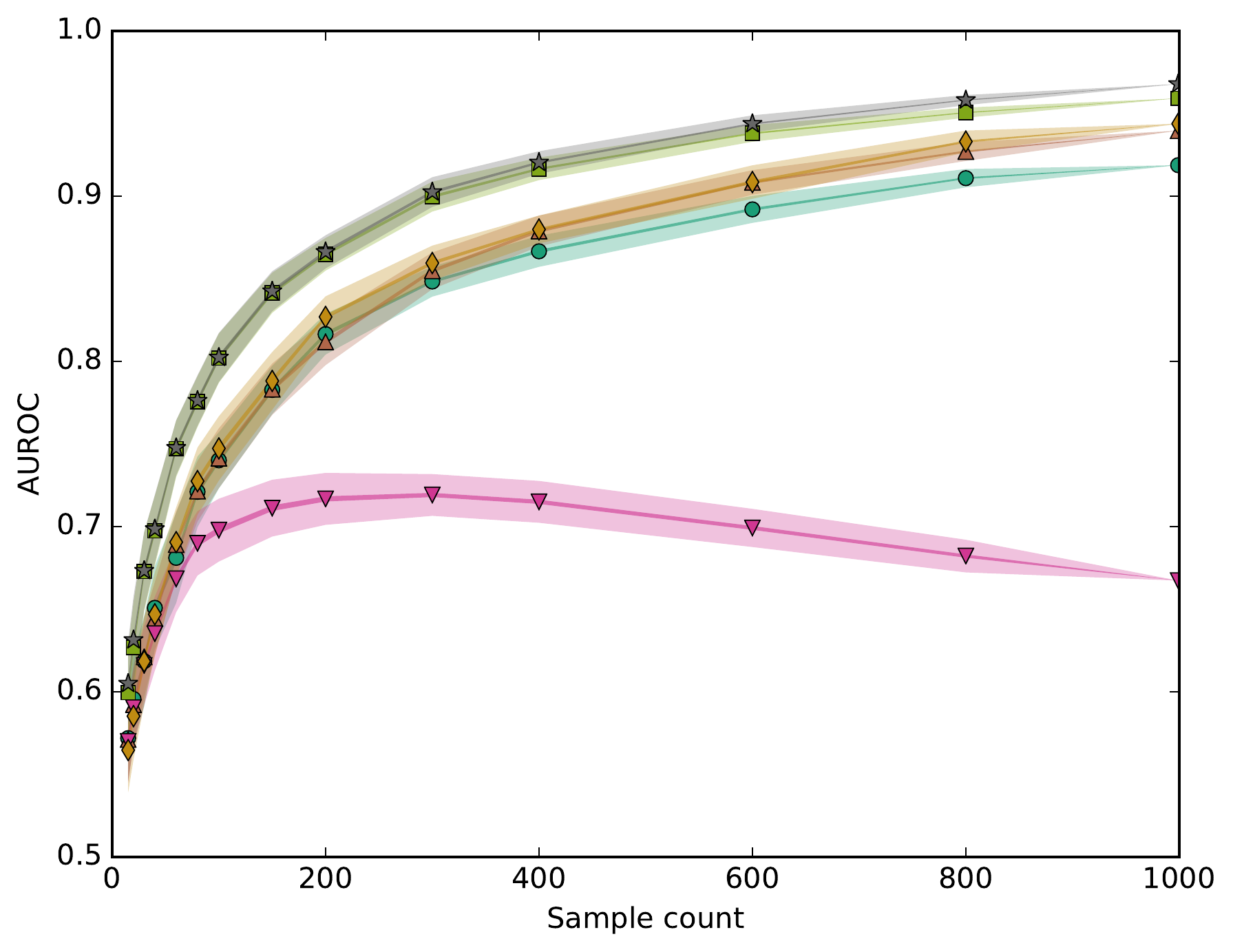}}
\subfigure{\includegraphics[width=0.37\textwidth]{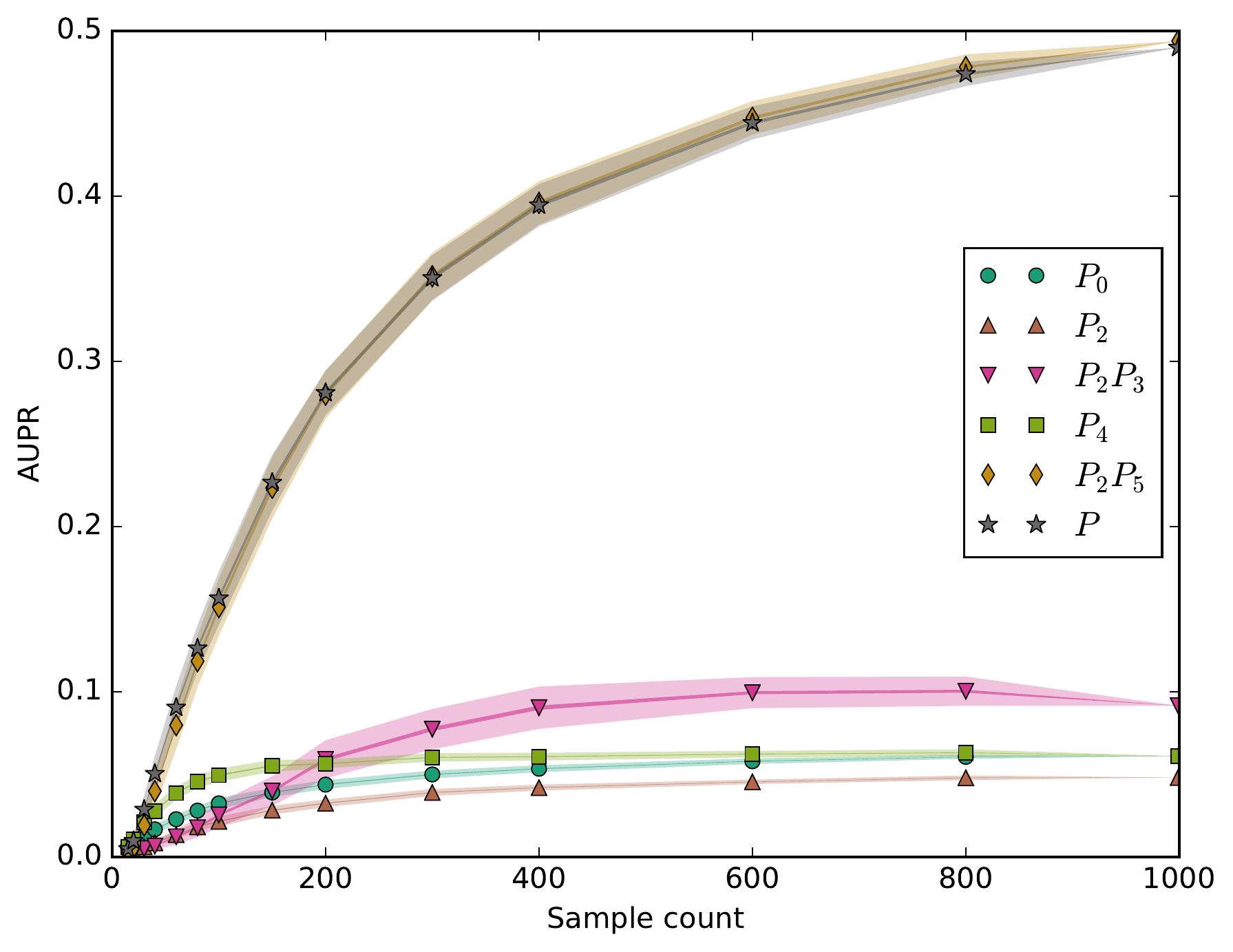}}\\
\subfigure{\includegraphics[width=0.37\textwidth]{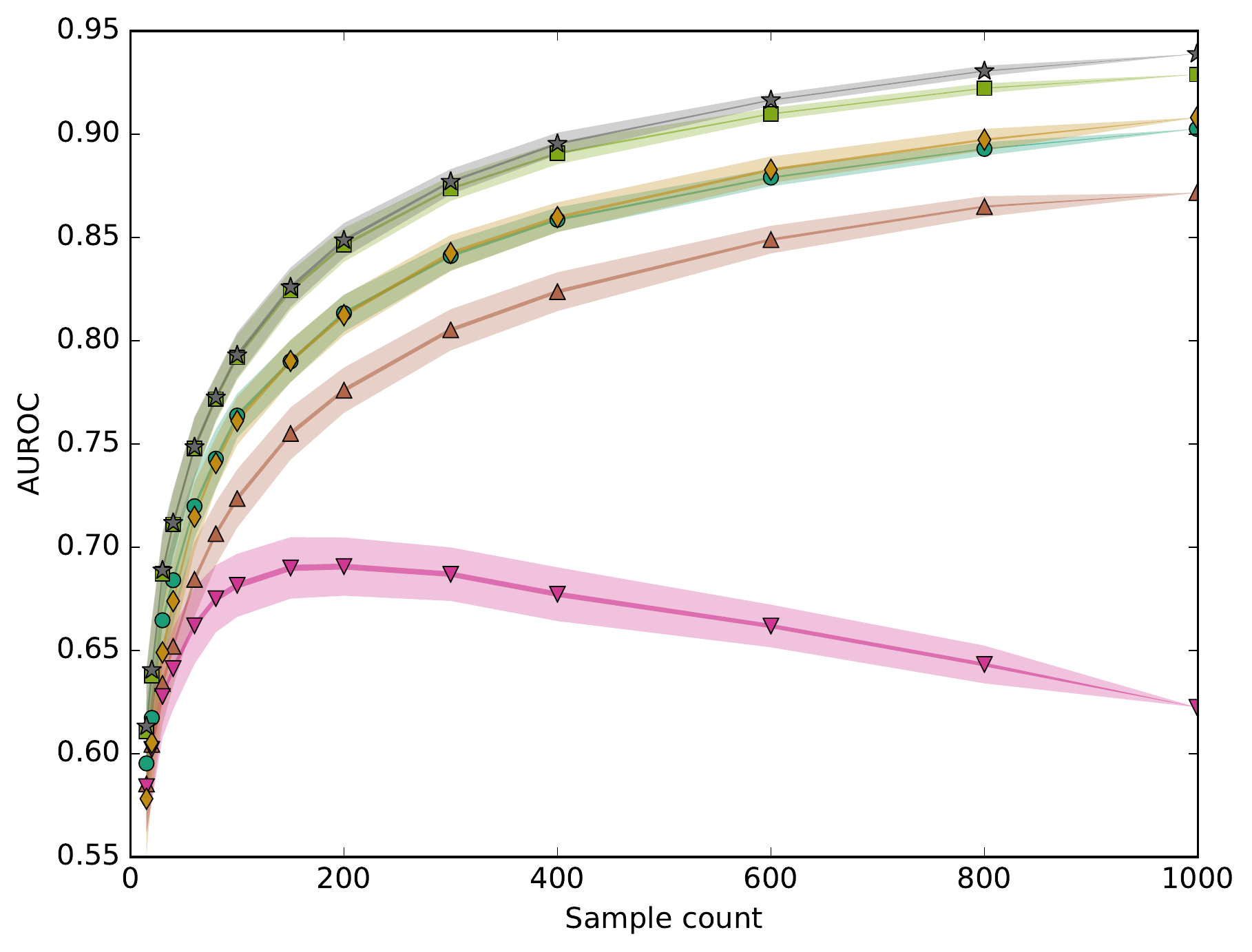}}
\subfigure{\includegraphics[width=0.37\textwidth]{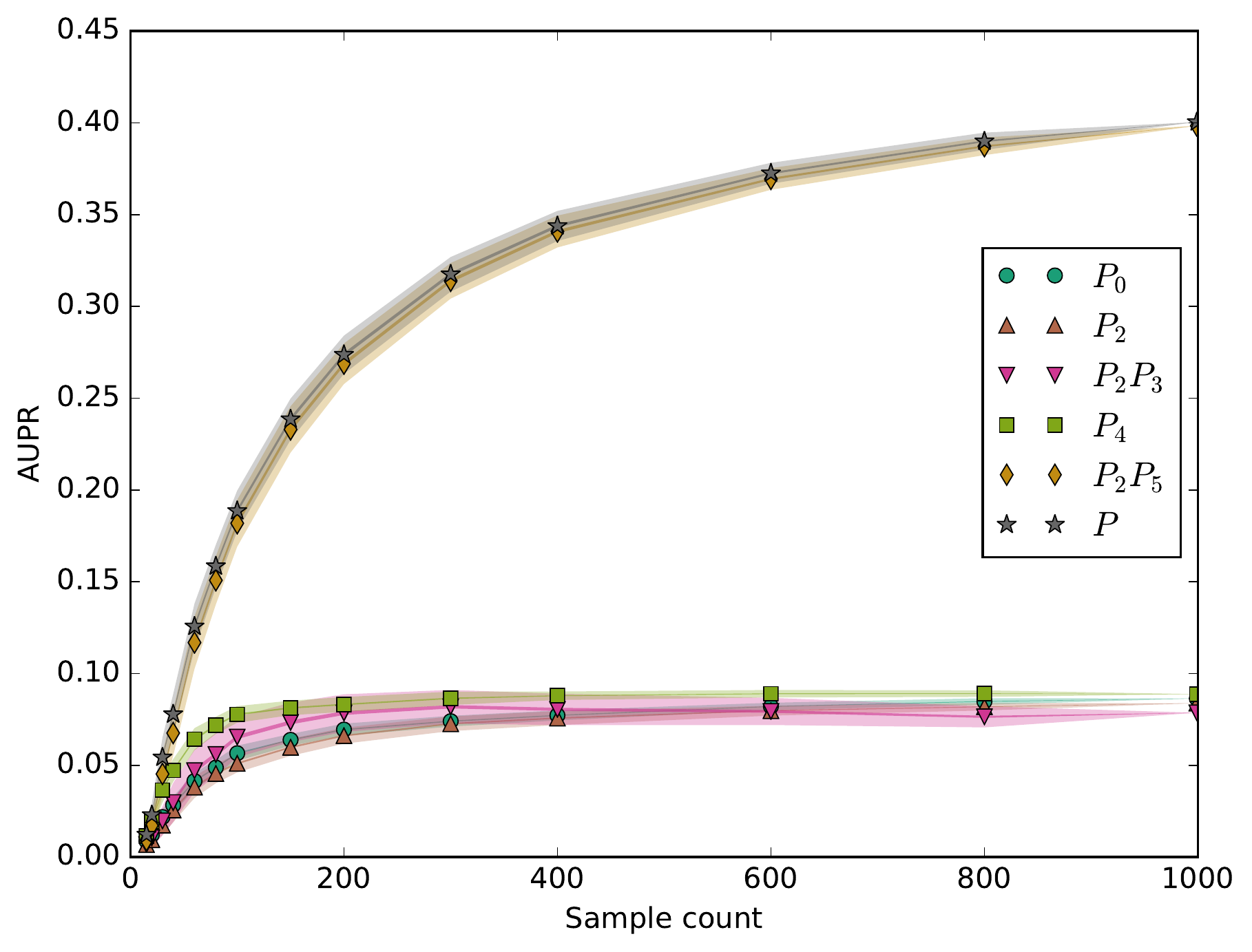}}\\
\subfigure{\includegraphics[width=0.37\textwidth]{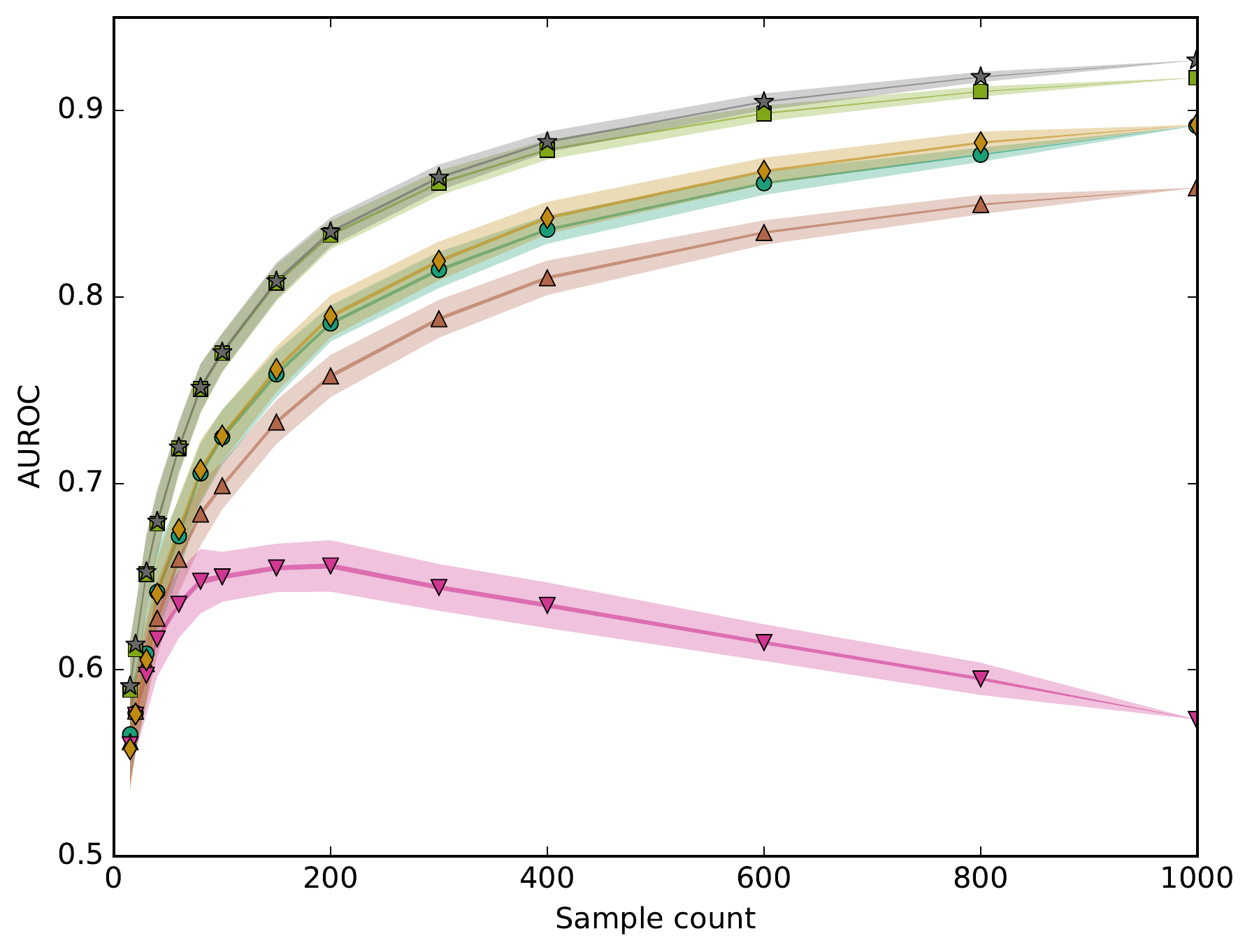}}
\subfigure{\includegraphics[width=0.37\textwidth]{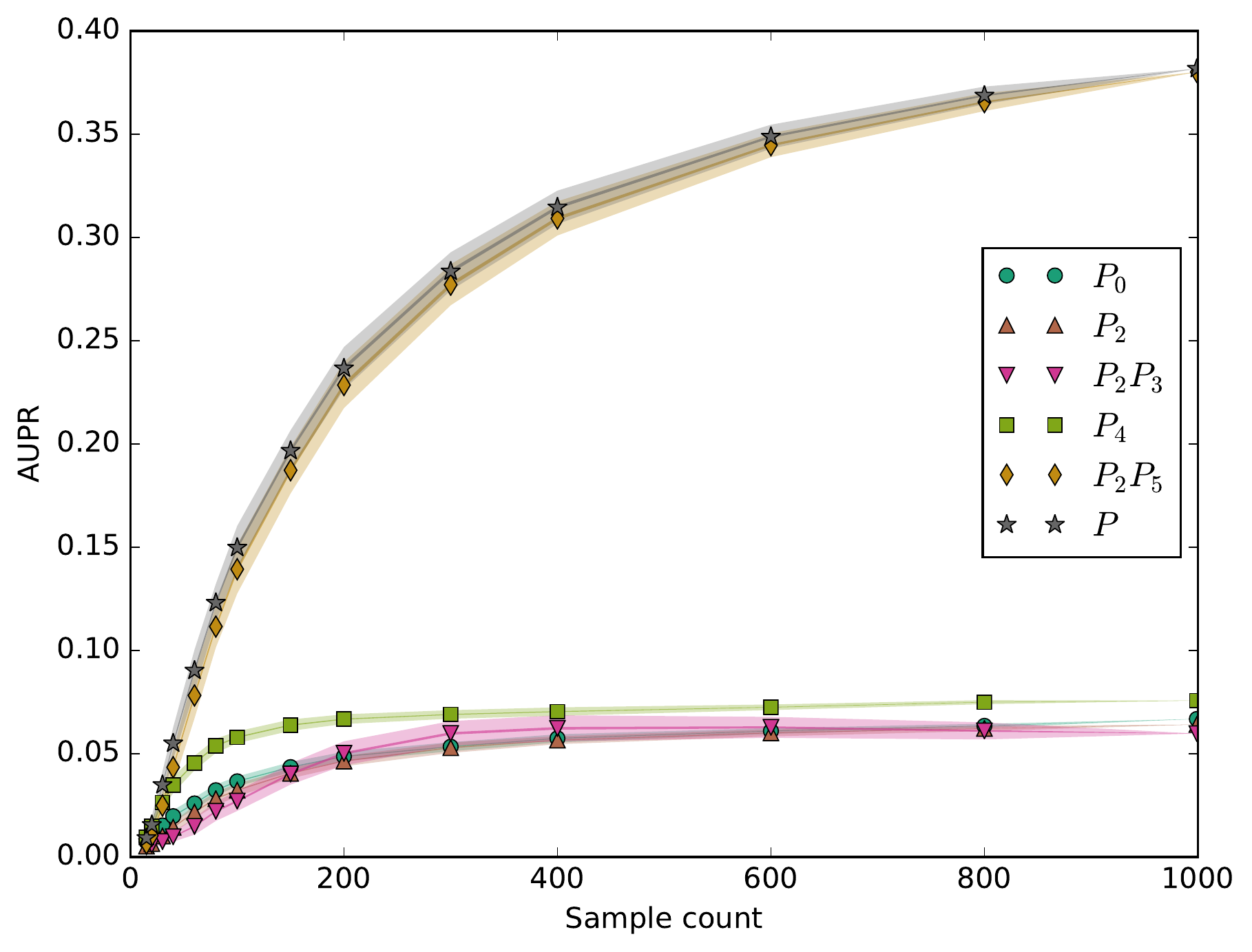}}\\
\subfigure{\includegraphics[width=0.37\textwidth]{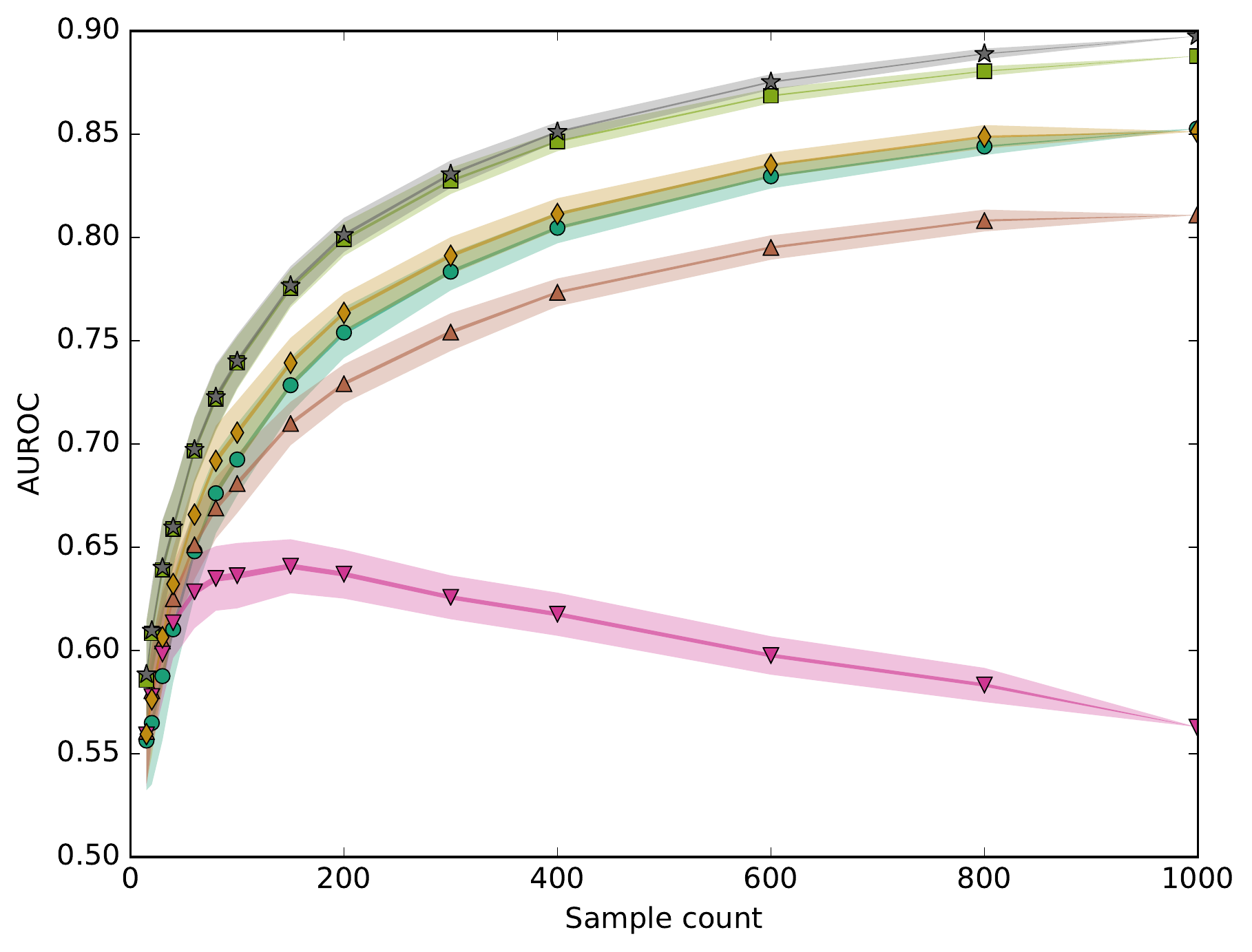}}
\subfigure{\includegraphics[width=0.37\textwidth]{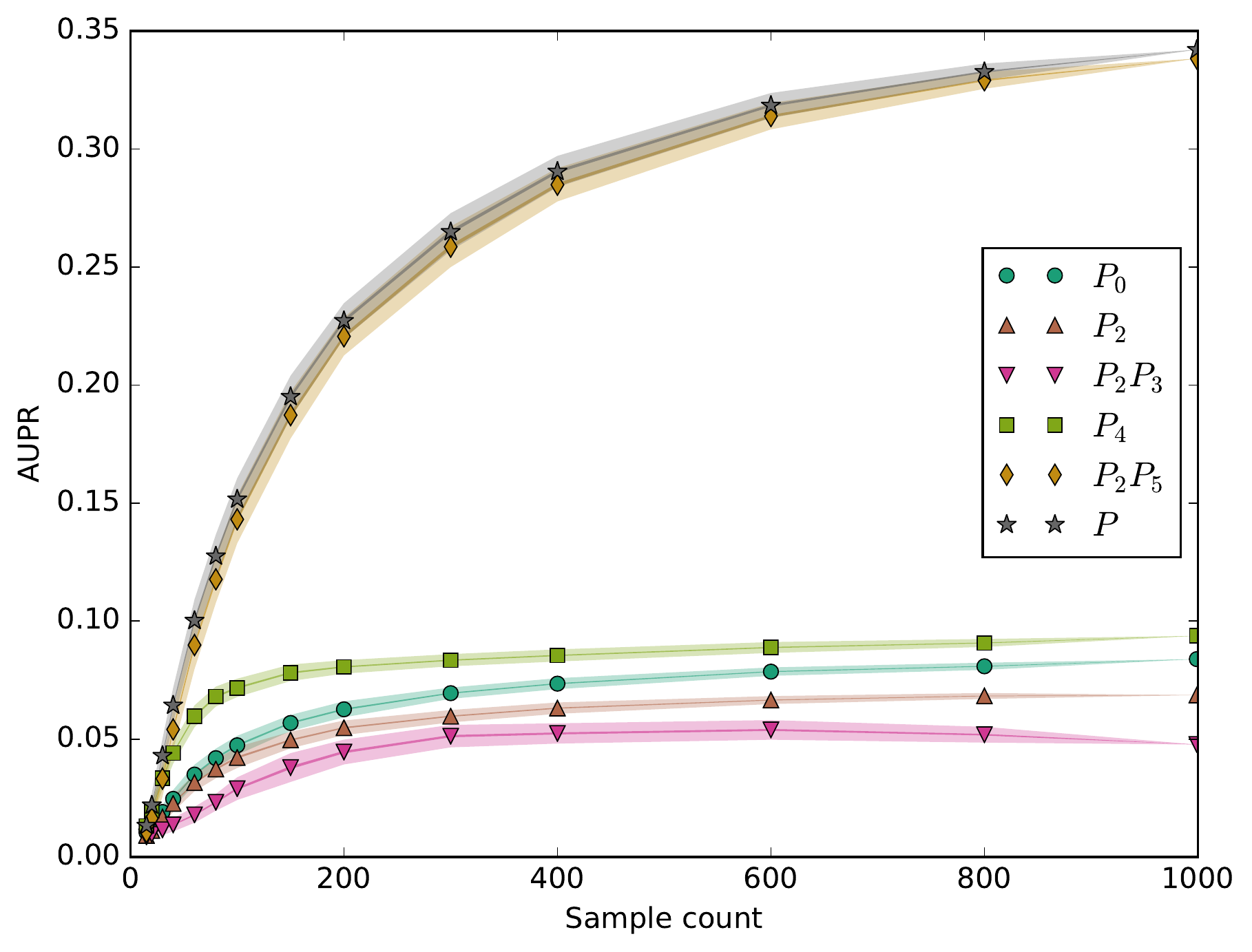}}\\
\caption{The mean AUROC and AUPR on subsampled data are shown for causal inference with traditional and new tests, together with the baseline correlation test. Every marker corresponds to the average AUROC or AUPR at specific sample sizes. At every sample size we performed 100 subsampling. Half widths of the lines and shades are the standard errors and standard deviations respectively, of AUROC or AUPR. Figures from top to bottom correspond to datasets 1, 2, 3, 5. For dataset 4, see \refig{dcomb}.\label{fig-dcomb-other}}\end{figure}

DREAM challenge contains five datasets that have 999 samples. With numbering 1 to 5, they each contain different number of true regulations, from $\sim1000$ to $\sim5000$ incrementally, for the purpose of characterizing regulatory networks of different complexity. Performances on the fourth dataset are shown in the main article, and the rest here in \refig{dcomb-other}.

We compared the performances of \pkg's new test ($P$), \pkg's correlation test ($P_0$), \pkg's traditional causal inference test ($P_T$), and CIT on all 15 datasets against the published leaderboard of the DREAM5 Systems Genetics Challenge \cite{DREAM:5}. \pkg's new test achieved highest AUROC and highest AUPR on all 15 datasets (\reftab{dream}). It is important to note that best performers can differ on different datasets or on AUROC and AUPR. For instance, the challenge winner \cite{vignes2011gene} attained best AUROC on 6/15 datasets and best AUPR on 5. When compared to other inference methods that also reported improvements \cite{ackermann2012teamwork,flassig2013effective,vananh2014gene}, \pkg{} demonstrates additional virtues besides the inference accuracy. Its nonparametric nature ensures robust performances across datasets without parameter tuning. Its pairwise computation scales linearly in time with the number of regulators, targets, and samples, as opposed to multivariate regression methods, providing scalability to datasets that are orders of magnitude larger.

\subsection{\pkg{} achieves best performance on miRNA target predictions from Geuvadis dataset \label{ssec-vag-sup}}

We compared the performance of \pkg{} on miRNA target prediction from the Geuvadis dataset with a suite of network inference methods that are based on gene expression data and, for some, genotype information. They include:
\begin{itemize}
\item All methods in the miRLAB package \cite{Le:2015}:
\begin{itemize}
\item\textbf{Correlation methods:} Pearson correlation, Spearman correlation, Kendal correlation, Distance correlation (dcov), Hoeffding's D measure (hoeffding), Randomized Dependence Coefficient (rdc), and Mutual Information (mi).
\item\textbf{Regression methods:} Lasso, and Elastic-net (elastic).
\item\textbf{Other methods:} Z-score, and Roleswitch (promise).
\item\textbf{Failed method:} Intervention calculus (ida) method failed due to excessive memory usage (greater than 16GB) and hence is excluded from comparison.
\end{itemize}
\item GENIE3 \cite{Huynh-Thu:2010} which utilizes random forests.
\item CIT \cite{Millstein:2009,Millstein:2016} which performs multiple causal inference tests with genotype data.
\item Multiple tests implemented in \pkg{}, including: traditional (\pkg-$P_T$), new (\pkg-$P$), and correlation (\pkg-$P_0$) tests.
\end{itemize}
The Trigger package\cite{Chen:2007} was not attempted because its eQTLs discovery routine exceeds both our memory and time limitations.

\begin{figure}[h!]
\center
\subfigure{\includegraphics[width=.8\textwidth]{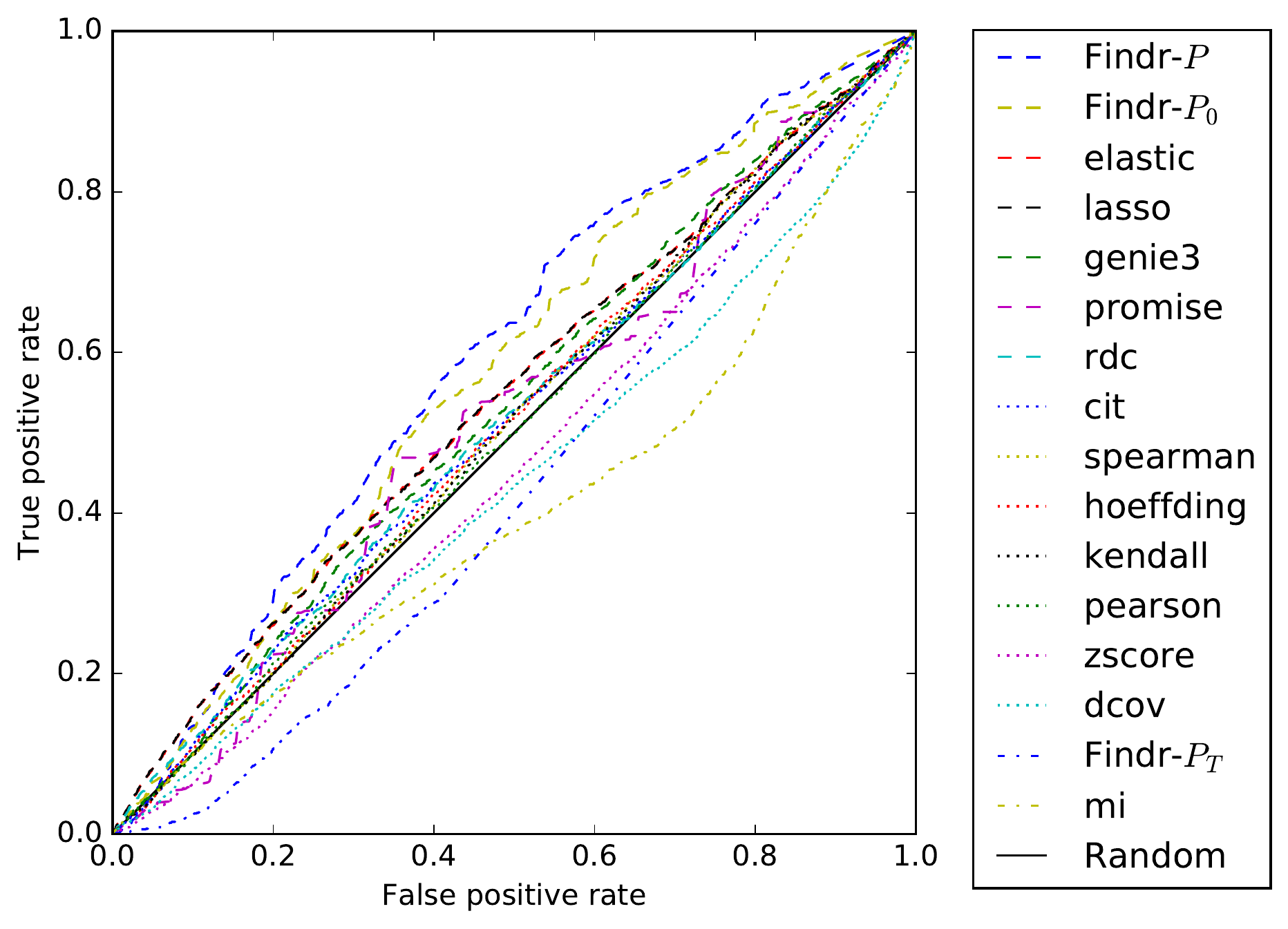}}\\
\subfigure{\includegraphics[width=.8\textwidth]{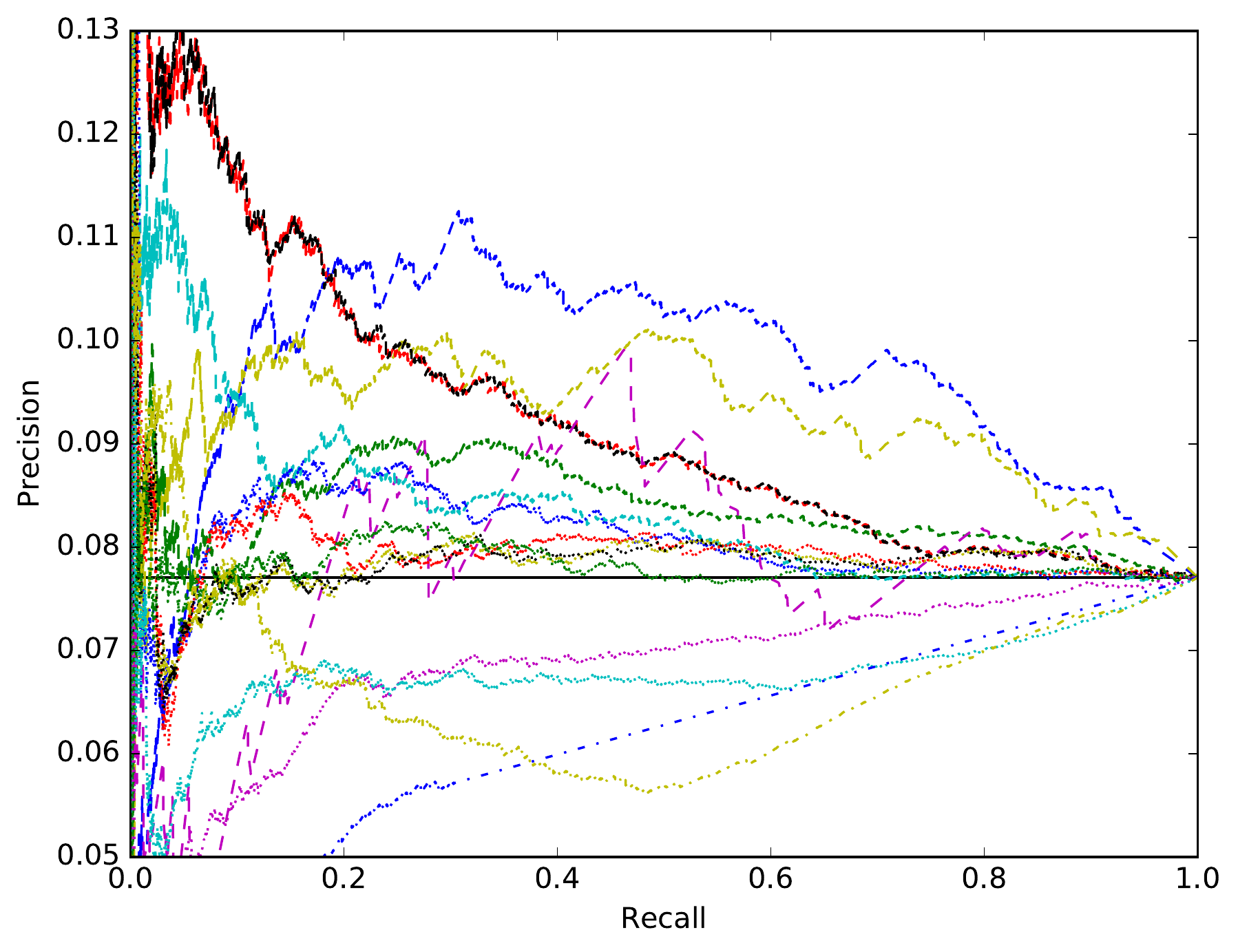}}\\
\caption{ROC (top) and PR (bottom) curves of miRNA target predictions were compared for \pkg's traditional, new, and correlation tests, GENIE3, CIT, and 11 methods in miRLAB, based on Geuvadis data. The solid black lines correspond to expected performances from random predictions. A higher curve indicates better prediction performance.\label{fig-Geuvadis}}
\end{figure}

Their AUROCs, AUPRs, and running times are presented in \reftab{Geuvadis}. The ROC and PR curves are shown in \refig{Geuvadis}. We observed the following results:
\begin{itemize}
\item The correlation test topped among methods without genotype information, and in particular performs much better than Pearson and Spearman correlations. The performance gain is due to Bayesian inference, which is able to account for different gene roles such as hubs. This suggests the possibility of replacing correlation based methods with their FPR estimation counterparts in future inference of genetic regulations.
\item The new test $P$ performed better than correlation test $P_0$. This is the first comparative study to demonstrate the effect of genotype information in the inference of gene regulatory relations.
\item The traditional causal inference test performed worse than random predictions. This confirms with real data that the indirect secondary test fails to identify true but weak regulations. The independence test had negligible effect as the sample size is small (not shown in \refig{Geuvadis}).
\item \pkg{} achieved higher AUROC and AUPR than all other methods attempted. It was also much faster than all other methods, especially CIT which also includes genotype data.
\item \pkg{} obtained a lower precision than lasso and elastic-net at small recalls. This might be explained by the fact that lasso and elastic-net are multi-variate methods which incorporate all other gene expression levels besides pairwise information, and therefore exclude indirect regulators better.
\end{itemize}

\subsection{\pkg{} predicts transcription factor targets with accurate FDR estimates\label{ssec-genegene}}
Since CIT is much slower than \pkg, with CIT we were only able to infer genetic regulations of the intersection set of the prediction and groundtruth datasets, as opposed to inferring all possible regulations with \pkg.

As shown in \refig{LCL}, AUPRs and AUROCs for TF target prediction with respect to known targets from siRNA silencing or TF-binding did not exhibit substantial differences, other than modest improvement over random predictions. We believe this is due to the unavoidable noise and size limitations in groundtruth data, which lead to large fluctuations in evaluation metrics and therefore could not compare methods perfectly. Furthermore, AUPR and AUROC test the entire ranked list of predictions for overlap with the groundtruth and will miss differences in enrichment for true positives between methods if they occur only among a small fraction of top-ranked predictions.

\begin{figure}
\center
\hspace{-0.0\textwidth}\subfigure{\includegraphics[width=\textwidth]{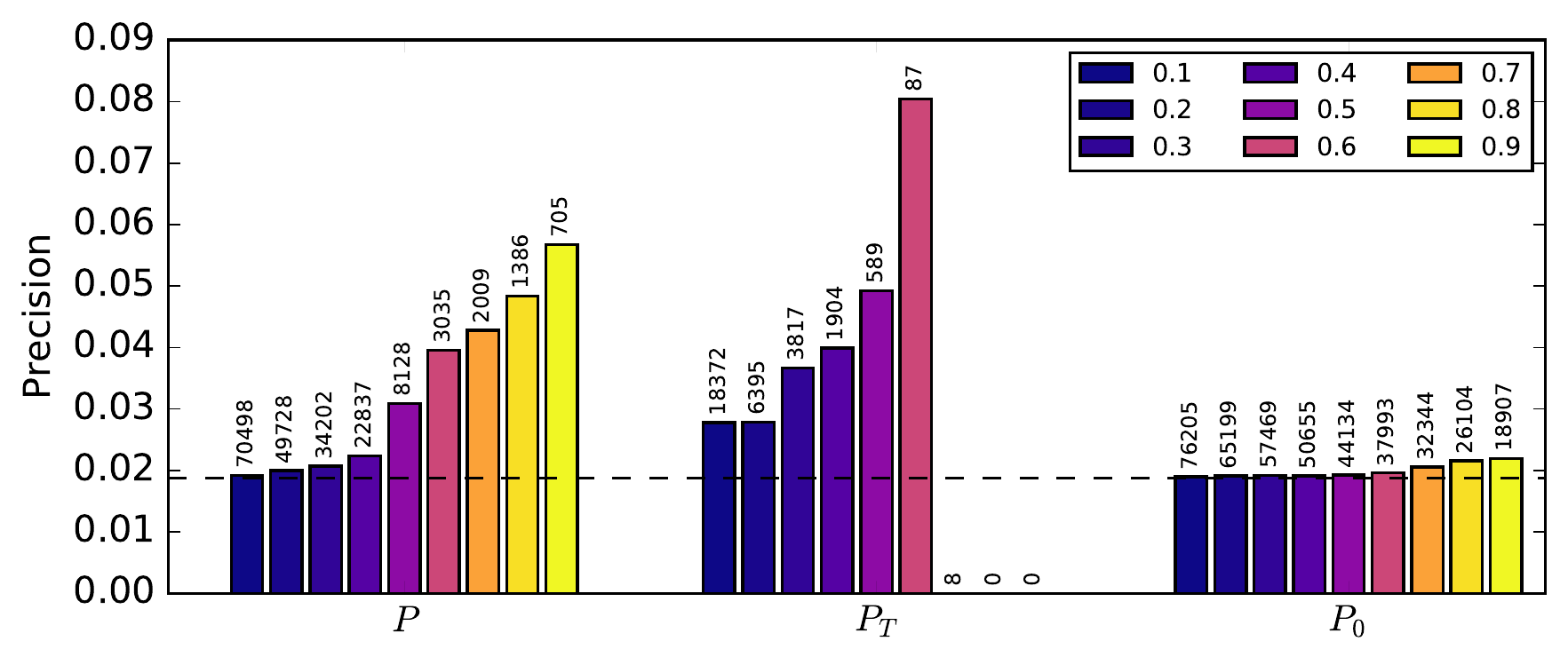}}
  \caption{Inference precision at estimated precision cutoffs 0.1 to 0.9 with respect to groundtruth network derived from TF binding of 14 TFs from ENCODE data \cite{gerstein2012architecture}. The number above each bar indicates the number of positive predictions at the corresponding threshold. The dashed line is precision from random predictions. \label{fig-encode}}
\end{figure}

The construction of gold standard regulatory networks from TF-binding data is dependent on how TF binding sites are mapped to target genes. Here a TF regulatory interaction was assumed if  a gene had at least two binding sites for a particular TF within 10kb of its transcription start site (TSS) \cite{cusanovich2014functional}. We repeated the analysis using the high-confidence (binding within 2.5kb) TF-target network derived from ENCODE from ChIP-sequencing data of 119 TFs in five cell types, including the lymphoblastoid cell line GM12787 \cite{gerstein2012architecture}. Fourteen TFs had a significant eQTL in the Geuvadis data. The analysis results are consistent (\refig{LCL} and \refig{encode}).

\section{Discussion}
\ed{Rapid advances in sequencing technology have made it possible to generate genome and transcriptome variation data in diverse cell types, tissues and organs across hundreds of individuals, and thereby allow to study the impact of genetic variation on complex human diseases and quantitative traits in exquisite detail. It has been known that by using genetic variation at eQTLs as a causal anchor or instrumental variable, causality between two genes can be inferred through a conditional independence test, provided they have no common upstream regulators (hidden confounders). However this assumption is at odds with the fact that genes are organized in complex hierarchical gene regulatory networks where common regulation of regulator-target pairs is an abundant feature. Despite of that, existing causal inference softwares rely on the conditional independence test under this assumption and consequently are unable to provide reliable inference.}

\begin{itemize}
\item \ea{\textbf{Existing softwares/methods not considered in this study:}}

\ed{Meanwhile, }No previous causal inference software has been able to handle the size of modern datasets. \ea{In} Trigger\cite{Chen:2007}, \ed{requires} eQTL discovery \ea{and causal inference are inextricably linked,} \ed{by itself} which is impractical for \ea{any sizeable dataset}, especially for mammalian species; \ea{much more efficient tools for eQTL discovery have meanwhile become available \cite{Shabalin:2012, Qi:2014}}. NEO\cite{Aten:2008} is similar to CIT\cite{Millstein:2016} in the tests employed, but even slower. \ed{in spite of being} It is able to account for multiple eQTLs\ea{, but} \ed{and} avoids permutations \ea{using asymptotic $\chi^2$ approximations, which result in} \ed{with} deflated estimations for null distributions \ea{as shown in } \refssec{match}. \ed{These computational limitations have prevented us from understanding the extent to which integration of genome and transcriptome variation data can inform the inference of causal gene networks. The community also lacks a software that is efficient, accurate, and robust for causal inference. Network} \ea{As stated in the discussion, multi-variate network} inference can be a downstream analysis based on pairwise causal inference, \ea{but evaluating such } \ed{whose} methods\cite{Liu:2008,Zhu:2008,Chaibub-Neto:2010} \ea{fell out of the scope of this paper.} \ed{did not participate our evaluation.} \ea{Finally, note that} CIT computes null distributions by permutation, separately for every gene pair, requiring months to tens of years of CPU time on a modern dataset. \ea{We were only able to include comparisons to CIT on the Geuvadis data by limiting it to the subset of gene pairs in the ground-truth tables, which correspond to 0.2\% of all gene pairs for the TF target prediction case.}

\item \textbf{\ea{Considerations on human datasets:}}

\ed{On the other hand, proving} \ea{Proving} in an easily reproducible manner that the\ed{se} results \ea{on simulated DREAM data} also hold on human data is challenging, due to the often restricted access to individual-level genotype data and the limited availability of reference databases of known interactions, especially when cell type specificity is taken into account. We performed our evaluation on the Geuvadis data, which provides transcriptome data in lymphoblastoid cells of nearly 400 individuals whose genotype data is publicly available through the 1000 Genomes project. We found that \pkg{} predicted miRNA targets more accurately than other causal inference methods and a panel of machine learning methods that used expression data alone. Although the absolute power to predict miRNA targets may appear modest, we were primarily interested in the relative performance of various methods, and did not incorporate information about known miRNA biology, such as a preference for negative correlations or the presence of miRNA seed sequences. Moreover, since miRNAs are frequently studied in the context of diseases such as cancer, the ground-truth set of experimentally confirmed targets may represent a biased set of interactions that are not necessarily present in the \ed{current} \ea{lymphoblastoid} cell type studied.

To address the issue of cell type specificity, we analysed the predicted targets of 25 transcription factors for which either functional targets from siRNA silencing experiments or DNA-binding targets from ChIP-sequencing or DNase footprinting in lymphoblastoid cells were available. For both data types we found that  \pkg's new test achieved a 2 to 5-fold enrichment for known TF targets compared to using TF-target coexpression alone, showing that causal inference is indeed able to prioritize highly probable causal interactions among coexpressed genes. Although \pkg's new test and conditional independence based causal inference tests resulted in similar performances in this case, the estimated FDRs of the traditional method were greatly inflated, such that enrichment for known interactions was only observed at estimated local FDR >40\%. This reaffirms the finding, consistently observed in all our analyses, that the conditional-independence methods are over conservative for calling causal interactions in the context of complex regulatory networks.
\end{itemize}

\ed{Here we developed a highly efficient software package \pkg{} (Fast Inference of Networks from Directed Regulations) implementing novel and traditional causal inference tests. Application of \pkg{} on synthetic gene networks with known genetic architecture from DREAM5 Systems Genetics challenge showed that traditional causal inference test underperformed the correlation test, due to failures of the secondary test to detect weak secondary linkage, and of the conditional independence test to account for hidden confounders. Accordingly, we proposed and implemented our novel test in \pkg, which consistently outperformed all participants of the DREAM5 Systems Genetics challenge as well as the correlation test on every single dataset.}




\newpage

\begin{figure}
\center
\hspace{-0.0\textwidth}\subfigure{\includegraphics[width=0.48\textwidth]{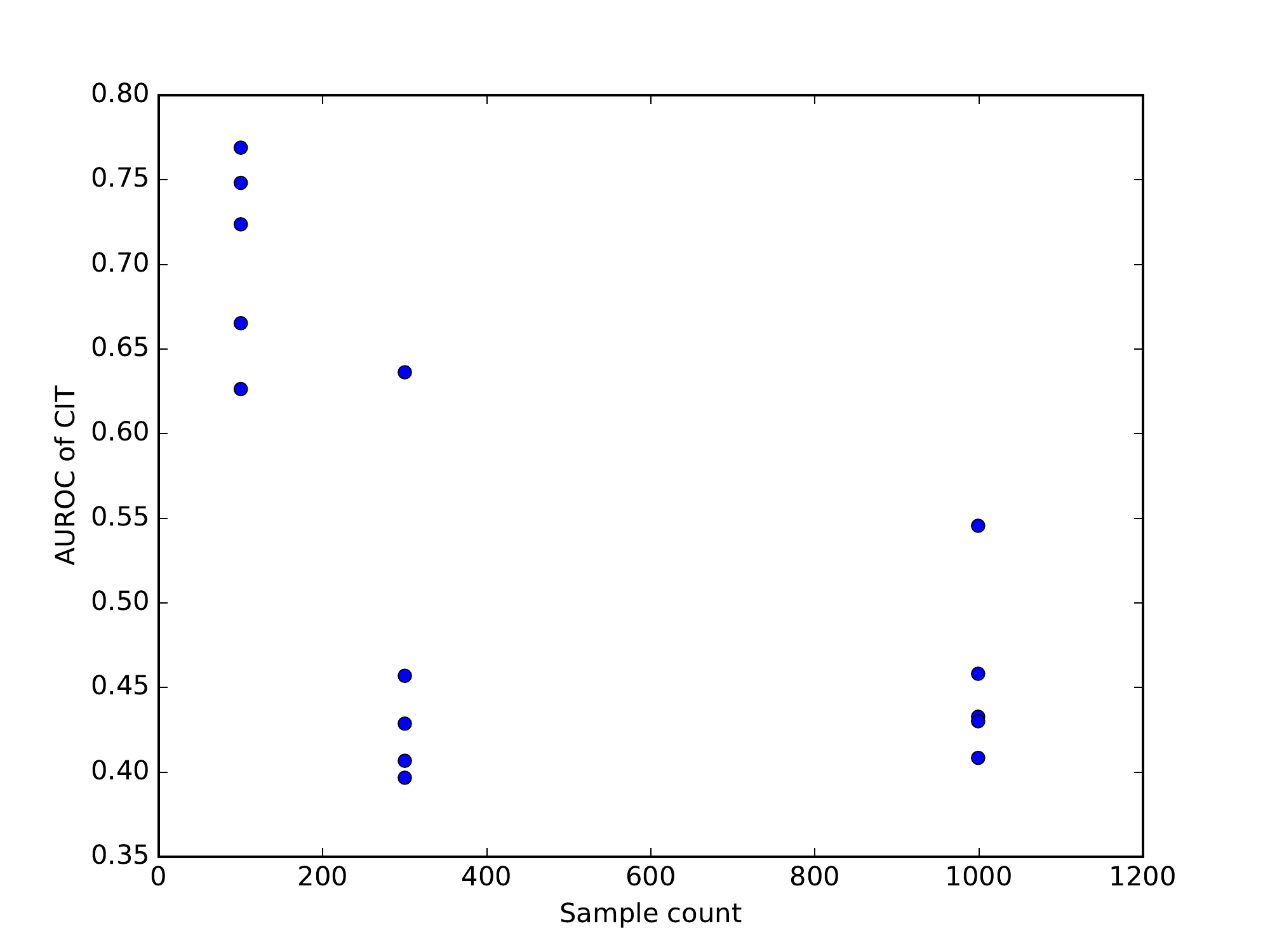}}\hspace{-0.02\textwidth}
\subfigure{\includegraphics[width=0.48\textwidth]{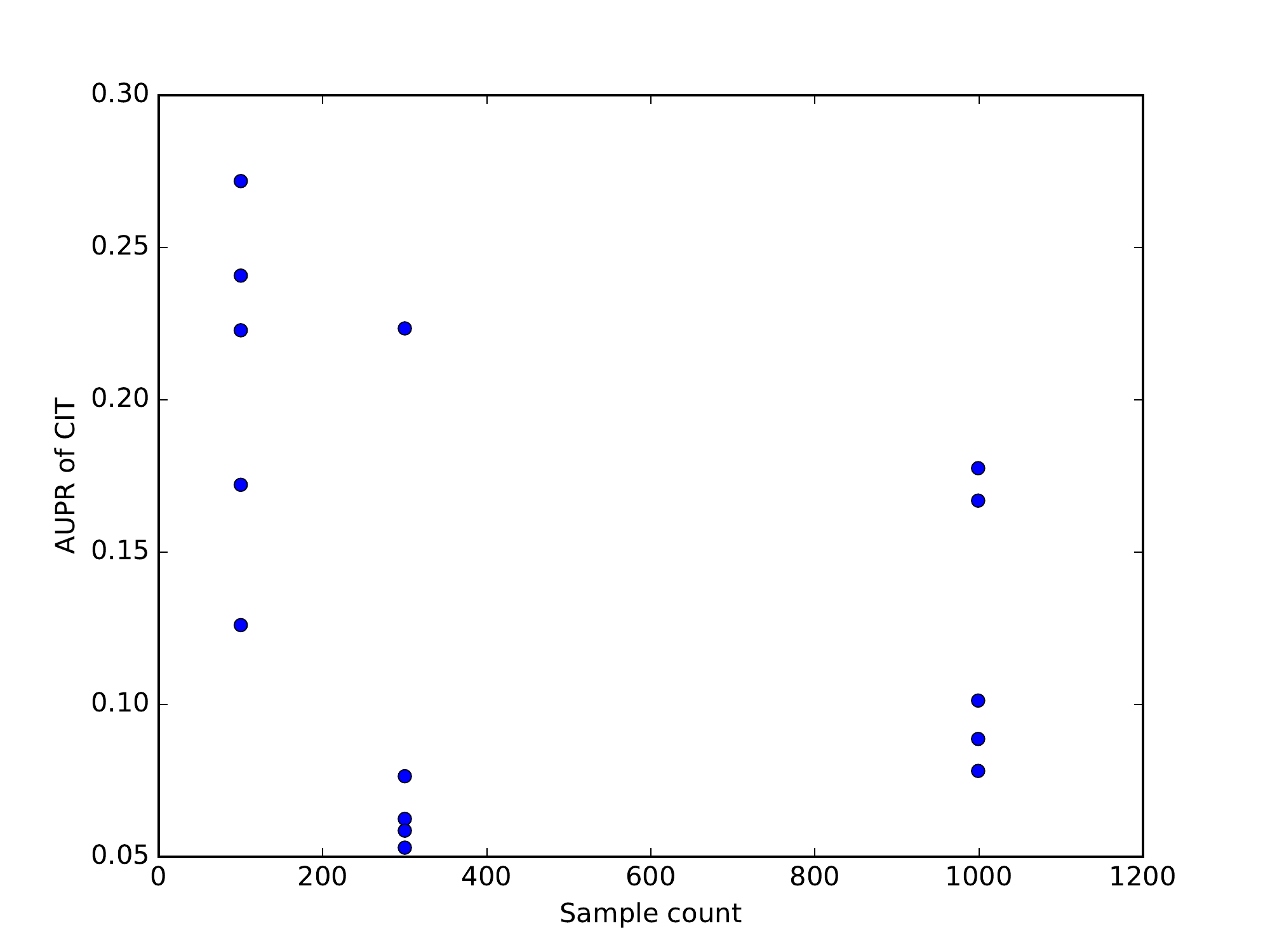}}
\caption{The AUROC and AUPR of CIT \cite{Millstein:2016} are shown for all 15 datasets of DREAM challenge. Every marker corresponds to the AUROC or AUPR of one dataset. CIT is an R package that includes the conditional independence test, along with tests 2 and 5, while also comparing $E\rightarrow A\rightarrow B$ against $E\rightarrow B\rightarrow A$. The subsampling analysis on CIT was not feasible due to its low speed.\label{fig-cit}}
\end{figure}

\begin{figure}
\begin{center}
\begin{tabular}{p{0em}p{0.43\linewidth}p{0em}p{0.43\linewidth}}
\vspace{0pt}\textbf{{\large A}} &\vspace{0pt}\includegraphics[width=\linewidth]{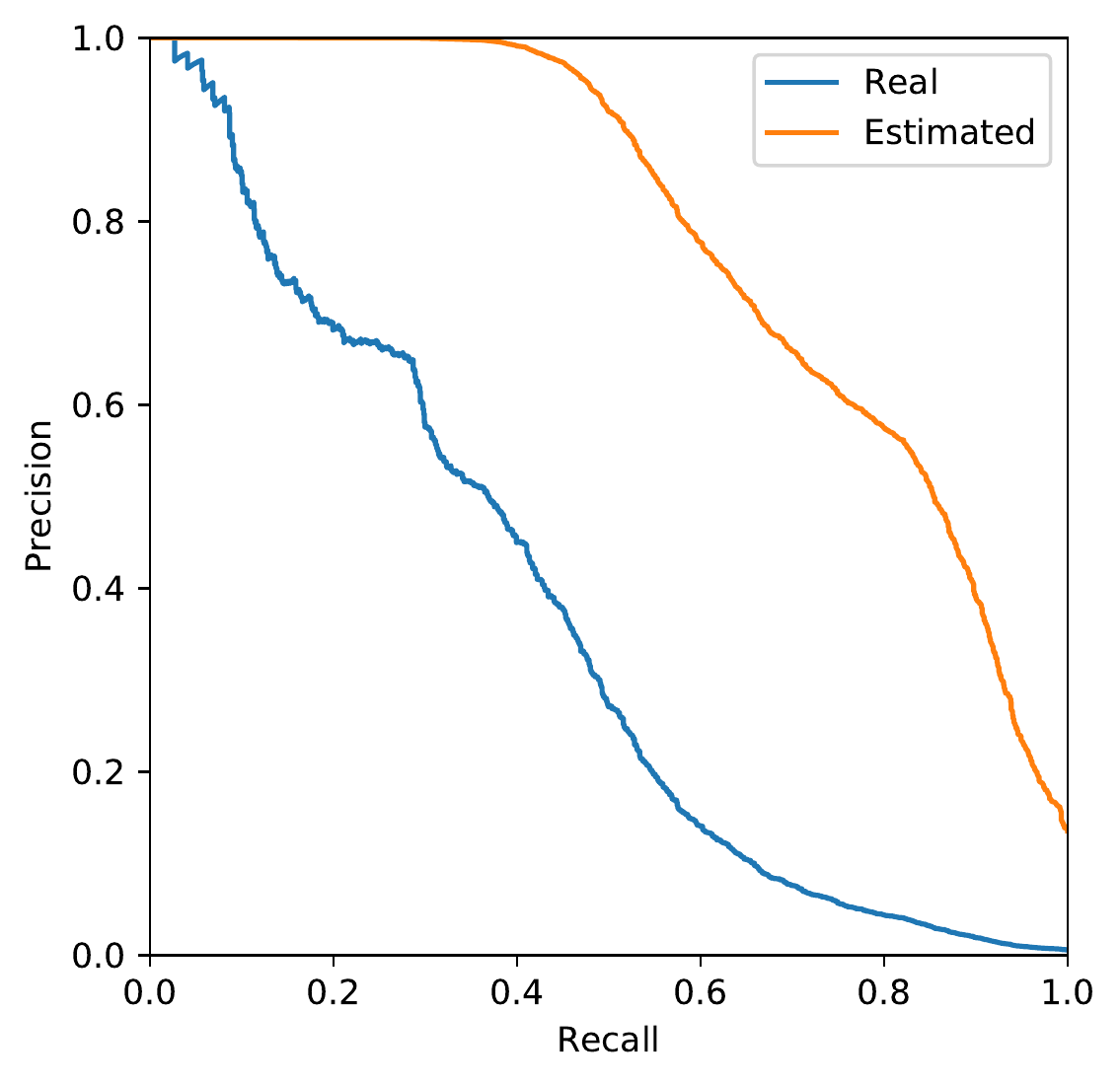}&
\vspace{0pt}\textbf{{\large B}} &\vspace{0pt}\includegraphics[width=\linewidth]{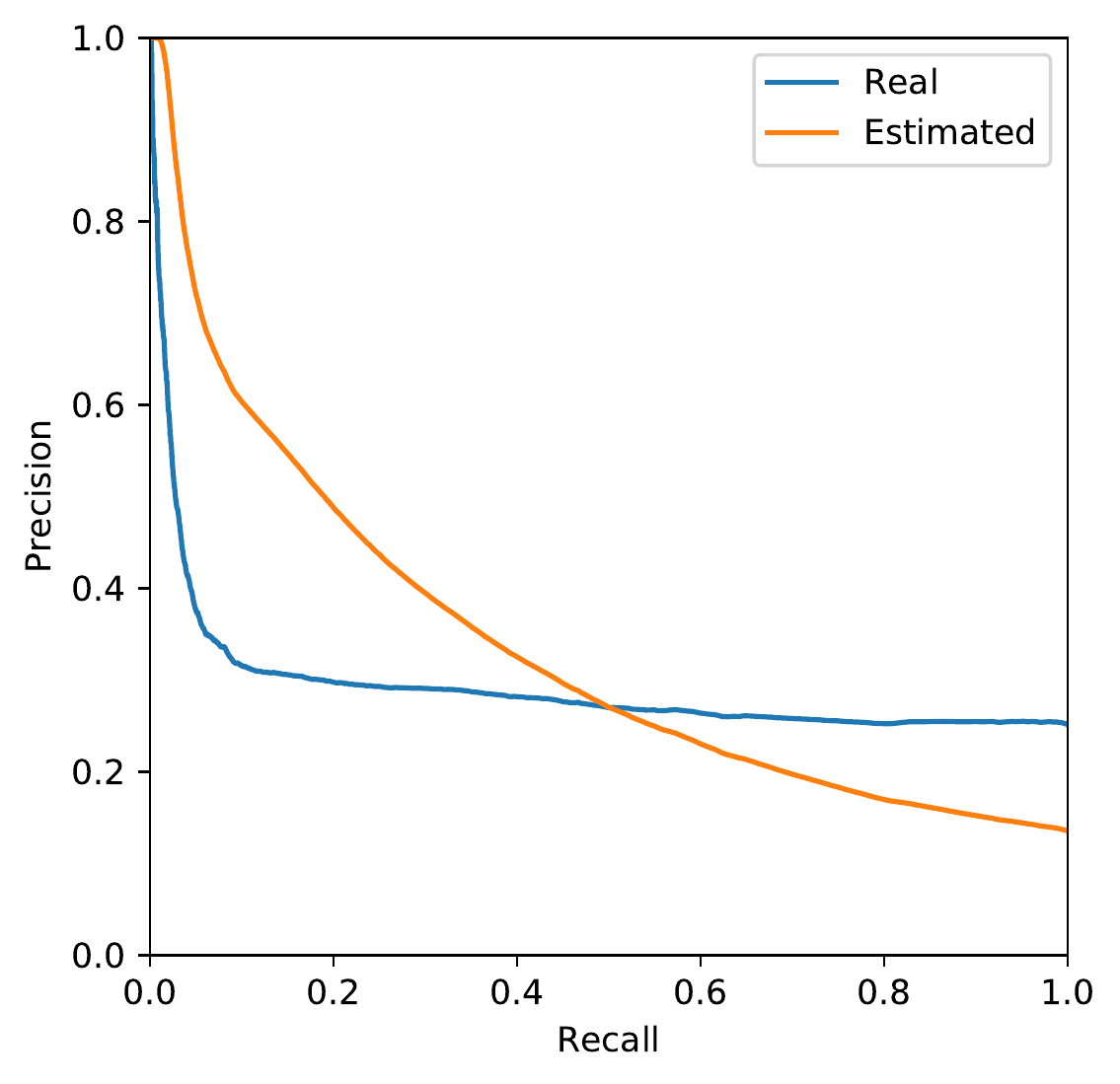}\\
\vspace{0pt}\textbf{{\large C}} &\vspace{0pt}\includegraphics[width=\linewidth]{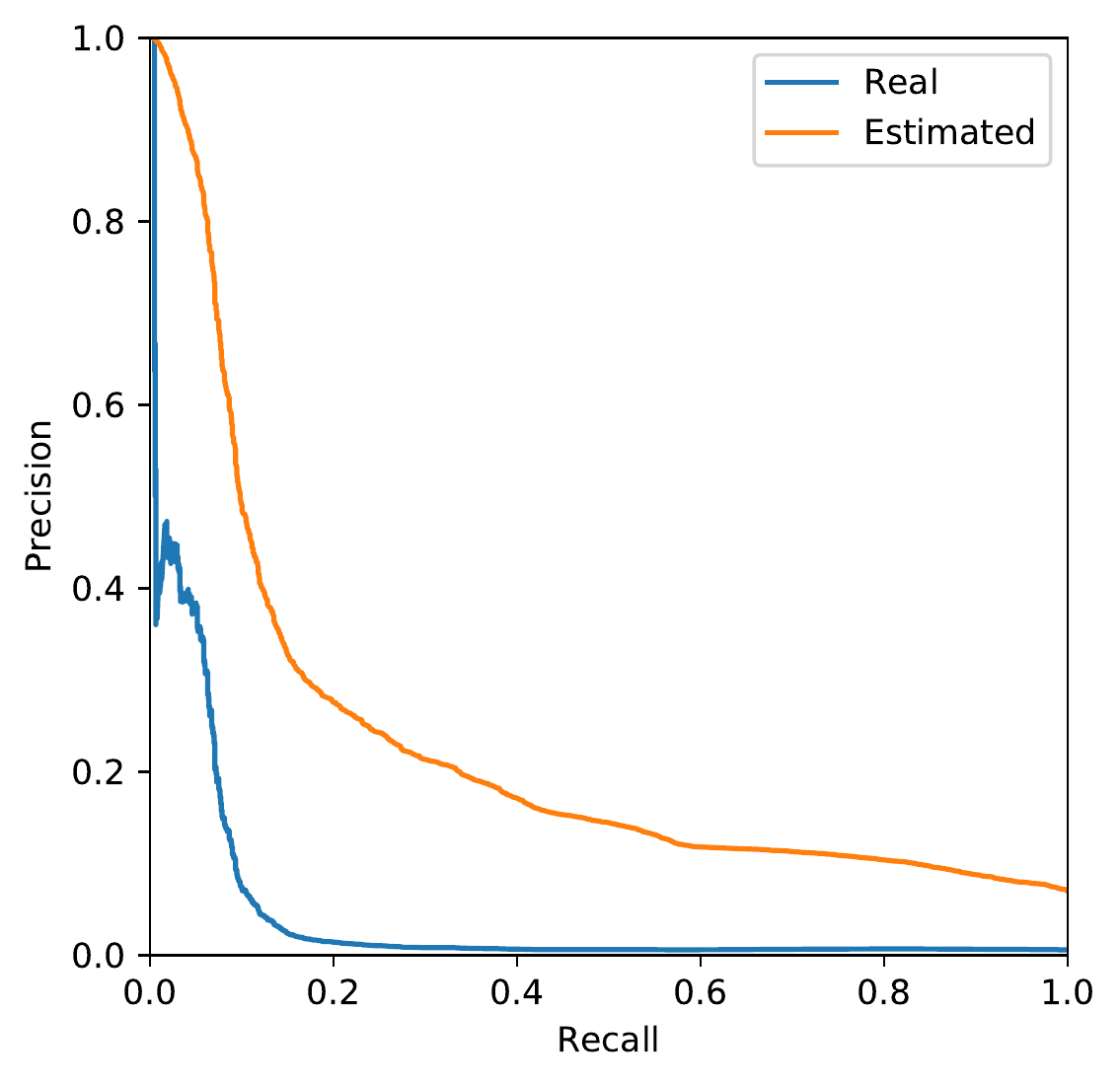}&
\vspace{0pt}\textbf{{\large D}} &\vspace{0pt}\includegraphics[width=\linewidth]{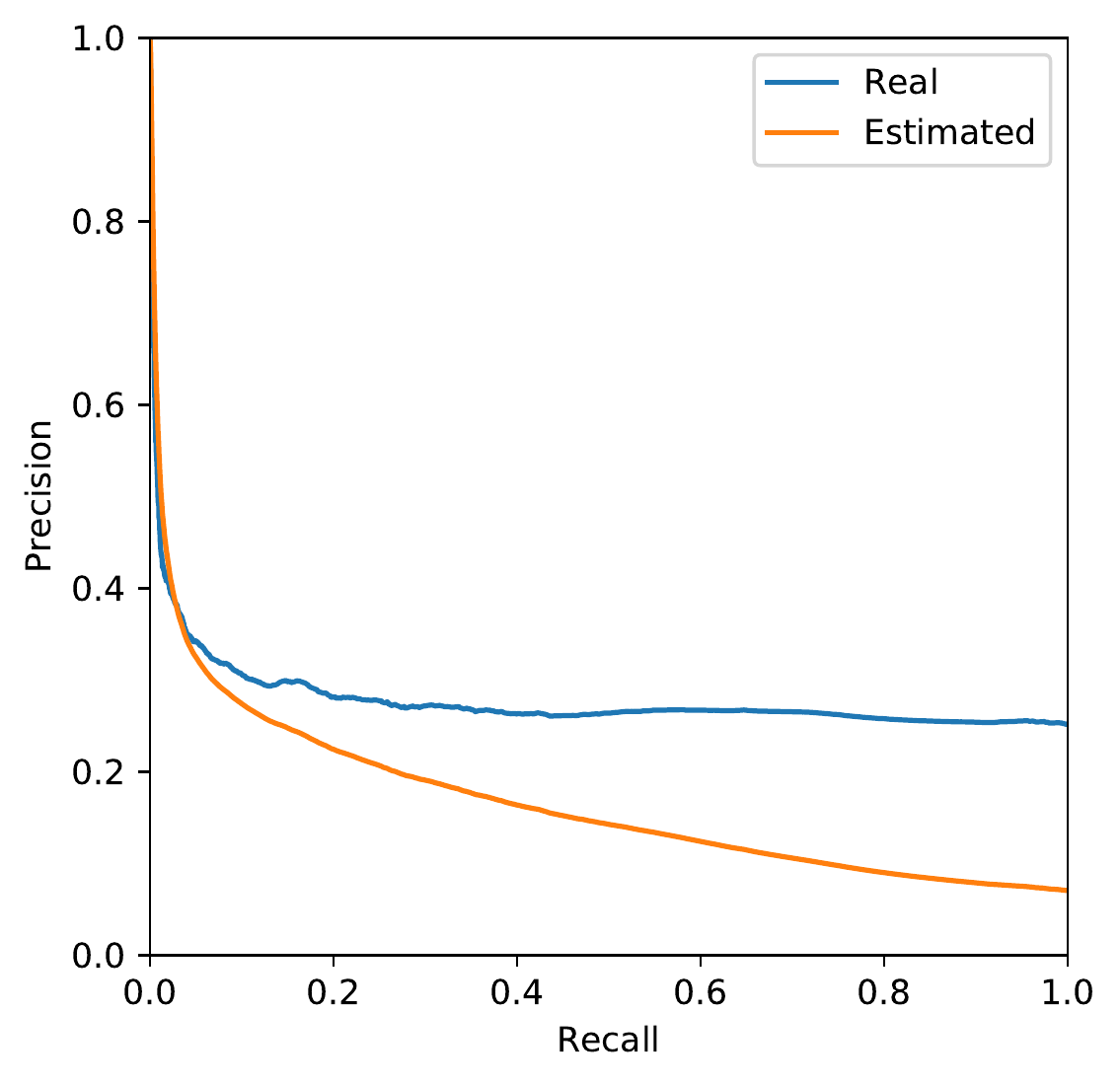}
\end{tabular}
\end{center}
\caption{\ea{Estimated and real precision-recall curves for dataset 4 of the DREAM challenge. The real precision was computed according to the groundtruth, whilst the estimated precision was obtained from the estimated FDR from the respective inference method (precision $=1-$FDR). Only genes with cis-eQTLs were considered as primary targets in prediction and validation. Both the novel (\textbf{A, B}) and the traditional (\textbf{C, D}) tests were evaluated. In \textbf{A, C} the original groundtruth table was used to validate predictions, whereas in \textbf{B, D} an extended groundtruth was used that also included indirect regulations at any level based on the original groundtruth.}\label{fig-dreampr}}
\end{figure}

\begin{figure}
\center
\subfigure[With confounder before conditioning on A \label{fig-indep-cf1}]{\includegraphics[width=0.3\textwidth]{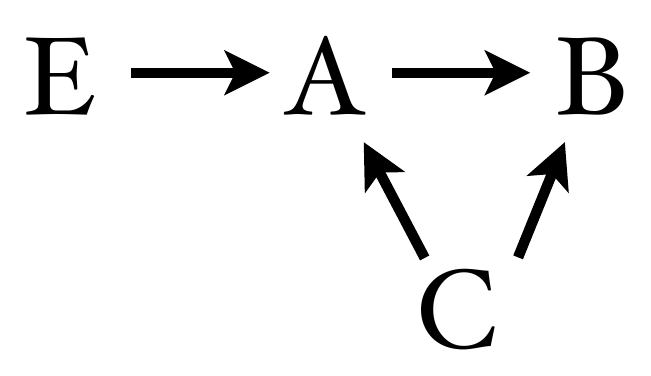}}
\hspace{0.05\textwidth}
\subfigure[With confounder after conditioning on A\label{fig-indep-cf2}]{\includegraphics[width=0.3\textwidth]{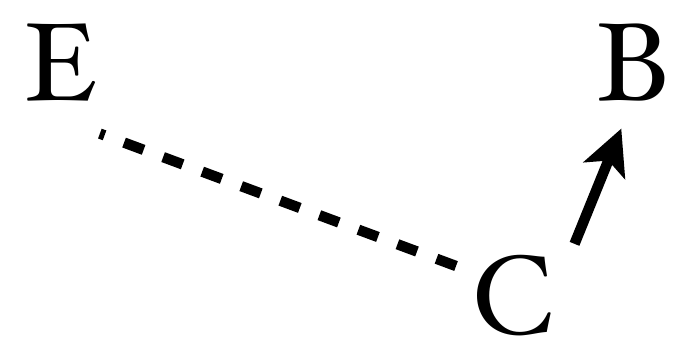}}
\caption{The conditional independence test fails in the presence of hidden confounders. When $A$ and $B$ are both regulated by a hidden confounder $C$, which is independent of $E$ (left), \ea{$A$ becomes a collider and} conditioning on $A$ would introduce inter-dependency between $E$ and $C$, which maintains $E\rightarrow B$ regulation (right). \label{fig-indep}}
\end{figure}

\begin{figure}
\begin{center}
\subfigure{\includegraphics[width=.4\textwidth]{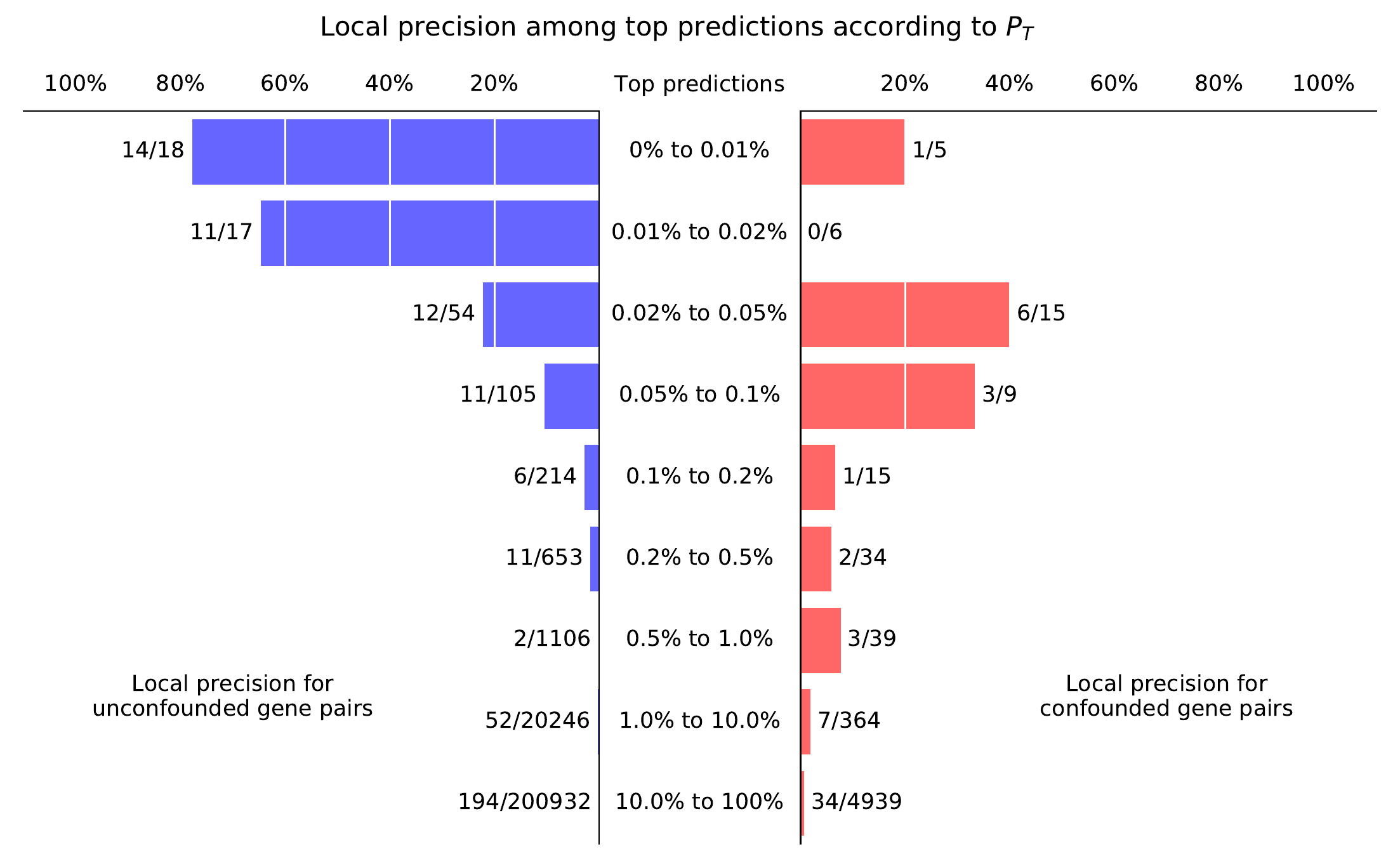}}
\subfigure{\includegraphics[width=.4\textwidth]{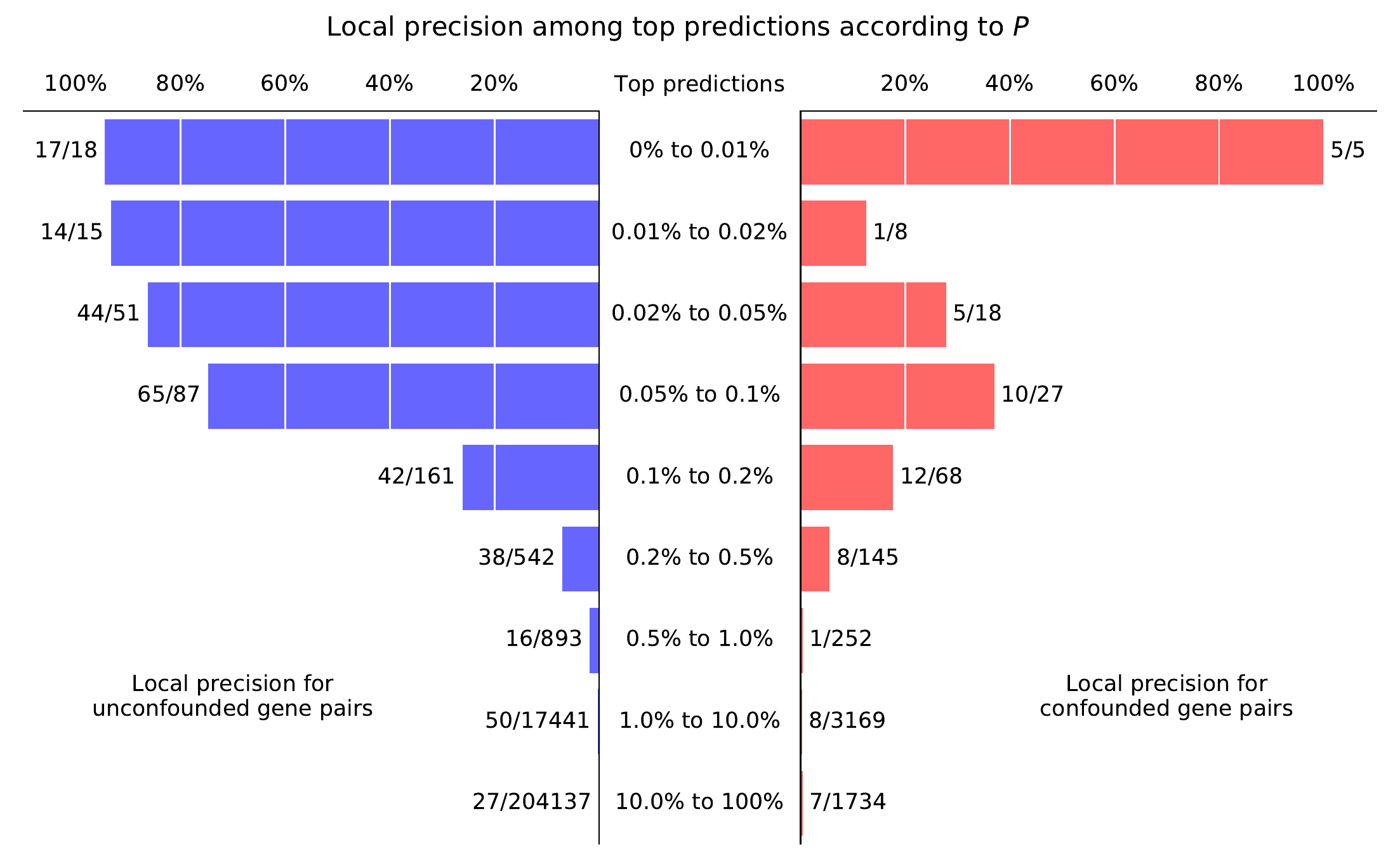}}\\
\subfigure{\includegraphics[width=.4\textwidth]{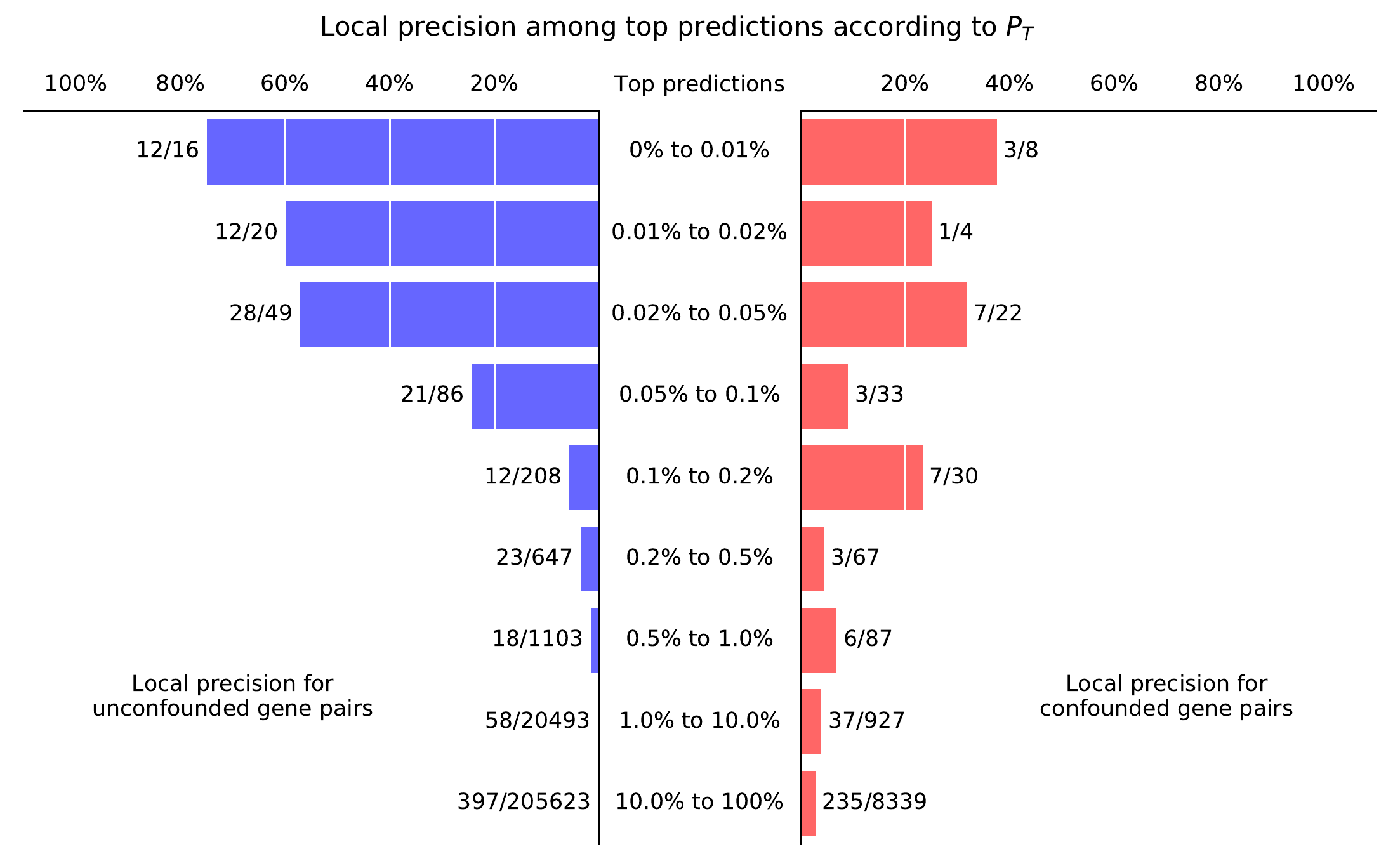}}
\subfigure{\includegraphics[width=.4\textwidth]{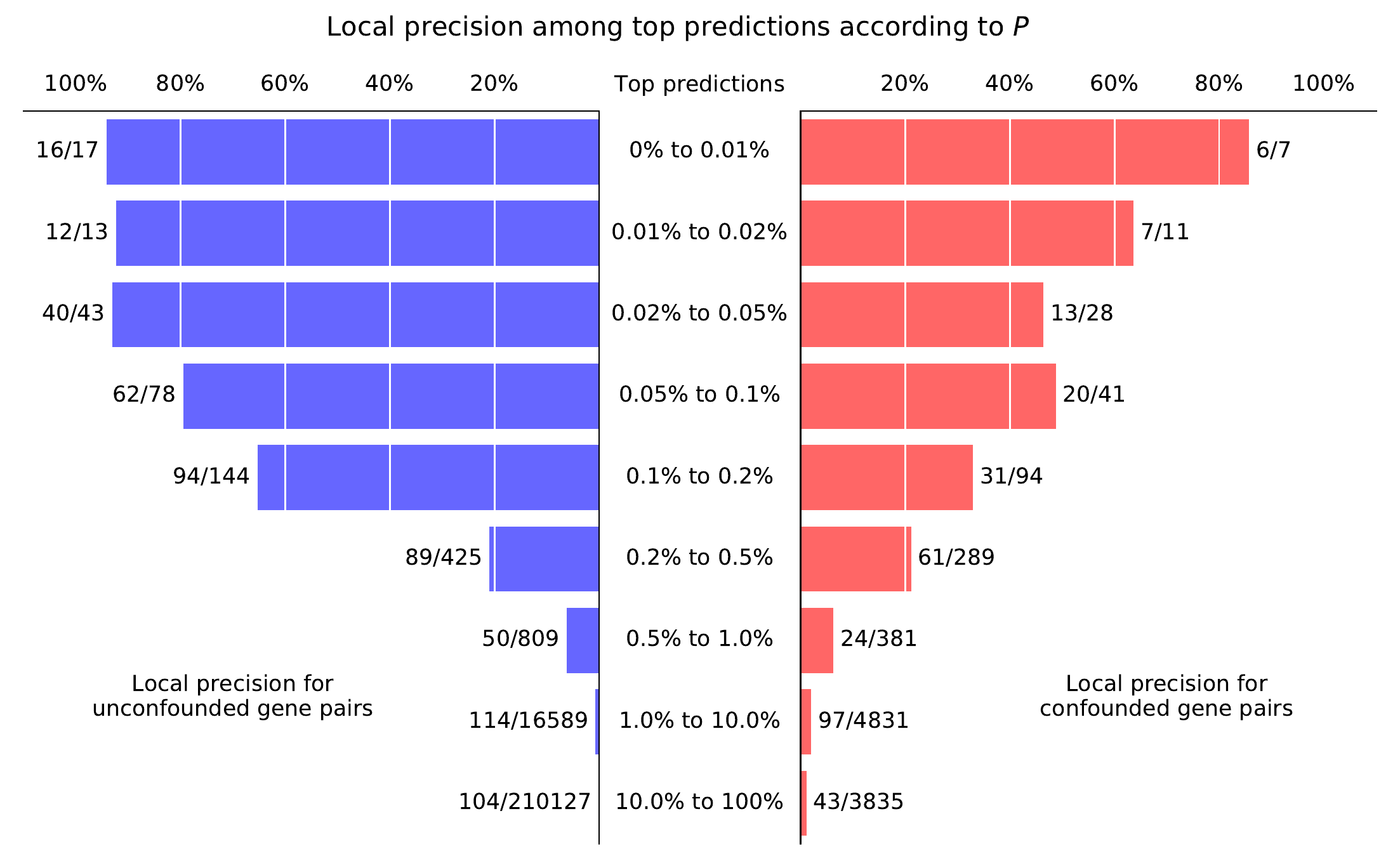}}\\
\subfigure{\includegraphics[width=.4\textwidth]{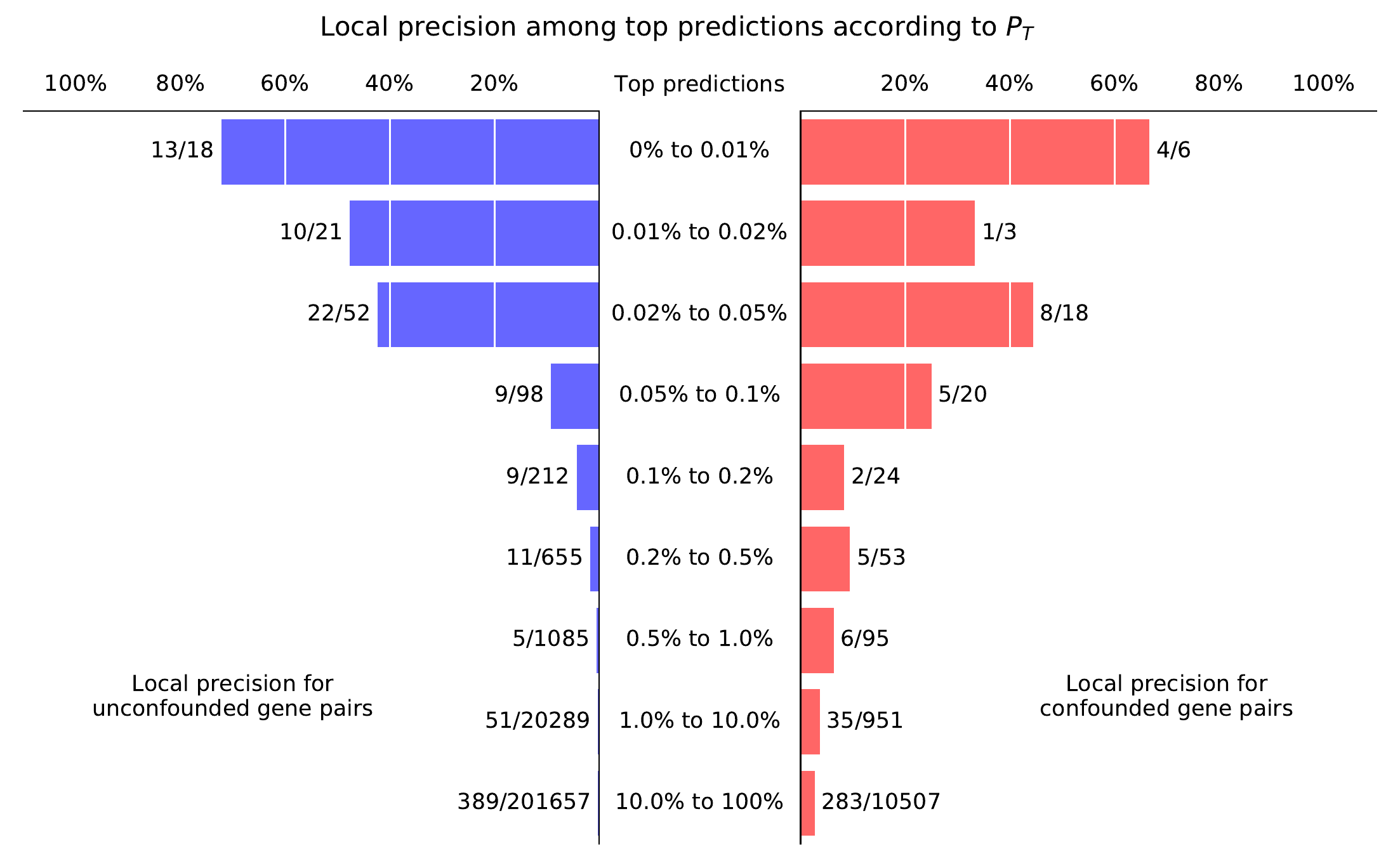}}
\subfigure{\includegraphics[width=.4\textwidth]{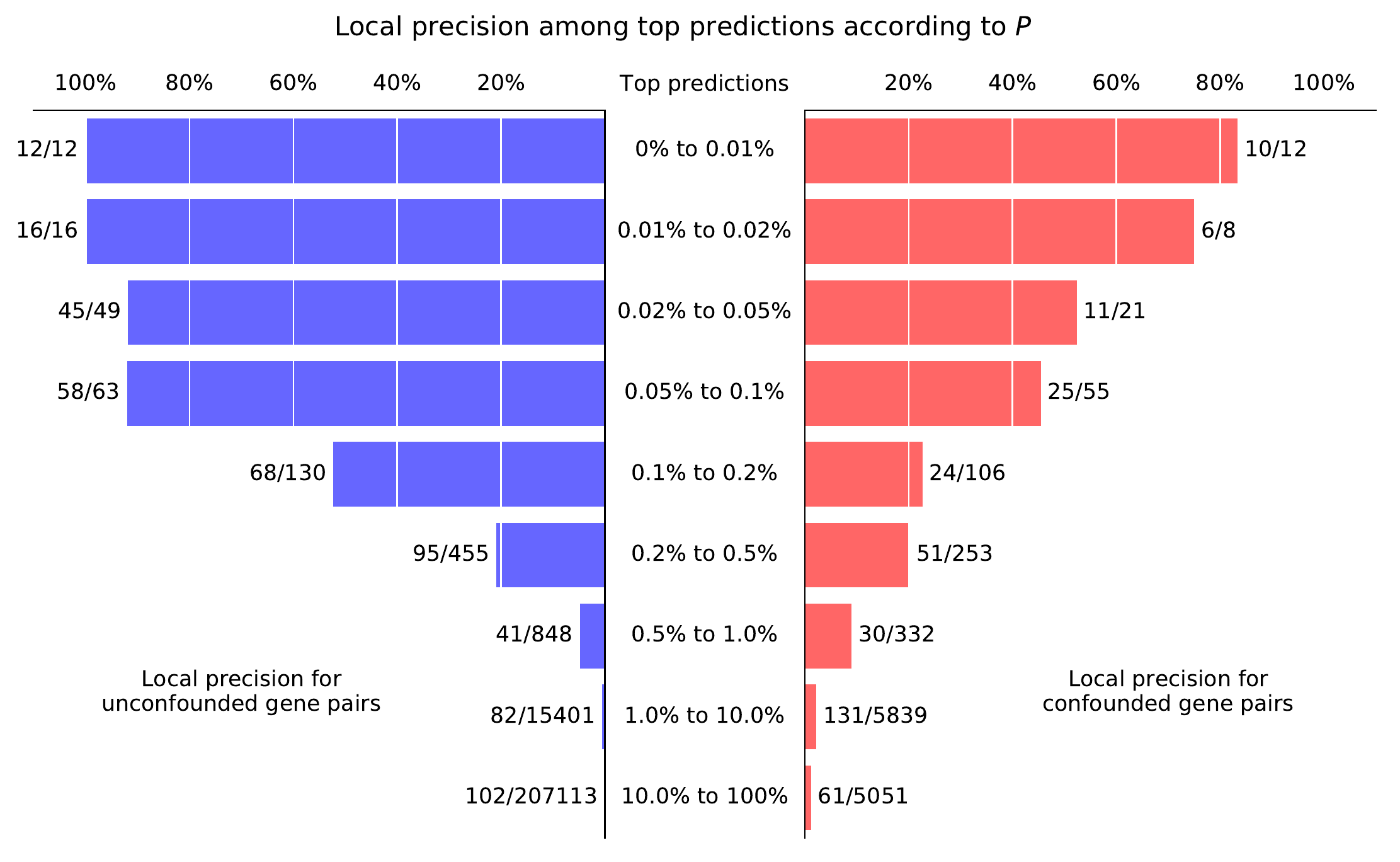}}\\
\subfigure{\includegraphics[width=.4\textwidth]{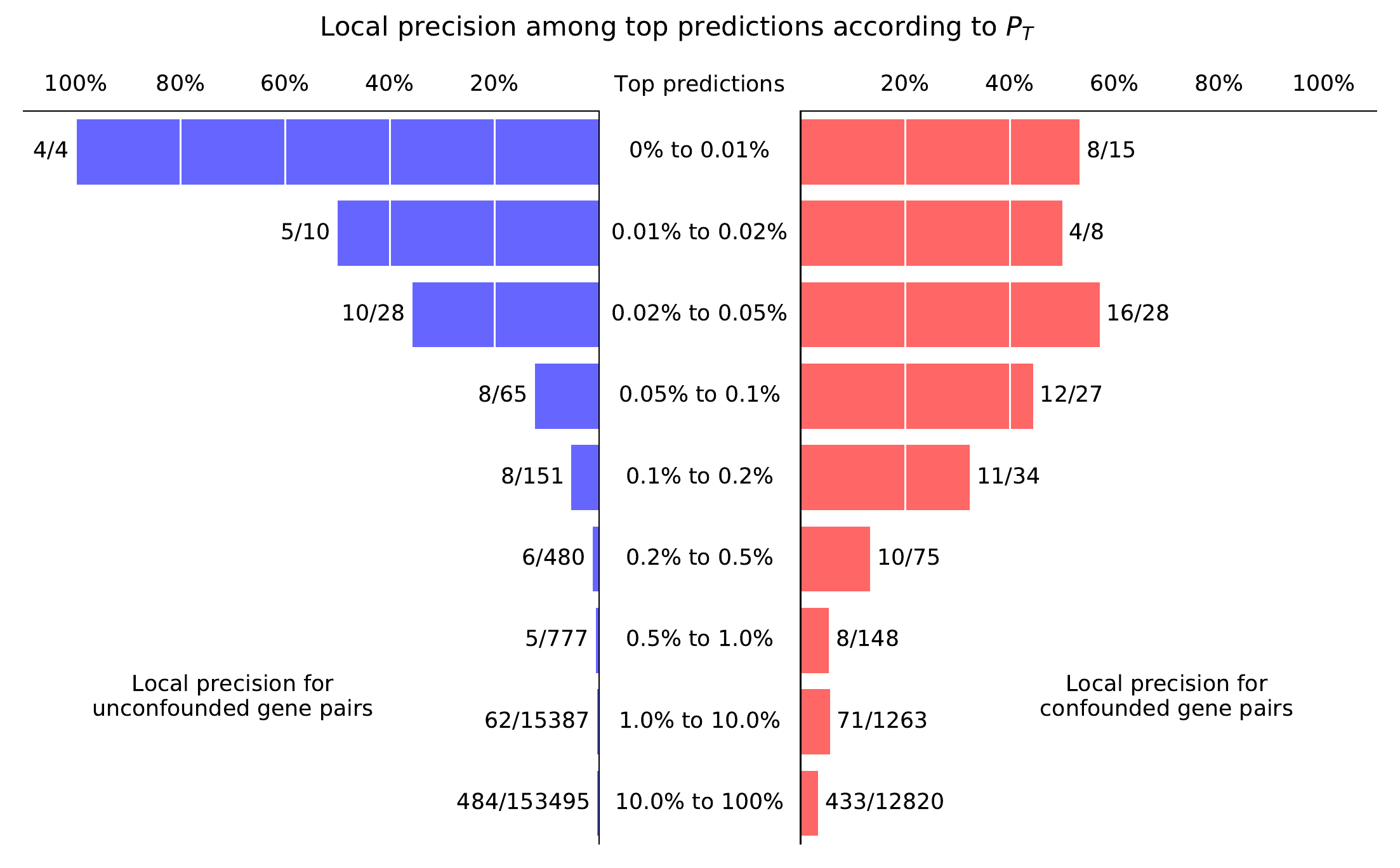}}
\subfigure{\includegraphics[width=.4\textwidth]{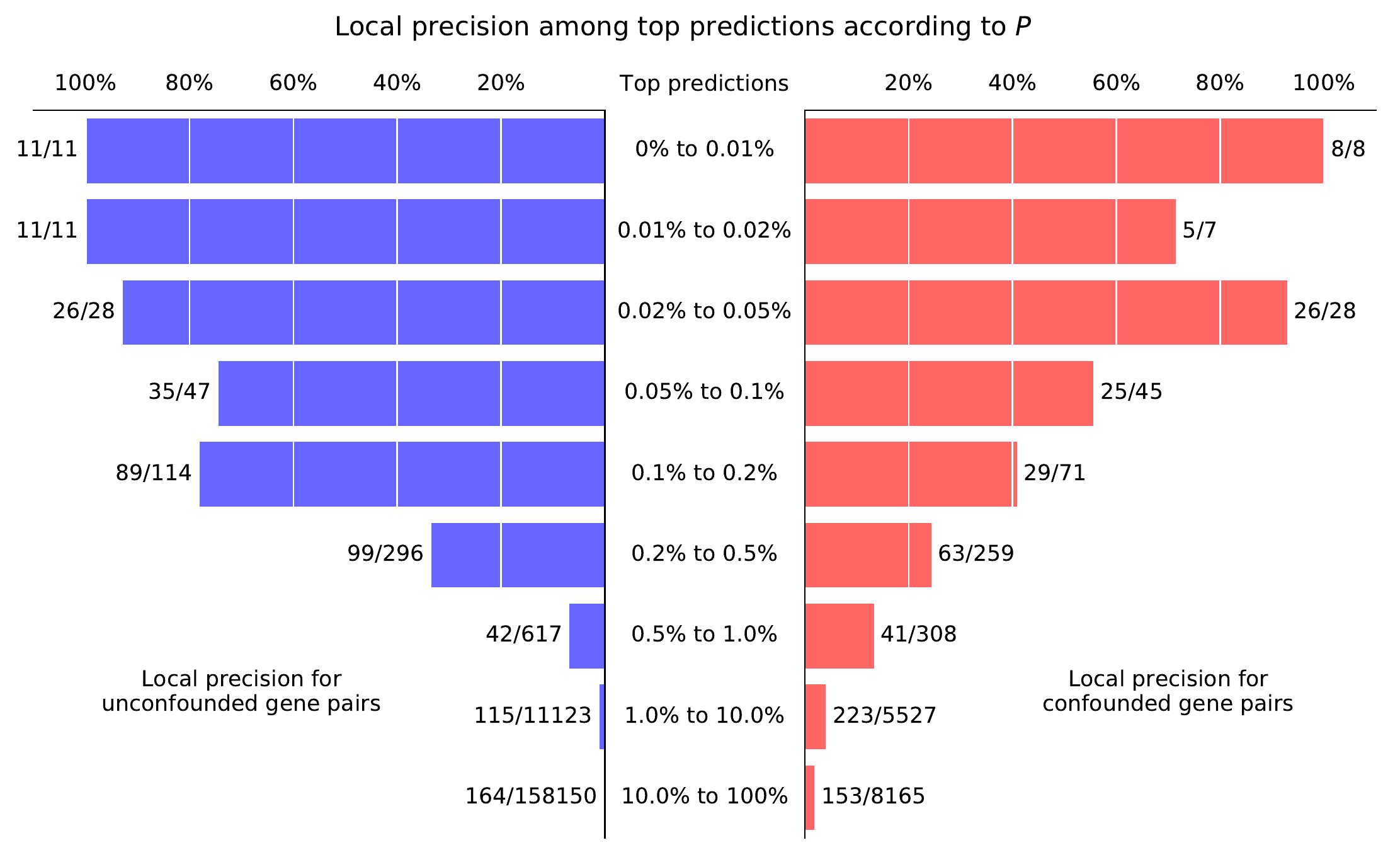}}\\
\end{center}
\caption{\ea{Local precision of top predictions for the traditional (left) and novel (right) tests for datasets (top to bottom) 1, 2, 3, and 5 of the DREAM challenge.}\label{fig-cmpbars}}
\end{figure}

\begin{figure}
\begin{center}
\begin{tabular}{p{0em}p{0.43\linewidth}p{0em}p{0.43\linewidth}}
\vspace{0pt}\textbf{{\large A}} &\vspace{0pt}\includegraphics[width=\linewidth]{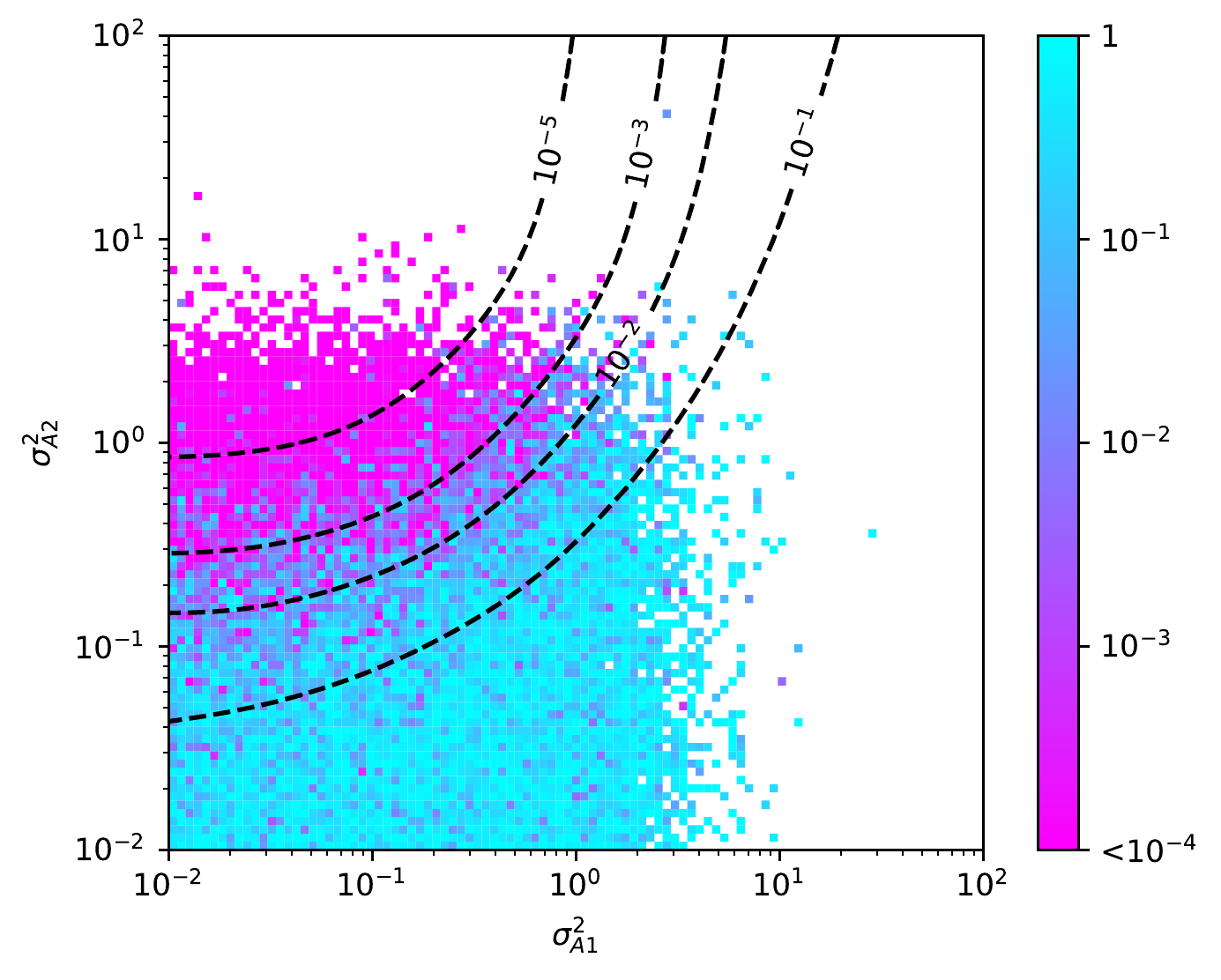}&
\vspace{0pt}\textbf{{\large B}} &\vspace{0pt}\includegraphics[width=\linewidth]{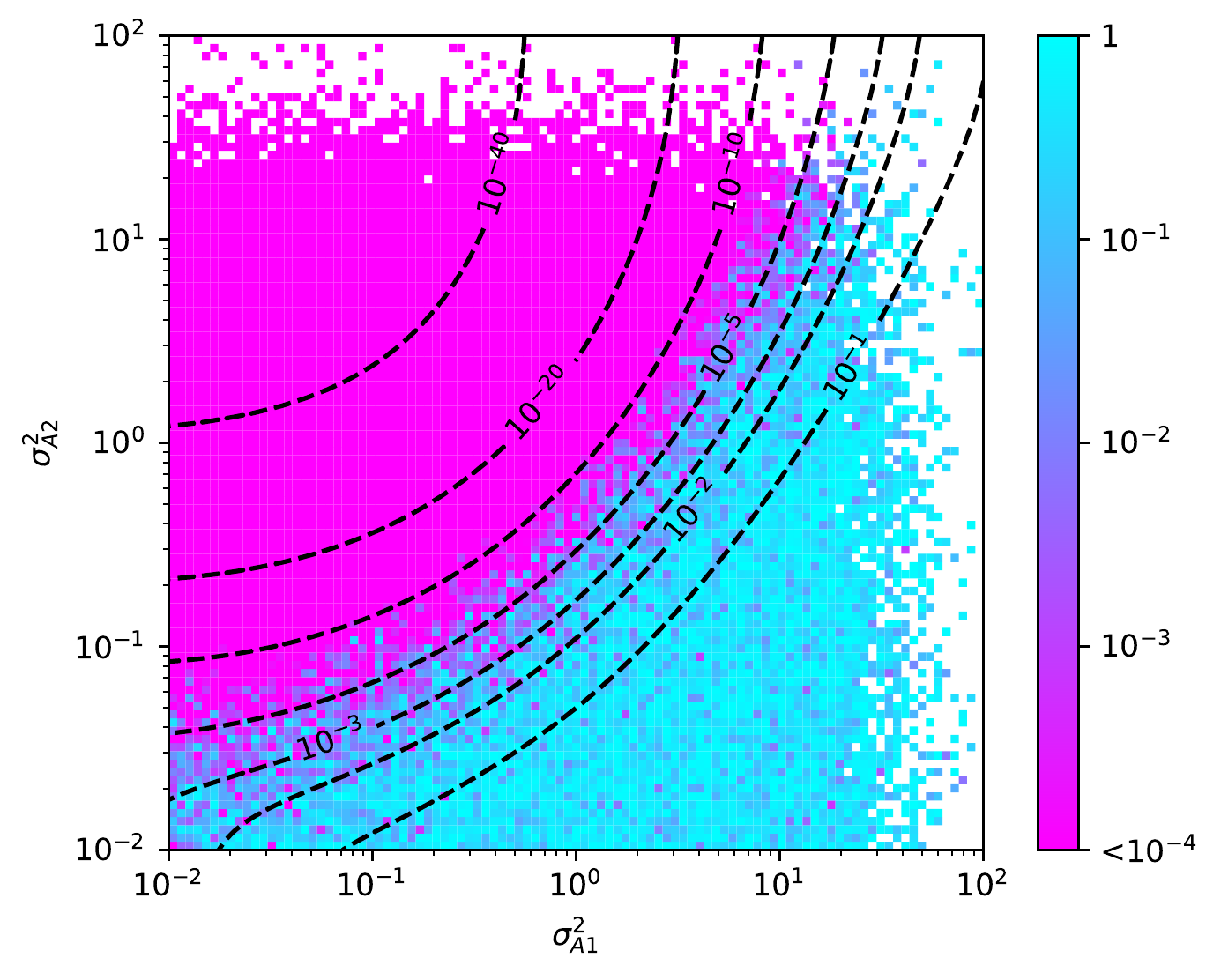}\\
\vspace{0pt}\textbf{{\large C}} &\vspace{0pt}\includegraphics[width=\linewidth]{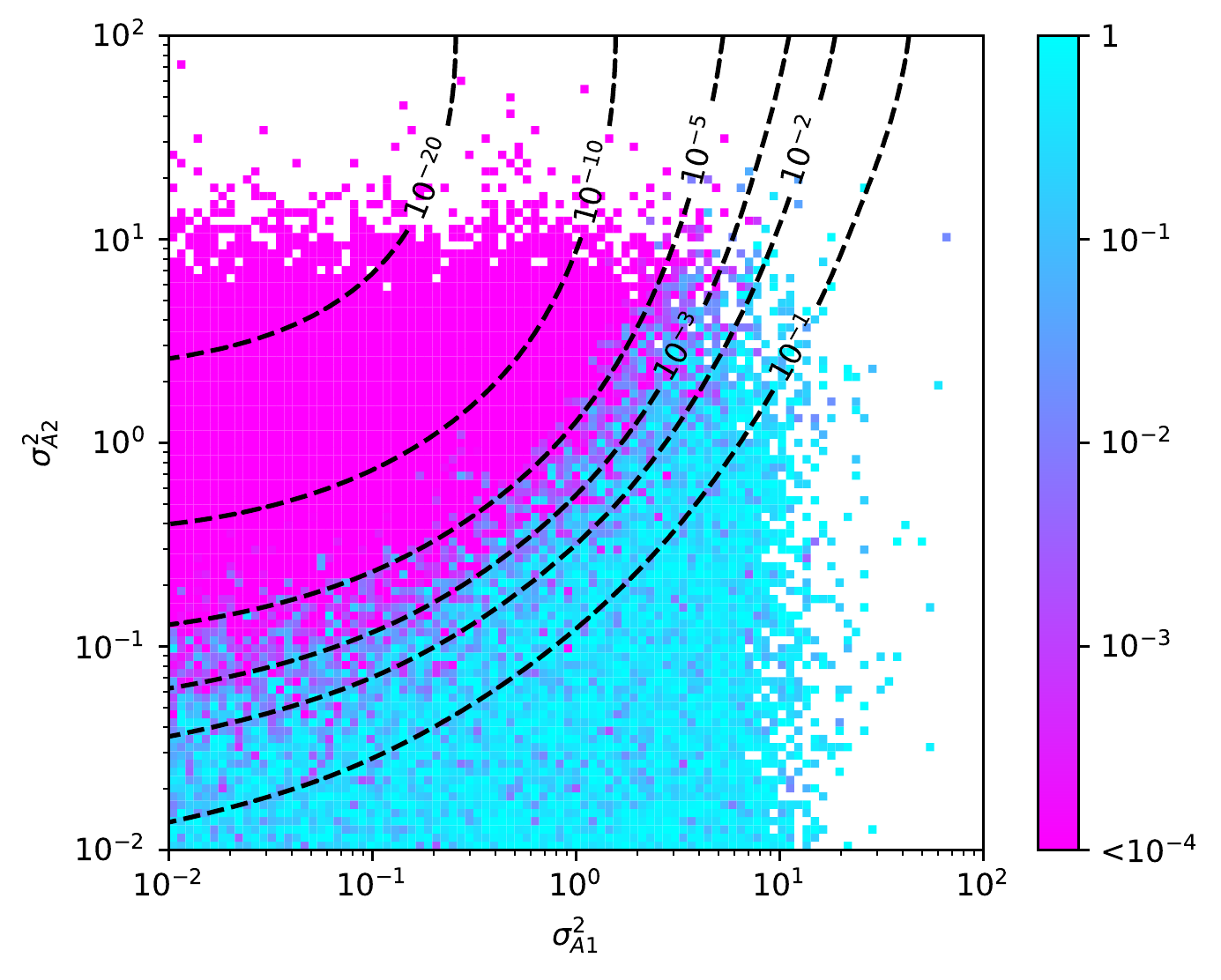}&
\vspace{0pt}\textbf{{\large D}} &\vspace{0pt}\includegraphics[width=\linewidth]{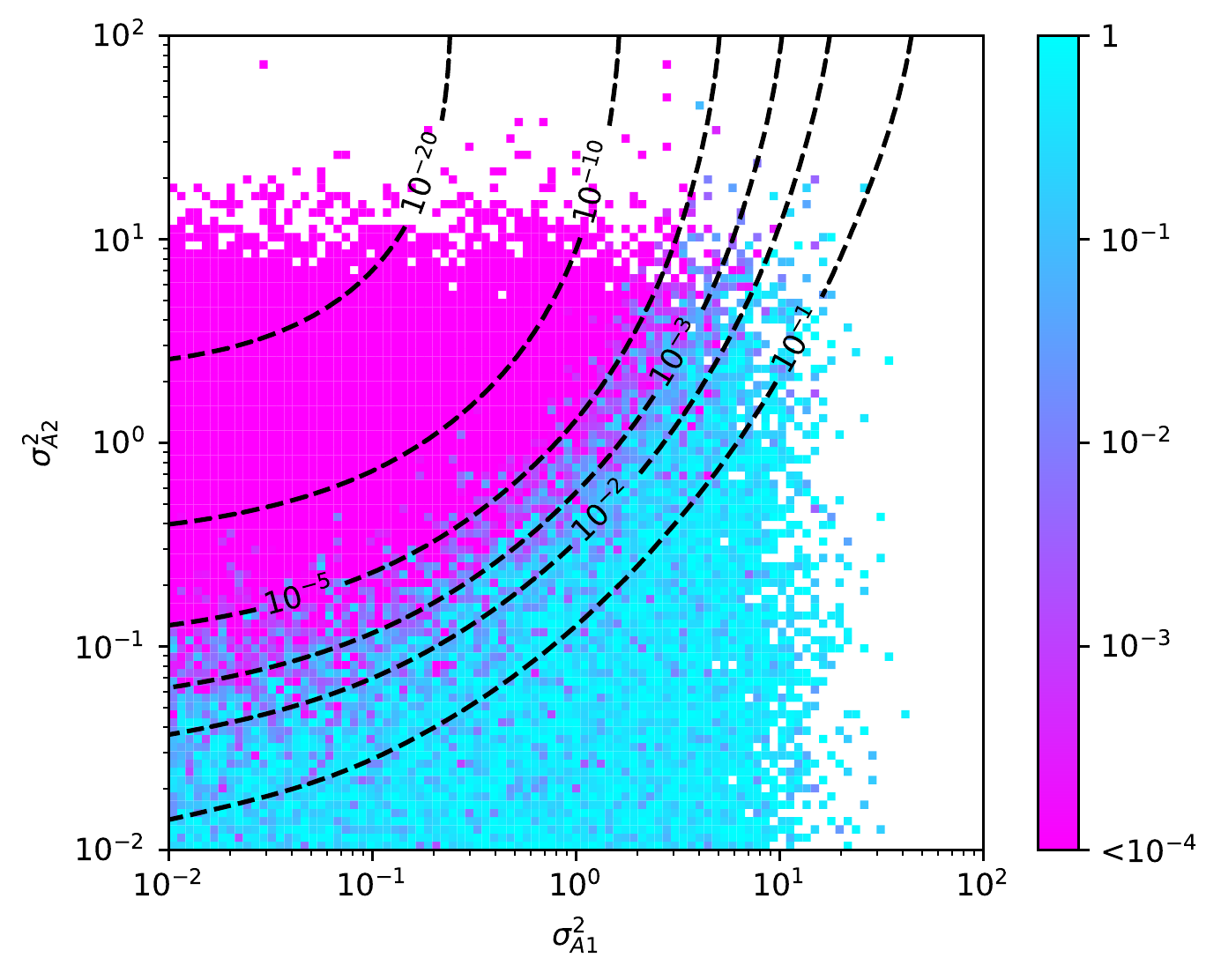}\\
\vspace{0pt}\textbf{{\large E}} &\vspace{0pt}\includegraphics[width=\linewidth]{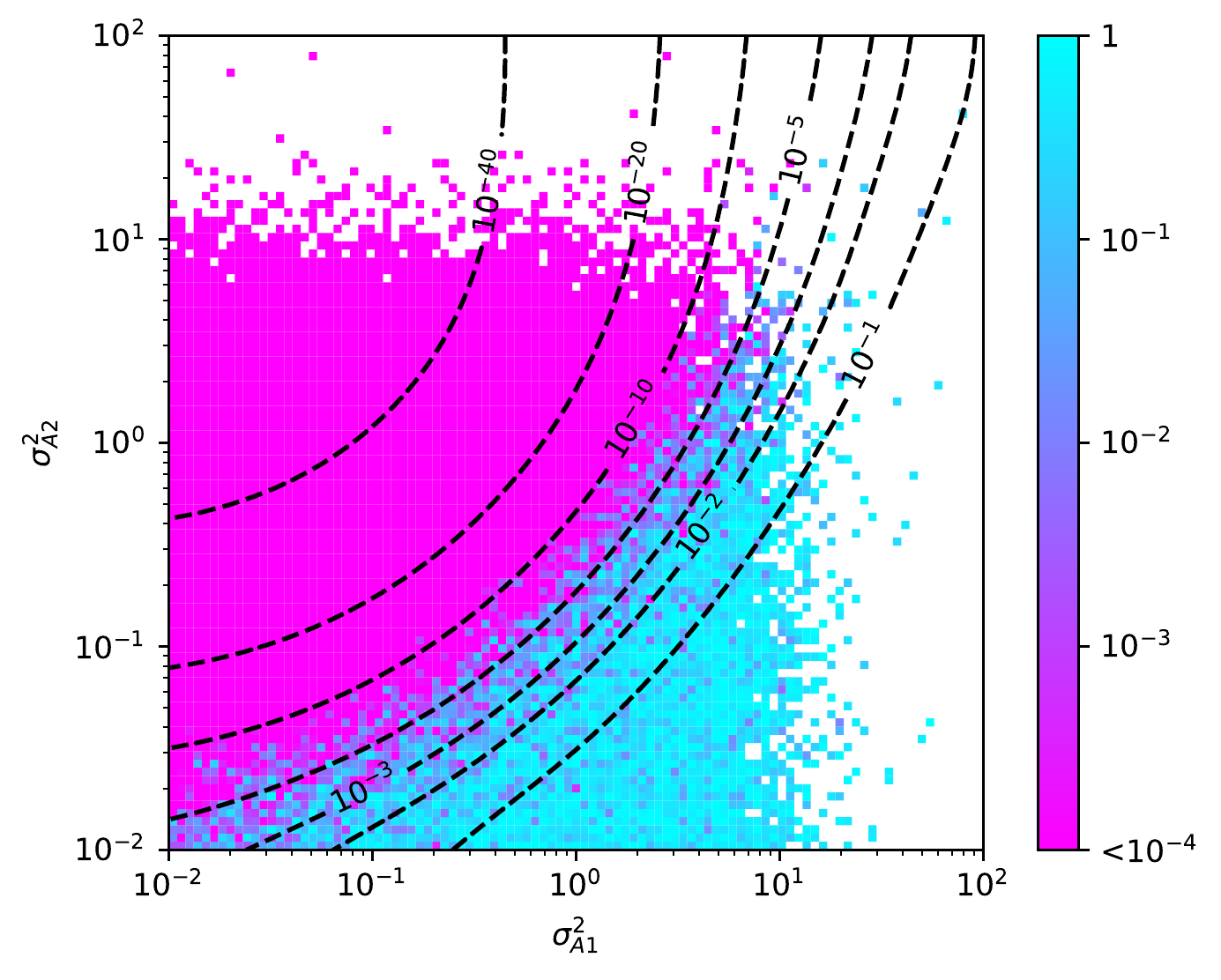}&
\vspace{0pt}\textbf{{\large F}} &\vspace{0pt}\includegraphics[width=\linewidth]{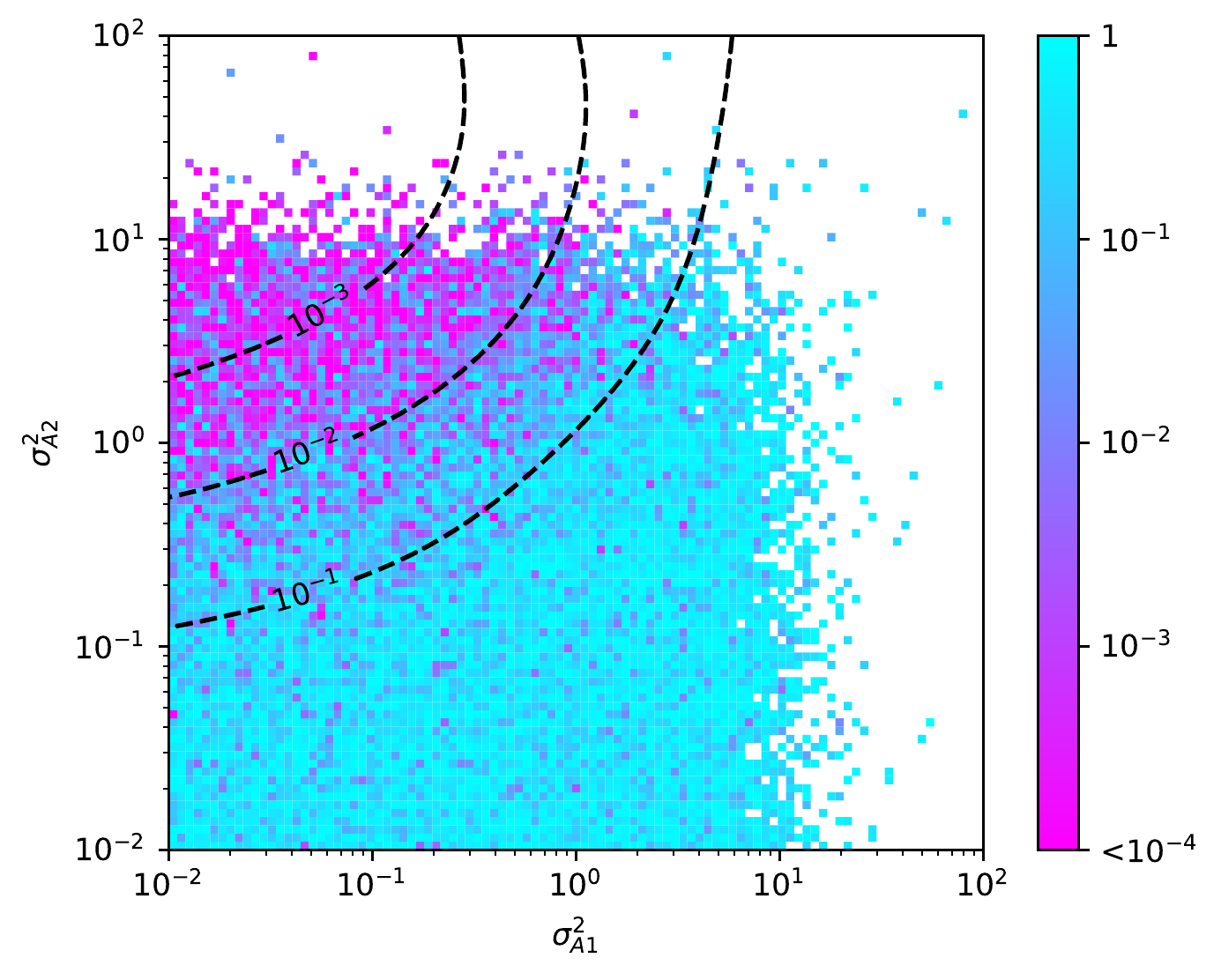}\\
\end{tabular}
\end{center}
\caption{\ea{Null hypothesis p-values of the conditional independence test on simulated data from the ground truth model $E\rightarrow A\sups{(t)}\rightarrow B$ with $A\sups{(t)}\rightarrow A$ under parameter settings other than \refig{sim}. \textbf{(A,B)} 100 \textbf{(A)} or 999 \textbf{(B)} samples. \textbf{(C,D)} Minor allele frequency is $0.05$ \textbf{(C)} or $0.3$ \textbf{(D)}. \textbf{(E,F)} Regarding $B$'s variance from $A\sups{(t)}\rightarrow B$ as unit variance, $B$'s variance from other  sources including measurement errrors is $0.2$ \textbf{(E)} or $20$ \textbf{(F)}. Unmentioned parameters remain the same as in \refig{sim}.}\label{fig-sims}}
\end{figure}

\begin{figure}
\center
\subfigure{\includegraphics[width=\textwidth]{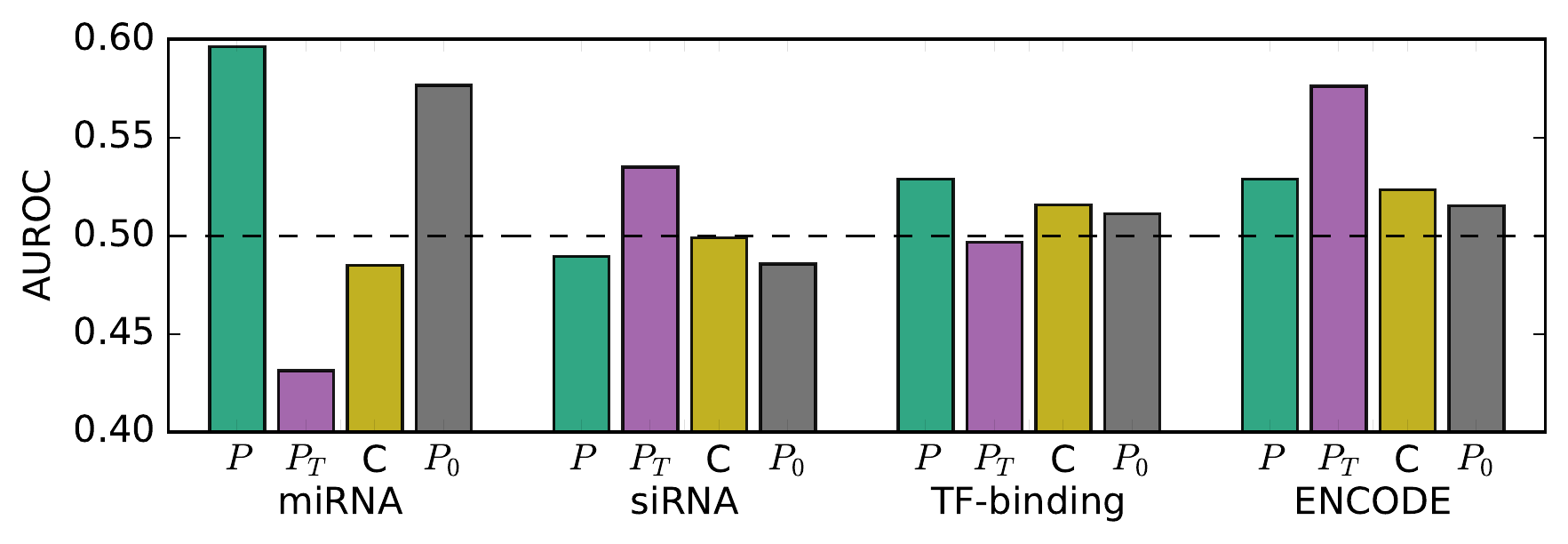}}\\
\subfigure{\includegraphics[width=\textwidth]{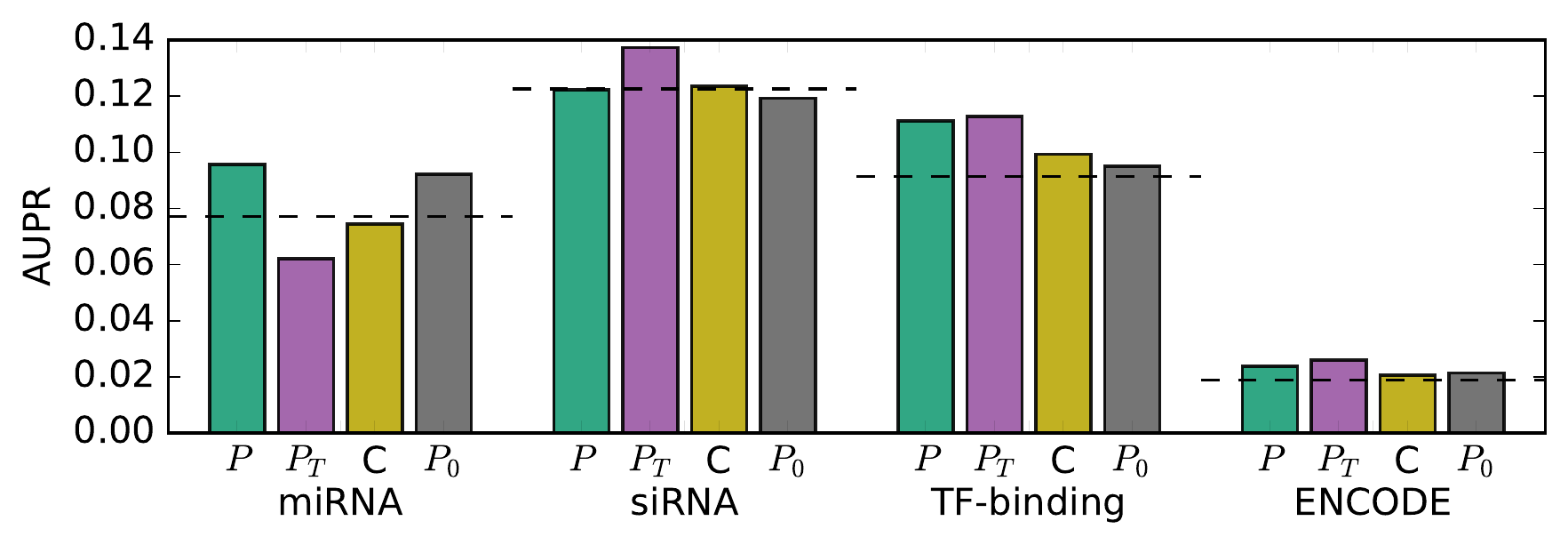}}
\caption{Three methods of causal inference were evaluated and compared against the baseline correlation test method ($P_0$): \pkg's new test ($P$), traditional causal inference test in \pkg{} ($P_T$), and CIT (C). AUROC and AUPR metrics are measured for three inference tasks (\Refssec{datasets}). MiRNA compares miRNA target predictions based on Geuvadis miRNA and mRNA expression levels against groundtruths from miRLAB. SiRNA and TF-binding compares gene-gene interaction predictions based on Geuvadis gene expression levels against groundtruths from siRNA silencing and TF-binding measurements\cite{cusanovich2014functional} respectively. ENCODE compares the same gene-gene interaction predictions against TF-binding networks derived from ENCODE data \cite{gerstein2012architecture}. Dashed lines indicate expected performances from random predictions. \label{fig-bar}}
\end{figure}

\begin{sidewaystable}[h!]
\center
\caption{Predictions from \pkg's new ($P$), traditional ($P_T$), and correlation ($P_0$) tests, and CIT were compared against DREAM challenge leaders on AUROC and AUPR for all 15 DREAM datasets. All cis- and trans-genes are included. DREAM challenge constrained the maximum number of submitted regulations by 100,000, which were also applied in our evaluation. \pkg's new test consistently obtained higher AUROC and AUPR than all other methods, including the leaders of DREAM challenge. \label{tab-dream}}\vspace{1em}
\begin{tabular}{c|ccccc|ccccc|ccccc}
Sample count&\multicolumn{5}{c|}{100 samples}&\multicolumn{5}{c|}{300 samples}&\multicolumn{5}{c}{999 samples}\\
\hline
Test id&1&2&3&4&5&1&2&3&4&5&1&2&3&4&5\\
\hline
$P$ AUROC&\textbf{0.772}&\textbf{0.750}&\textbf{0.737}&\textbf{0.736}&\textbf{0.719}&\textbf{0.882}&\textbf{0.842}&\textbf{0.839}&\textbf{0.825}&\textbf{0.797}&\textbf{0.941}&\textbf{0.899}&\textbf{0.882}&\textbf{0.867}&\textbf{0.848}\\
$P_T$ AUROC&0.617&0.609&0.597&0.574&0.572&0.647&0.594&0.581&0.592&0.573&0.616&0.611&0.570&0.616&0.565\\
$P_0$ AUROC&0.709&0.706&0.706&0.699&0.700&0.843&0.798&0.803&0.792&0.766&0.905&0.870&0.850&0.837&0.813\\
CIT AUROC&0.585&0.582&0.571&0.548&0.569&0.630&0.586&0.575&0.574&0.566&0.614&0.640&0.594&0.614&0.577\\
Leader AUROC&0.754&0.718&0.699&0.694&0.688&0.861&0.793&0.799&0.769&0.757&0.933&0.885&0.845&0.828&0.813\\
\hline
$P$ AUPR&\textbf{0.222}&\textbf{0.183}&\textbf{0.172}&\textbf{0.161}&\textbf{0.155}&\textbf{0.421}&\textbf{0.326}&\textbf{0.279}&\textbf{0.264}&\textbf{0.258}&\textbf{0.547}&\textbf{0.368}&\textbf{0.366}&\textbf{0.342}&\textbf{0.333}\\
$P_T$ AUPR&0.044&0.048&0.042&0.023&0.041&0.109&0.054&0.051&0.054&0.047&0.070&0.068&0.042&0.070&0.049\\
$P_0$ AUPR&0.051&0.049&0.040&0.051&0.055&0.084&0.053&0.063&0.078&0.057&0.093&0.072&0.070&0.076&0.077\\
CIT AUPR&0.075&0.066&0.060&0.031&0.050&0.162&0.075&0.074&0.080&0.067&0.168&0.149&0.098&0.168&0.096\\
Leader AUPR&0.103&0.072&0.067&0.068&0.067&0.309&0.243&0.191&0.182&0.191&0.358&0.258&0.195&0.183&0.178
\end{tabular}
\end{sidewaystable}

\begin{table}
\center
\caption{AUROCs and AUPRs of miRNA target predictions were compared for \pkg's traditional, new, and correlation tests, GENIE3, CIT, and 11 methods in miRLAB, based on Geuvadis data. Higher AUROC and AUPR values signify stronger predictive power. Program running times have units in seconds (s), minutes (m), hours (h), or days (d). \pkg{} outperformed other methods in statistical power and speed, with or without genotype information.\label{tab-Geuvadis}}\vspace{1em}
\begin{tabular}{c|cccccc}
&\pkg-$P$&\pkg-$P_0$&elastic&lasso&genie3&promise\\
\hline
AUROC&0.60&0.58&0.55&0.54&0.53&0.52\\
AUPR&0.096&0.092&0.092&0.092&0.083&0.078\\
Time&0.88s&0.30s&4.53m&4.47m&12.1h&2.37m\\
\hline
\hline
&rdc&cit&spearman&hoeffding&kendall&pearson\\
\hline
AUROC&0.52&0.51&0.51&0.51&0.51&0.50\\
AUPR&0.083&0.080&0.078&0.079&0.078&0.078\\
Time&50.3m&7.5d&2.37m&16.3m&41.9m&2.27m\\
\hline
\hline
&random&zscore&dcov&\pkg-$P_T$&mi\\
\hline
AUROC&0.50&0.46&0.44&0.43&0.41\\
AUPR&0.077&0.068&0.068&0.062&0.068\\
Time&-&2.90m&4.42h&0.84s&23.1m\\
\end{tabular}
\end{table}

\end{document}